\documentclass[aps,nofootinbib]{revtex4}
\usepackage{amsmath,epsfig}
\usepackage{color}
\pdfoutput=1

\begin{document}
\vspace*{-2.5cm}
\vspace*{0.5cm}
\begin{flushright}
  TTK-20-43, P3H-20-078, CAVENDISH-HEP-20/14
\end{flushright}
\vspace{0.cm}
\title{\boldmath   NLO QCD corrections to off-shell
    ${t\bar{t}W^\pm}$   production at the LHC:
    Correlations and Asymmetries}
  \author{Giuseppe Bevilacqua${}^{(a)}$\footnote{e-mail:
      giuseppe.bevilacqua@science.unideb.hu},  Huan-Yu
    Bi${}^{(b)}$\footnote{e-mail: bihy@physik.rwth-aachen.de}, 
    Heribertus Bayu Hartanto${}^{(c)}$\footnote{e-mail:
      hbhartanto@hep.phy.cam.ac.uk}, Manfred
    Kraus${}^{(d)}$\footnote{e-mail: mkraus@hep.fsu.edu},
    Jasmina Nasufi${}^{(b)}$\footnote{e-mail:
      jasmina.nasufi@rwth-aachen.de} and Malgorzata
    Worek${}^{(b)}$\footnote{e-mail: worek@physik.rwth-aachen.de
      (corresponding author)}
}                     
%
%
\affiliation{{\phantom.}\\
  $^{(a)}$ \mbox{MTA-DE Particle Physics Research Group, University of
  Debrecen, H-4010 Debrecen, PBox 105, Hungary}\\
$^{(b)}$ \mbox{Institute for Theoretical Particle Physics and
  Cosmology, RWTH Aachen University, D-52056 Aachen, Germany}\\
$^{(c)}$ \mbox{Cavendish Laboratory, University of Cambridge,
J.J. Thomson Avenue, Cambridge CB3 0HE, United Kingdom}\\
$^{(d)}$ \mbox{Physics
Department, Florida State University, Tallahassee, FL 32306-4350, USA}
}

\begin{abstract}
Recent discrepancies between theoretical predictions and experimental
data in multi-lepton plus $b$-jets analyses for the $t\bar{t}W^\pm$
process, as reported by the ATLAS collaboration, have indicated that
more accurate theoretical predictions and high precision observables
are needed to constrain numerous new physics scenarios in this
channel. To this end we employ NLO QCD computations with full
off-shell top quark effects included to provide theoretical
predictions for the ${\cal R}=
\sigma_{t\bar{t}W^+}/\sigma_{t\bar{t}W^-}$ cross section ratio at the
LHC with $\sqrt{s}=13$ TeV. Depending on the transverse momentum cut
on the $b$-jet we obtain $2\% -3 \%$ theoretical precision on ${\cal
R}$, which should help to shed some light on new physics effects that
can reveal themselves only once sufficiently precise Standard Model
theoretical predictions are available. Furthermore, triggered by these
discrepancies we reexamine the charge asymmetry of the top quark and
its decay products in the $t\bar{t}W^\pm$ production process. In the
case of charge asymmetries, that are uniquely sensitive to the chiral
nature of possible new physics in this channel, theoretical
uncertainties below $15\%$ are obtained.  Additionally, the impact of
the top quark decay modelling is scrutinised by  explicit
comparison with predictions in the narrow-width approximation.
\end{abstract}
\maketitle
%
%
\section{Introduction}
\label{sec:intro}
%

The Large Hadron Collider (LHC) with the Run II energy of
$\sqrt{s}=13$ TeV has opened up the possibility of studying various
top quark production and decay mechanisms at larger mass scales than
previously explored in any experiment.  The $t\bar{t}$
pair production associated with the $W^\pm$ gauge boson is among the
most interesting signatures that can be studied with high precision at
the LHC. It is a key process to constrain top quark intrinsic
properties, which might be modified in the presence of new
physics. Moreover, the process can be used in the framework of the
Standard Model Effective Field Theory (SMEFT), where the effects of
potential new particles can be systematically included in terms of
higher-dimensional operators, see
  e.g. \cite{Hagiwara:1993ck,Giudice:2007fh,Grzadkowski:2010es,
Corbett:2012ja,Brivio:2017vri}. The latter are   suppressed by a  
sufficiently large new physics energy scale $\Lambda$. The framework
relies on the idea that new physics is too heavy to be directly
produced and observed at the LHC, thus, only deviations from the
Standard Model (SM) can be probed in various ATLAS and CMS top quark
measurements.  Compared with top quark pair production and single top
quark production, the associated $t\bar{t}W^\pm$ process does not
bring sensitivity to new operators, however, it helps to resolve blind
directions in the SMEFT parameter space that occur in the current LHC
fits. On top of that $t\bar{t}W^\pm$ can probe operators that are
difficult to access in other channels. For example, since the $W^\pm$
gauge boson is radiated from the initial state, $t\bar{t}W^\pm$ is
sensitive to a subset of the possible four-quark operators only. In
the SM, $t\bar{t}W^\pm$ is dominated by quark-antiquark interactions,
while $t\bar{t}$ is dominated by the $gg$ initial state. This means
that relative to the SM contribution the four-quark operators would
give sizeable effects in the $t\bar{t}W^\pm$ production
process. Consequently, $t\bar{t}W^\pm$ production is often included in
the global SMEFT analysis of LHC top quark measurements, see
e.g. Ref.~\cite{Brivio:2019ius}.

In addition, the $t\bar{t}W^\pm$ process plays an important role in 
studies of the top quark charge asymmetry denoted as $A_c^t$
\cite{Maltoni:2014zpa,Maltoni:2015ena}. Also in this case the lack of
the symmetric $gg$ initial state and the emission of the $W^\pm$ gauge
boson from the initial states contribute to a substantially larger top
quark charge asymmetry than that measured in the $t\bar{t}$
process. Furthermore, the asymmetry of the top quark decay products,
i.e the charged lepton $(A_c^\ell)$ and the $b$-jet $(A_c^b)$ are very
large and already present at the LO due to the polarisation of the
initial fermionic line by the $W^\pm$ emission. These asymmetries are
an interesting playground for various beyond the SM (BSM) theories, as
$A_c^t$, $A_c^\ell$ and $A_c^b$ are uniquely sensitive to the chiral
nature of possible new physics that might directly affect such
measurements.

Last but not least, $t\bar{t} W^\pm$ production is a background
process in the multi-lepton final state with two same-sign leptons,
accompanied by missing transverse momentum and $b$-jets
\cite{Aad:2016tuk,Khachatryan:2016kod,Sirunyan:2017uyt,Aaboud:2017dmy}.
Even though same-sign leptons are a relatively rare phenomenon in the
SM, as they only appear in processes with a rather small cross
section, they have been extensively exploited in various models of new
physics. The same-sign lepton signature is present, among others, in
models with supersymmetry, universal extra dimensions, top-quark
partners and an extended Higgs boson sector
\cite{Barnett:1993ea,Guchait:1994zk,Baer:1995va,Maalampi:2002vx,
Cheng:2002ab,Dreiner:2006sv,Contino:2008hi,vonBuddenbrock:2016rmr,
vonBuddenbrock:2017gvy,vonBuddenbrock:2018xar,
vonBuddenbrock:2019ajh}. Besides, same-sign leptons are considered a
key feature in searches for heavy Majorana neutrinos as well as for
$tt$ and $\bar{t}\bar{t}$ resonances \cite{Almeida:1997em,Bai:2008sk}.

Finally, the $pp\to t\bar{t}W^\pm$ process is the main background in
SM measurements involving final states with multiple leptons and
$b$-jets.  This is the case, for example, for the measurement of the
associated production of the SM Higgs boson with top quarks
\cite{ATLAS:2019nvo}. The $pp\to t\bar{t}W^\pm$ process has also
played a crucial role in the announcement of strong evidence for the
production of four top quarks, an analysis, which has been
recently performed by the ATLAS Collaboration \cite{Aad:2020klt}.

The direct measurement of $pp\to t\bar{t}W^\pm$ production in
multi-lepton final states has already been carried out at
$\sqrt{s}=13$ TeV by the ATLAS and CMS collaborations
\cite{Aaboud:2016xve,Sirunyan:2017uzs,Aaboud:2019njj}. In the recent
measurements of $t\bar{t}H$ and $t\bar{t}W^\pm$ production in
multi-lepton final states \cite{ATLAS:2019nvo} the resulting
$t\bar{t}W^\pm$ normalisation has been found to be higher than the
theoretical prediction provided by multipurpose Monte Carlo (MC)
generators, which are currently employed by the ATLAS
collaboration. Apart from the $t\bar{t}W^\pm$ normalisation, a tension
in the modelling of the final state kinematics in the phase space
regions dominated by $t\bar{t}W^\pm$ production, has been
observed. From the experimental point of view such an accurate study of
$pp\to t\bar{t}W^\pm$ production in the same-sign lepton final state
has become feasible thanks to the increasing amount of data collected
at the LHC with $\sqrt{s}=13$ TeV. This increased integrated
luminosity has significantly raised the need for more precise
theoretical predictions.  The latter should include higher order QCD
corrections both to the production and decays of top quarks and $W$
gauge bosons as well as $t\bar{t}$ spin correlations at
the same level of accuracy.

The first calculations for the $pp\to t\bar{t}W^\pm$ process, that
meet the mentioned conditions, have been carried out in the
narrow-width approximation (NWA) within the \textsc{Mcfm} framework
\cite{Campbell:2012dh}. The first full NLO QCD computations, which
include complete top quark off-shell effects for the $pp\to
t\bar{t}W^\pm$ process in the multi-lepton channel, have been recently 
presented  in Ref.~\cite{Bevilacqua:2020pzy}. In these
computations, obtained with the help of \textsc{Helac-NLO}, off-shell
top quarks have been described by Breit-Wigner propagators,
furthermore, double-, single- as well as non-resonant top-quark
contributions along with all interference effects have been
consistently incorporated at the matrix element level. Independent
computations for $t\bar{t}W^+$ production have been obtained very
recently within the \textsc{MoCaNLO+Recola} framework
\cite{Denner:2020hgg}. They not only confirmed the results presented
in Ref.~\cite{Bevilacqua:2020pzy} but also performed a comparison
between the full results and those obtained with the help of the
double-pole approximation.

In Ref.~\cite{Bevilacqua:2020pzy} results at NLO QCD accuracy have
been presented in the form of fiducial integrated and differential
cross sections for two selected renormalisation and factorisation
scale choices (a fixed and a dynamical one) and three different PDF
sets. Detailed studies of the scale dependence of the NLO predictions
have been carried out together with calculations of PDF uncertainties.
Furthermore, the impact of the top quark off-shell effects on the
$pp\to t\bar{t}W^\pm$ cross section has been examined by an explicit
comparison with the results in the NWA. In the current paper we will
move away from the technical aspects of higher order calculations and
the estimation of the residual theoretical uncertainties and go
towards more phenomenological studies for the $pp\to t\bar{t}W^\pm$
process. Specifically, the purpose of this paper is twofold. First, we
would like to provide a systematic analysis of the two processes
$pp\to t\bar{t} W^+$ and $pp \to t\bar{t}W^-$ in the multi-lepton
decay channel to extract the most accurate NLO QCD predictions for the
${\cal R}=\sigma^{\rm NLO}_{t\bar{t}W^+}/\sigma^{\rm
NLO}_{t\bar{t}W^-}$ cross section ratio. Generally, cross section
ratios are more stable against radiative corrections than  absolute
cross sections, assuming that the two processes are correlated. They
have smaller theoretical uncertainties as various uncertainties tend
to cancel in a cross section ratio. Consequently, such precise
theoretical predictions have enhanced predictive power and should be
used in indirect searches for new physics at the LHC. Let us add here,
that the ${\cal R}=\sigma^{\rm
NLO}_{t\bar{t}W^+}/\sigma^{\rm NLO}_{t\bar{t}W^-}$ cross section ratio
has recently been studied in Ref. \cite{Frederix:2020jzp} in the
context of parton shower. The NLO QCD and subleading electroweak
corrections for the $t\bar{t}W^\pm$ process were matched, using the
\textsc{MC@NLO} matching scheme
\cite{Frixione:2002ik,Frixione:2003ei}, to the parton
shower using the PYTHIA8 framework \cite{Sjostrand:2014zea} and the
\textsc{MadGraph5}${}_{-}$\textsc{aMC@NLO} system
\cite{Alwall:2014hca,Frederix:2018nkq}.  The top
quark and the $W^\pm$ gauge boson decays were realised within the
\textsc{MadSpin} framework \cite{Artoisenet:2012st} in order to fully
keep the LO spin correlations.  The ${\cal R}$ ratio has been
considered for the two following signatures: the same sign di-lepton
and the multi lepton channels.  The scale uncertainties
were taken to be correlated, although, correlations were not studied
in any detail.  In addition, the PDF uncertainties were
not discussed at all.

The second goal
of the paper is to study separately the intrinsic properties of
$t\bar{t}W^+$ and $t\bar{t}W^-$ production. More specifically, we
shall use state-of-the-art NLO QCD theoretical predictions for the
$t\bar{t}W^\pm$ process to re-examine the top quark charge asymmetry
and asymmetries of the top quark decay products both at the integrated
and differential level. Likewise, in this case, the polarisation and
asymmetry effects in the $pp \to t\bar{t}W^\pm$ production process can
be employed to constrain new physics effects that might occur in this
channel.  Furthermore, for both the cross section ratio and the top
quark (decay products) charge asymmetry, the impact of the modelling
of top quark production and decays will be studied.

We note here, that state-of-the art theoretical predictions at NLO
in QCD with complete top quark off-shell effects included are also
available for other processes at the LHC.  Such effects, for example,
have been incorporated for $pp\to t\bar{t}$
\cite{Denner:2010jp,Bevilacqua:2010qb,Denner:2012yc,Denner:2017kzu},
$pp\to t\bar{t}j$ \cite{Bevilacqua:2015qha,Bevilacqua:2016jfk}, $pp\to
t\bar{t}H$ \cite{Denner:2015yca}, $pp\to t\bar{t}\gamma$
\cite{Bevilacqua:2018woc} and for $t\bar{t}Z(Z\to \nu_\ell\nu_\ell)$
\cite{Bevilacqua:2019cvp}. They have also been
incorporated for the $pp\to t\bar{t}b\bar{b}$ process
\cite{Denner:2020orv}. We additionally add, that
continuous efforts have been devoted to improve the theoretical
modeling of hadronic observables for $t\bar{t}W^\pm$ at NLO through
matching with parton shower and multi-jet merging
\cite{Garzelli:2012bn,Frederix:2020jzp,vonBuddenbrock:2020ter,
Cordero:2021iau}. A further step towards a more precise modelling of
on-shell $t\bar{t}W^\pm$ production with stable top quarks and $W^\pm$
gauge boson has been achieved by including either NLO electroweak
corrections \cite{Frixione:2015zaa} and the subleading electroweak
corrections \cite{Dror:2015nkp,Frederix:2017wme} or by incorporating
soft gluon resummation effects with next-to-next-to-leading
logarithmic accuracy \cite{Li:2014ula,Broggio:2016zgg,Kulesza:2018tqz,
  Broggio:2019ewu,Kulesza:2020nfh}.  Very recently, NLO QCD and
electroweak corrections to the full off-shell $t\bar{t}W^+$ production
at the LHC have been combined for the first time for the
three-charged-lepton channel \cite{Denner:2021hqi}.

The paper is organised as follows. In section \ref{sec:setup} the
\textsc{Helac-NLO} computational framework and input parameters used
in our studies are briefly described. In section \ref{sec:correl} correlations
between $t\bar{t}W^+$ and $t\bar{t}W^-$ are examined.  The results for
the cross section ratio ${\cal R}=\sigma^{\rm
NLO}_{t\bar{t}W^+}/\sigma^{\rm NLO}_{t\bar{t}W^-}$ are provided in
section \ref{sec:ratio}.  The integrated top quark charge asymmetry as
well as asymmetries of the top quark decay products are studied in
section \ref{sec:assym}. Results for the differential and cumulative
$A_c^\ell$ asymmetry  are provided in section
\ref{sec:diff}. In section \ref{sec:summary} the results are
summarised and our conclusions are provided. Finally, in  appendix
\ref{appendix} we discuss the Kolmogorov-Smirnov  test that may be
used to provide a quantitative measure of similarity between the two
given processes.

%
\section{Computational Framework and Input Parameters}
\label{sec:setup}
%

All our results both for the full off-shell and NWA computations have
been obtained with the help of the \textsc{Helac-NLO} Monte Carlo
framework \cite{Bevilacqua:2011xh}. The calculation was performed
using \textsc{Helac-1Loop} \cite{vanHameren:2009dr,Ossola:2007ax} for
the virtual corrections and \textsc{Helac-Dipoles}
\cite{Czakon:2009ss,Bevilacqua:2013iha} for the real emission
part. The integration over the phase space has been achieved with the
help of \textsc{Kaleu} \cite{vanHameren:2010gg}. In our studies we
keep the Cabibbo-Kobayashi-Maskawa mixing matrix diagonal and neglect
the Higgs boson contributions. Following recommendations of the
PDF4LHC Working Group for the usage of PDFs suitable for applications
at the LHC Run II \cite{Butterworth:2015oua} we employ the NNPDF3.0
PDF set \cite{Ball:2014uwa}. In particular, we use
\texttt{NNPDF30-nlo-as-0118} with $\alpha_s(m_Z) = 0.118$
(\texttt{NNPDF30-lo-as-0130} with $\alpha_s(m_Z) = 0.130$) at NLO
(LO). The running of the strong coupling constant $\alpha_s$ with
two-loop accuracy at NLO is provided by the LHAPDF interface
\cite{Buckley:2014ana}. The number of active flavours is set to $N_F =
5$. We employ the following SM input parameters
\begin{equation}
\begin{array}{lll}
 G_{ \mu}=1.166378 \cdot 10^{-5} ~{\rm GeV}^{-2}\,, & \quad \quad
                                                      \quad\quad
&   m_{t}=172.5 ~{\rm GeV} \,,
\vspace{0.2cm}\\
 m_{W}=80.385 ~{\rm GeV} \,, &
&\Gamma^{\rm NLO}_{W} = 2.09767 ~{\rm GeV}\,, 
\vspace{0.2cm}\\
  m_{Z}=91.1876  ~{\rm GeV} \,, &
  &\Gamma^{\rm NLO}_{Z} = 2.50775 ~{\rm GeV}\,, \vspace{0.2cm}\\
  \Gamma_{t}^{\rm NLO} = 1.33247  ~{\rm GeV} \,,
&&\Gamma^{\rm NLO}_{t,{\rm NWA}} = 1.35355 ~{\rm GeV}\,.
\end{array}
\end{equation}
For the $W$ and $Z$ gauge boson widths we use the NLO QCD values as
calculated for $\mu_R=m_W$ and $\mu_R=m_Z$ respectively. All other
partons, including bottom quarks, as well as leptons are treated as massless
particles. The LO and NLO top quark widths are calculated according to
Ref.~\cite{Denner:2012yc}. The top quark width is treated
as a fixed parameter throughout this work. Its value corresponds to a
fixed scale $\mu_R=m_t$.  The electromagnetic coupling $\alpha$ is
calculated from the Fermi constant $G_\mu$, i.e. in the
$G_\mu-$scheme, via
\begin{equation}
\alpha_{G_\mu}=\frac{\sqrt{2}}{\pi} \,G_\mu \,m_W^2  \,\sin^2\theta_W
\,,
\end{equation}
where $\sin^2\theta_W$ is defined according to  
\begin{equation}
\sin^2\theta_W = 1-\frac{m_W^2}{m_Z^2}\,.
\end{equation}
We use kinematic-dependent factorisation and renormalisation
scales $\mu_R = \mu_F = \mu_0$ with the central value $\mu_0=H_T/3$
where $H_T$ is the scalar sum of all transverse momenta in the event,
including the missing transverse momentum.  The latter is constructed
from the three neutrinos $\nu_e$, $\nu_e$ and $\nu_\mu$. The
additional light jet, if resolved, is not included in the definition
of $H_T$.  In various comparisons we also use a fixed scale defined as
$\mu_0=m_t+m_W/2$. Jets are constructed out of all final-state partons
with pseudo-rapidity $|\eta| <5$ via the {\it anti}$-k_T$ jet
algorithm \cite{Cacciari:2008gp} with the separation parameter
$R=0.4$.  We require exactly two $b$-jets and three charged leptons,
two of which are same-sign leptons. All final states have to fulfil
the following selection criteria that mimic very closely the ATLAS
detector response \cite{ATLAS:2019nvo}
\begin{equation}
\begin{array}{ll l l  }
  p_{T,\, \ell} >25 ~{\rm GeV}\,,    &
                                     \quad\quad \quad\quad\quad \quad\quad&
                                                                  p_{T,\,b}
                                                                  >25 ~{\rm GeV}\,, 
\vspace{0.2cm}\\
 |y_\ell |<2.5\,,&& |y_b|<2.5 \,,   
\vspace{0.2cm}\\
 \Delta R_{\ell  \ell} > 0.4\,, &&
\Delta R_{\ell  b} > 0.4\,, 
\end{array}
\end{equation}
where $\ell$ stands for the charged lepton. We do not impose any
restrictions on the kinematics of the additional light jet and the
missing transverse momentum.

%
  \section{Correlations between 
    $\boldsymbol{t\bar{t}W^+}$ and
    $\boldsymbol{t\bar{t}W^-}$}
\label{sec:correl}
%
%

We start with the NLO QCD differential cross sections for $pp \to e^+
\nu_e \, \mu^- \bar{\nu}_\mu \, e^+ \nu_e \, b\bar{b} \,+X$ and $pp
\to e^- \bar{\nu}_e \, \mu^+ \nu_\mu \, e^- \bar{\nu}_e \, b\bar{b}
\,+X$.  They are obtained for the LHC Run II energy of $\sqrt{s}= 13$
TeV.  For brevity, we will refer to these reactions as $pp\to
t\bar{t}W^+$ and $pp \to t\bar{t}W^-$. In
Ref. \cite{Bevilacqua:2020pzy} we have shown that the NLO QCD effects
for $t\bar{t}W^+$ and $t\bar{t}W^-$ are very similar. Indeed, both
processes show alike ${\cal K}$-factors and dependencies on the
perturbative scales. Furthermore, the off-shell effects for both
processes are of the same order. Thus, it is highly probable that some
of the uncertainties cancel in the ratio of $t\bar{t}W^+$ and
$t\bar{t}W^-$ cross sections, and ${\cal R}= \sigma^{\rm
NLO}_{t\bar{t}W^+}/\sigma^{\rm NLO}_{t\bar{t}W^-}$ might exhibit an
enhanced perturbative stability. In the following we would like to
understand similarities and potential differences between the two
processes even further.  We note that, at the leading order the
production mechanism for $t\bar{t}W^+$ $(t\bar{t}W^-)$ is via the
scattering of up-type quark (anti-quark) and the corresponding
down-type anti-quark (quark), i.e. $u\bar{d}$ and $c\bar{s}$ for
$pp\to t\bar{t}W^+$ as well as $\bar{u}d$ and $\bar{c}s$ for $pp \to
t\bar{t}W^-$. The quark-gluon initial state opens up only at NLO in
QCD. Similarities with respect to higher order QCD corrections, in the
production mechanisms as well as in the kinematics of the final states
suggest that the two processes may be treated as correlated as far as
the choice of scales is concerned. In any case, as long as one is
interested in some specific observables listed below. To show this we
examine the common features in the kinematics of the final
states. Since we are interested in the shape differences/similarities
only and because the fiducial cross section for $pp \to t\bar{t}W^-$
is about a factor of two smaller than the one for the $pp \to
t\bar{t}W^+$ process we concentrate on the normalised NLO QCD
differential cross sections.
%
\begin{figure}[h!]
  \begin{center}
  \includegraphics[width=0.49\textwidth]{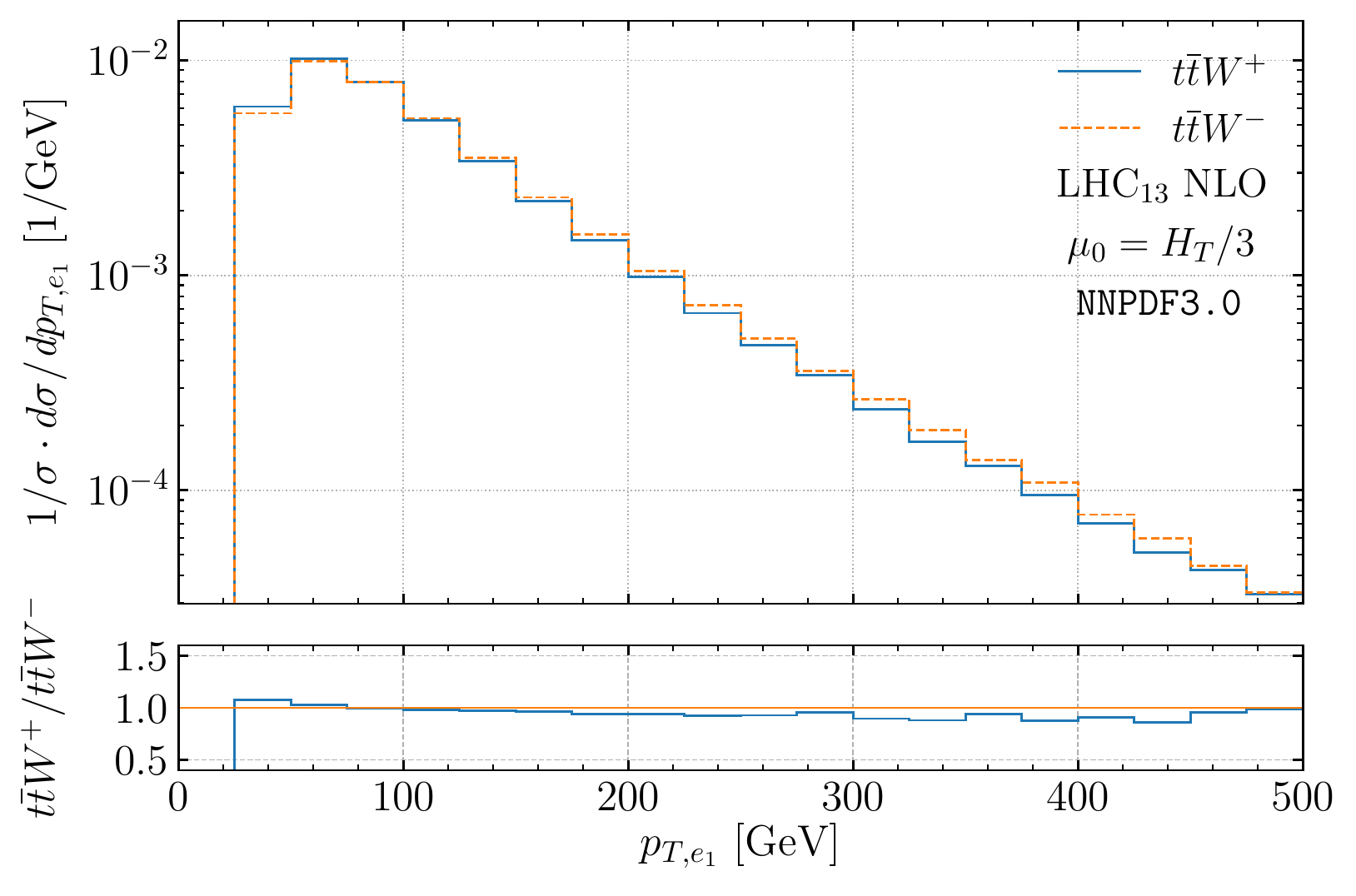}
  \includegraphics[width=0.49\textwidth]{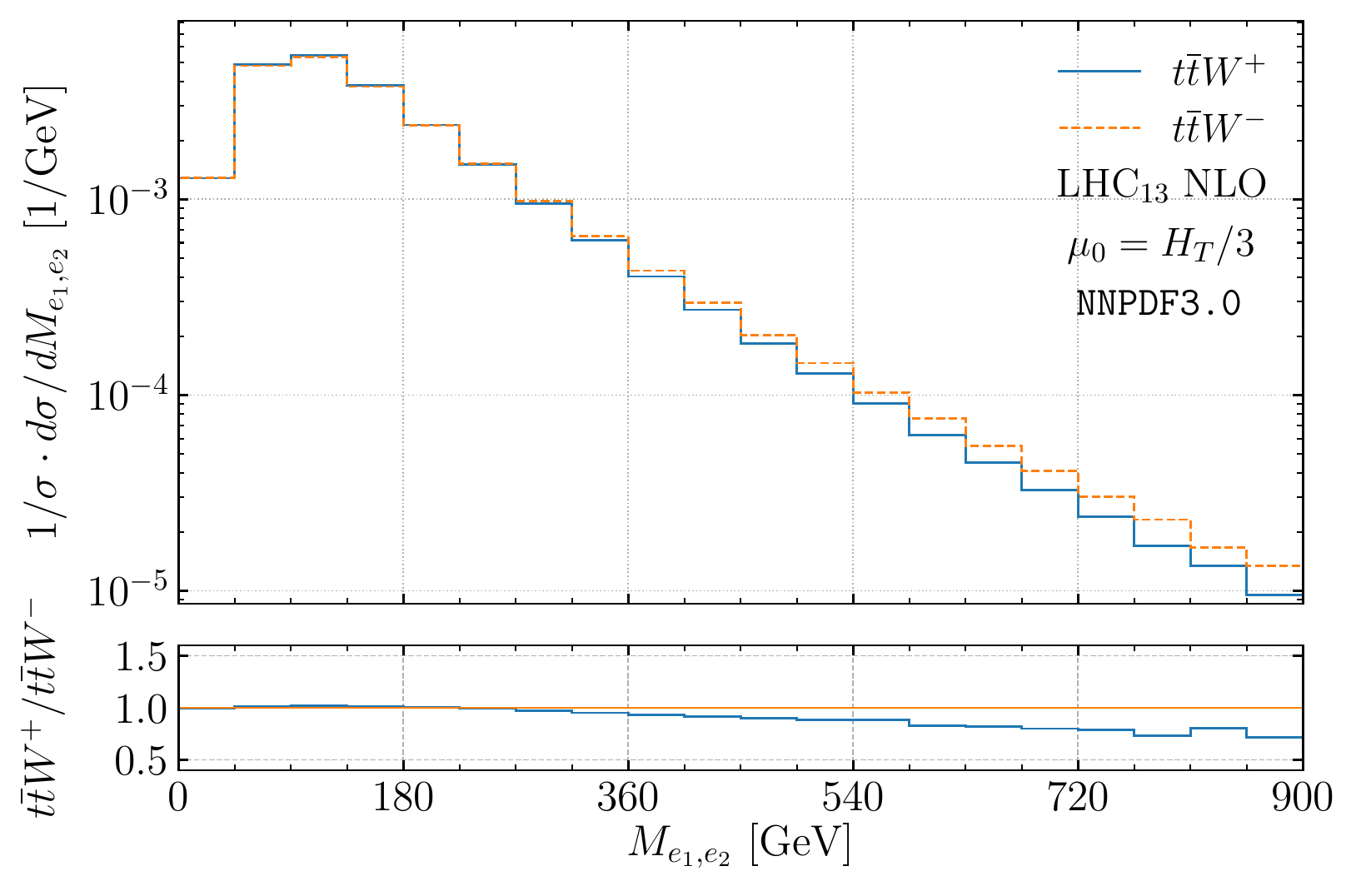}
  \includegraphics[width=0.49\textwidth]{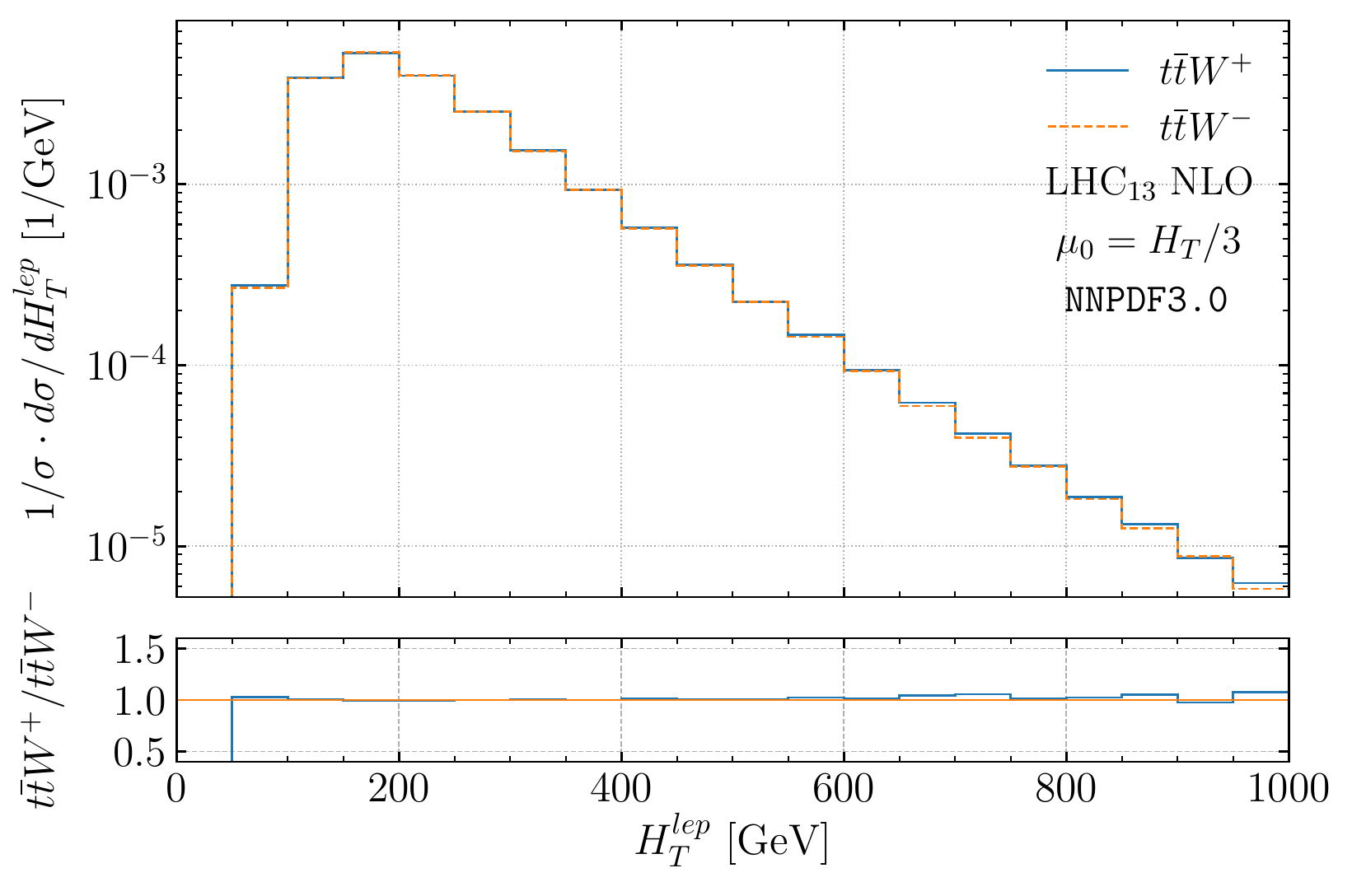}
  \includegraphics[width=0.47\textwidth]{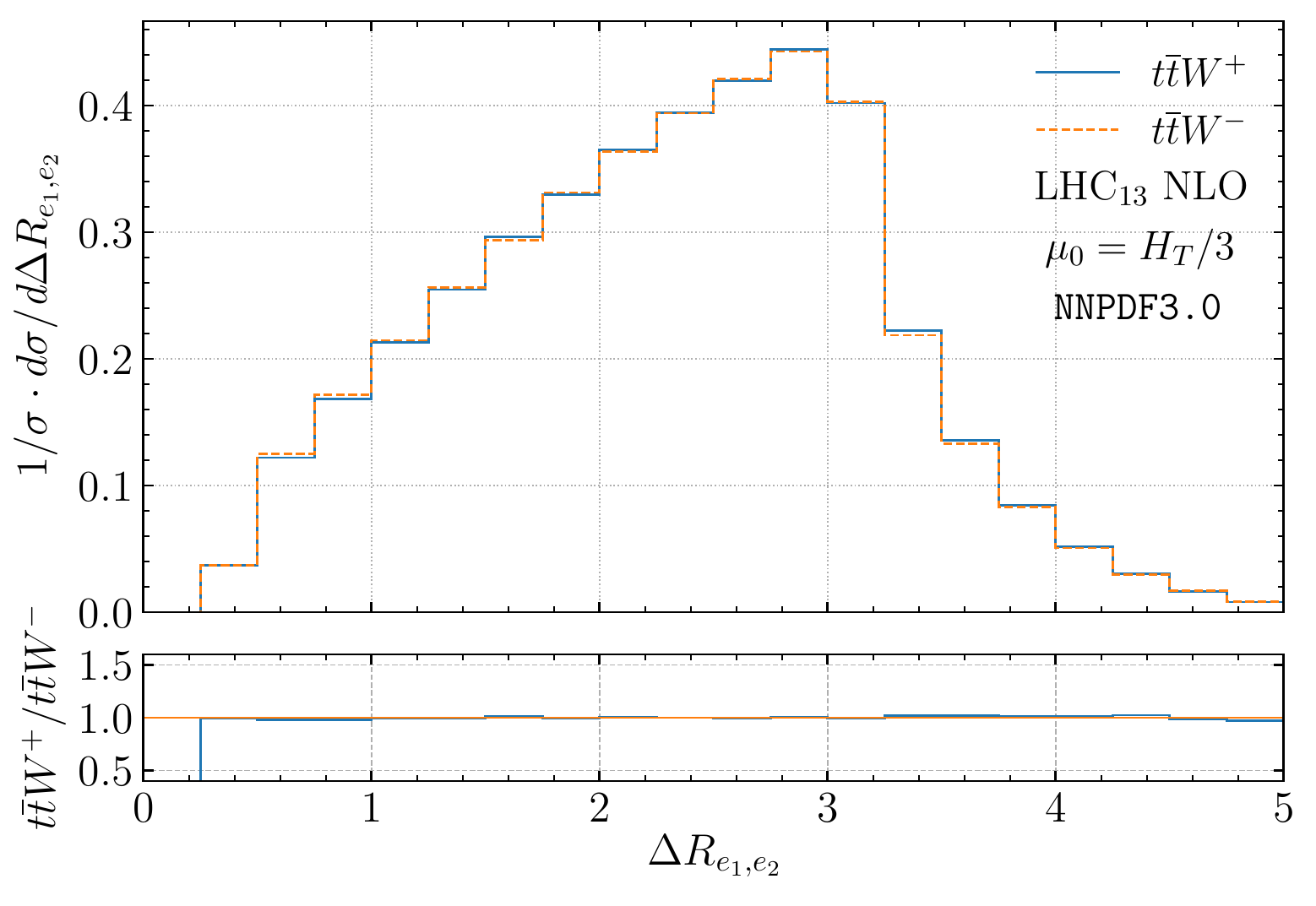}
\end{center}
\caption{\label{fig:leptonic} \it  Comparison of the normalised NLO QCD
differential cross sections for $pp \to t\bar{t}W^\pm$ in the multi-lepton
final state at the LHC with $\sqrt{s}=13$ TeV.  The transverse
momentum of the hardest same-sign lepton $(p_{T,\,e_1})$ and the
invariant mass of the two same-sign leptons $(M_{e_1 e_2})$
are presented. Also given are the scalar sum of the transverse momenta
of the leptons $(H_T^{lep})$ and the distance in the azimuthal
angle rapidity plane between the two same-sign leptons
$(\Delta R_{e_1 e_2})$.  The lower panels display  the
  ratio of the normalised distributions
  $t\bar{t}W^+/t\bar{t}W^-$. The NLO NNPDF3.0 PDF set is employed and
$\mu_R=\mu_F=H_T/3$ is used.}
\end{figure}
\begin{figure}[h!]
  \begin{center}
  \includegraphics[width=0.49\textwidth]{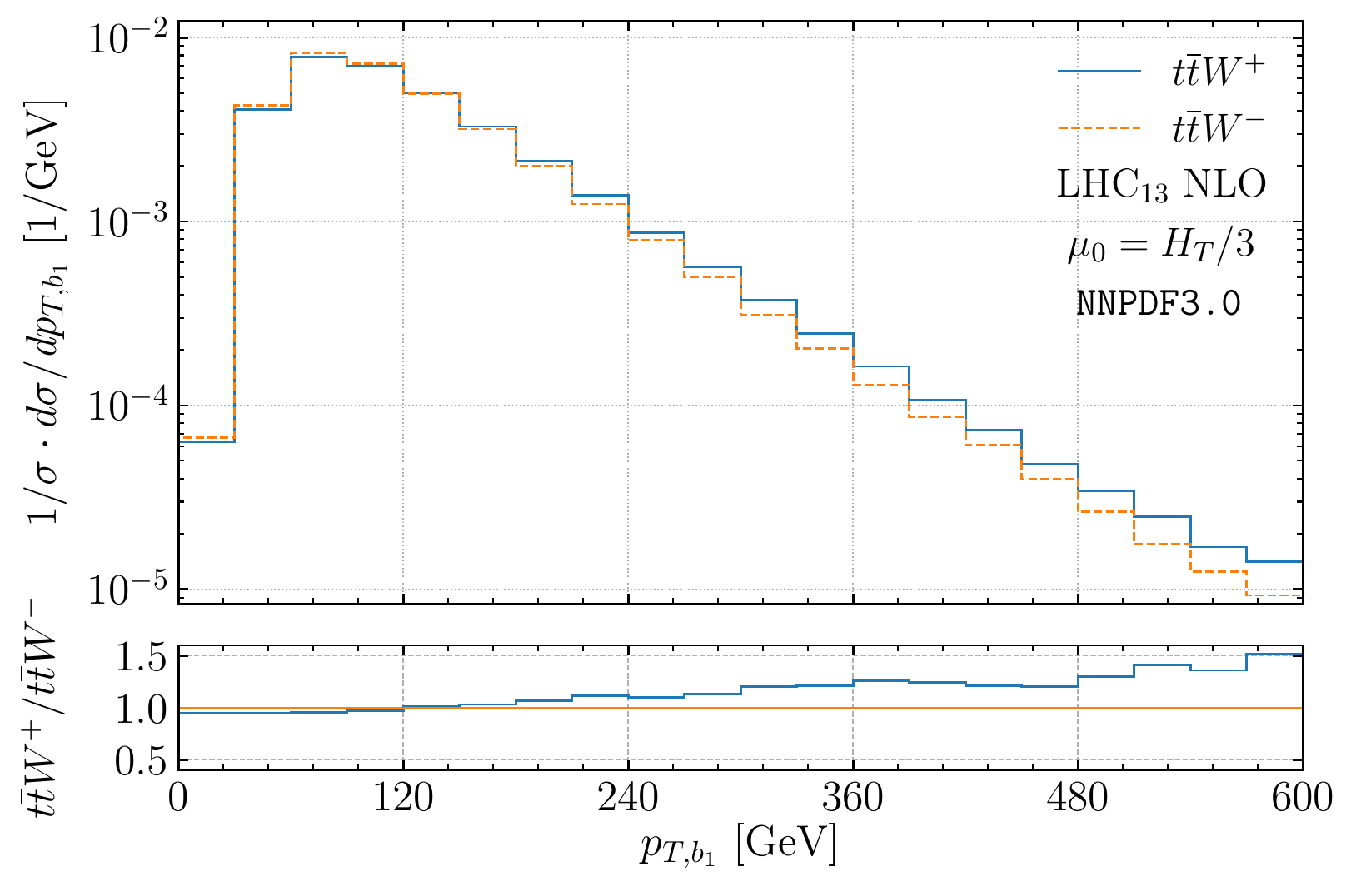}
  \includegraphics[width=0.49\textwidth]{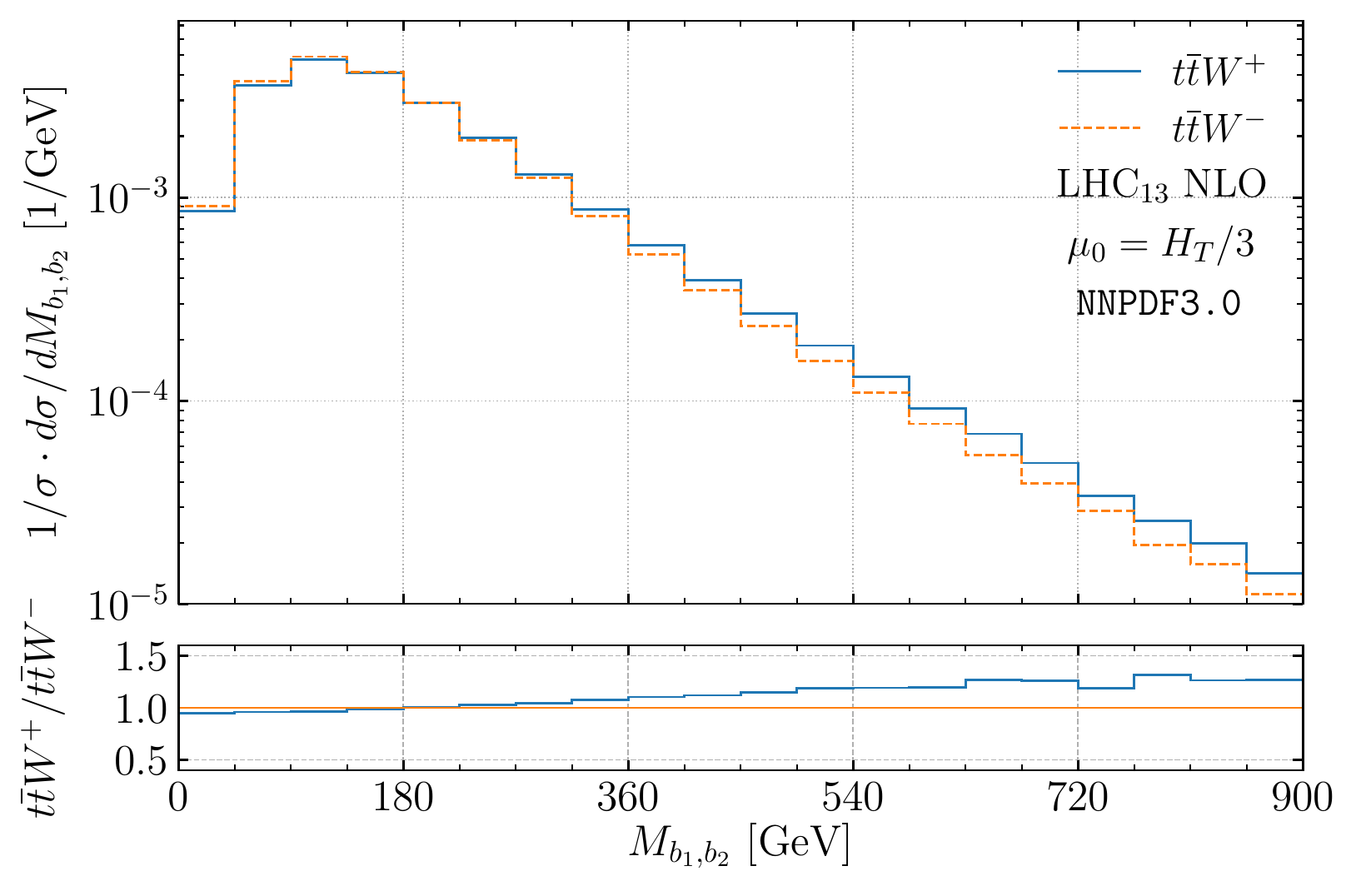}
  \includegraphics[width=0.49\textwidth]{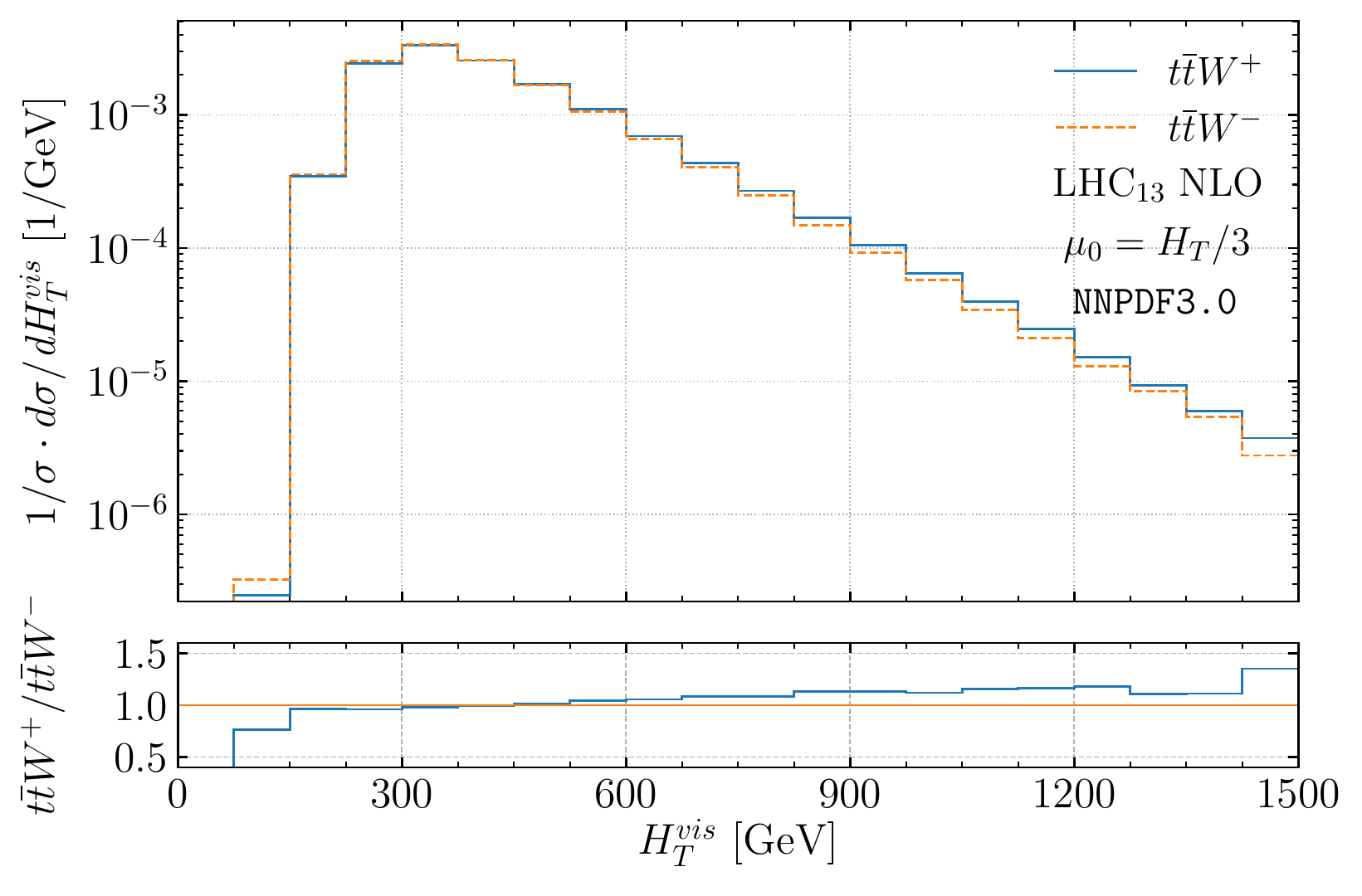}
  \includegraphics[width=0.47\textwidth]{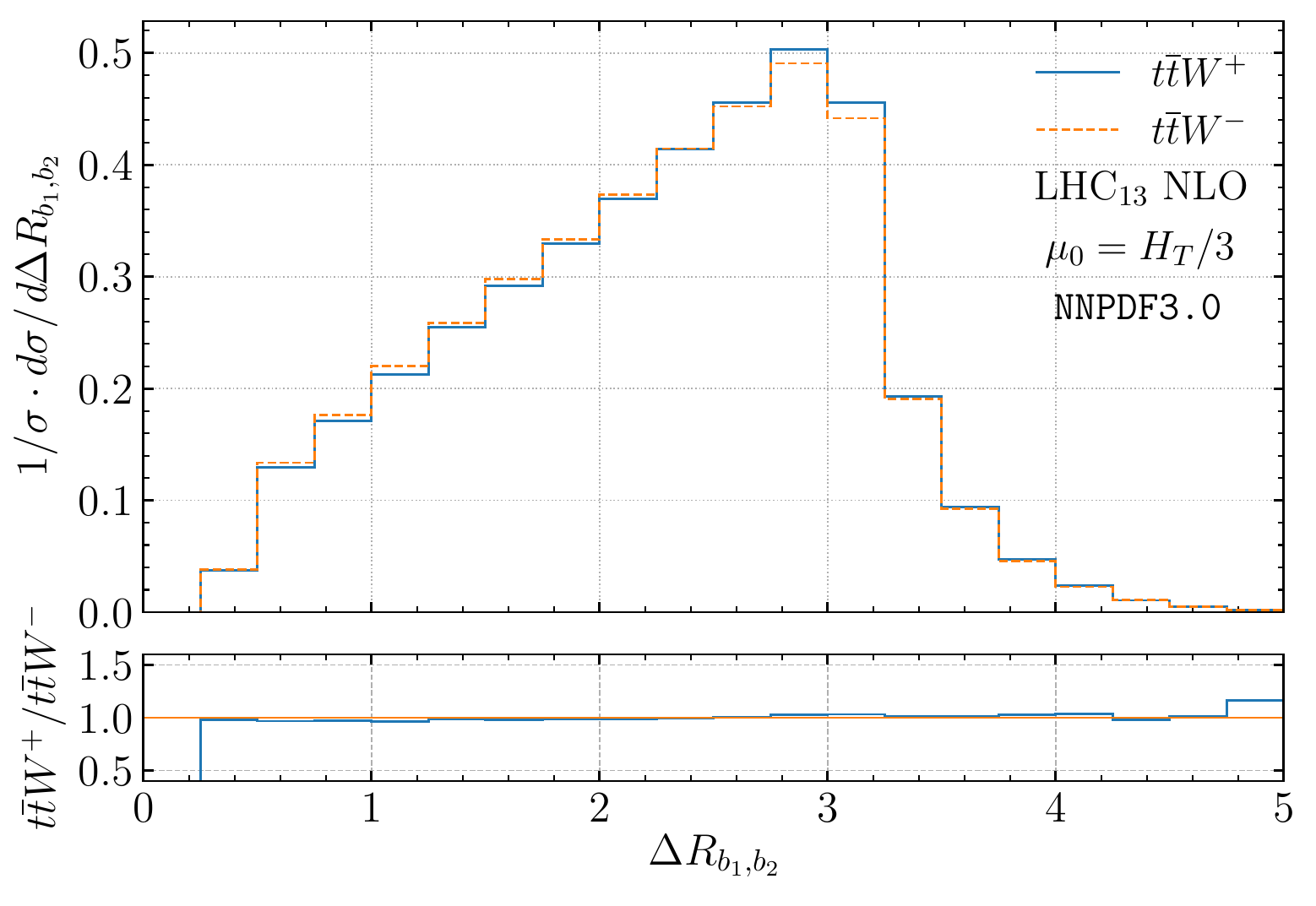}
\end{center}
\caption{\label{fig:bjet} \it  As in Figure \ref{fig:leptonic} but for 
the transverse momentum of the hardest $b$-jet $(p_{T,\,b_1})$, the
invariant mass of the two $b$-jets $(M_{b_1 b_2})$, the scalar sum of
the transverse momenta of the visible final states $(H_T^{vis})$ and
the distance in the azimuthal angle rapidity plane between the two
$b$-jets $(\Delta R_{b_1 b_2})$.}
\end{figure}

In the following the collection of leptonic observables will be
examined.  In the $pp\to t\bar{t}W^\pm$ process same-sign charged
leptons $e^\pm e^\pm$ occur.  In the case of final states with
identical leptons the ordering in $p_T$ has to be introduced to label
the particles. To this end, we denote the first and the second hardest
same-sign charged lepton as $e_1^\pm $ and $e_2^\pm$ respectively.  In
Figure~\ref{fig:leptonic} we present the NLO QCD differential cross
sections for $pp\to t\bar{t}W^+$ and $pp \to t\bar{t}W^-$ as a
function of the transverse momentum of $e^\pm_1$ $(p_{T,\,e_1})$, the
invariant mass of the $e_1^\pm e_2^\pm$ system $(M_{e_1 ,\,e_2})$ and
the scalar sum of the transverse momenta of the charged leptons
available in the given process $(H_T^{lep})$. The latter is defined as
\begin{equation}
  H_T^{lep}= p_{T,\,\mu^\mp} + p_{T,\,e^\pm_1}
+p_{T,\,e^\pm_2}\,.
\end{equation}
Also shown in Figure~\ref{fig:leptonic} is the distance in the
azimuthal angle rapidity plane between $e_1^\pm $ and $e_2^\pm$
$(\Delta R_{e_1 ,\, e_2})$. All differential cross sections shown are,
indeed, rather similar. This is particularly true for $H_T^{lep}$, 
$\Delta R_{e_1 ,\, e_2}$ and  $p_{T,\,e_1}$  but also for $M_{e_1
,\,e_2}$ at the beginning of the spectrum. Even-though
the latter distribution diverges substantially in the tails, the
contribution from these particular phase-space regions to the
integrated fiducial $pp\to t\bar{t}W^\pm$ cross section is
negligible.  To contrast these results, we refer the reader to
Ref. \cite{Bevilacqua:2014qfa} where the processes $pp\to
t\bar{t}b\bar{b}$ and $pp\to t\bar{t}jj$ demonstrate sizeable 
dissimilarities over the whole kinematics range.

In the next step we look at the $b$-jet kinematics. The two $b$-jets
are ordered according to their $p_T$. The hardest $(b_1)$ and the
softest $b$-jet $(b_2)$ kinematics are exhibited in
Figure~\ref{fig:bjet}. We note here, however, that the charge
identification of the $b$-jets is possible at the LHC, see e.g
\cite{Krohn:2012fg,Waalewijn:2012sv,Tokar:2017syr,ATLAS:2018lhe}.
Thus, one can distinguish between $b$-jets initiated by $b$ and
$\bar{b}$. In this work, however, we do not perform such $b$-jet
identification. We depict the NLO QCD differential cross sections as a
function of the transverse momentum of $b_1$ $(p_{T,\,b_1})$, the
invariant mass of the two $b$-jet system $(M_{b_1 ,\,b_2})$ and the
distance in the azimuthal angle rapidity plane between $b_1$ and $b_2$
$(\Delta R_{b_1, \,b_2})$. Also presented in Figure~\ref{fig:bjet} is
the scalar sum of the transverse momenta of all the visible final
states, denoted as $H_T^{vis}$. The latter is given by
\begin{equation}
H_T^{vis}= p_{T,\, b_1}+p_{T,\,{b_2}} + p_{T,\,\mu^\mp} + p_{T,\,e^\pm_1}
+p_{T,\,e^\pm_2}\,.
\end{equation}  
An interesting comment can be made here. Namely, that the $b$-jets are
preferably produced in back-to-back configurations. Hereby, $b$-jets
come more often from top quark decays rather than from the $g\to
b\bar{b}$ splitting. The latter configuration, which is produced in
the off-shell case where no top-quark resonances are present, would
manifest itself in the enhancement close to $\Delta R_{b_1,\,b_2}
\approx 0.4$. In the case of $b$-jet kinematics and for the
$H_T^{vis}$ observable we can see similarities between $pp\to
t\bar{t}W^+$ and $pp\to t\bar{t}W^-$.

To summarise this part, as anticipated both $t\bar{t}W^+$ and
$t\bar{t}W^-$ production processes show a good level of
similarity. In addition to what has already been
demonstrated in Ref.  \cite{Bevilacqua:2020pzy}, namely that the
dominant higher order QCD effects are alike for $t\bar{t}W^+$ and
$t\bar{t}W^-$, we have shown here that the kinematics (shapes of
various differential cross sections) of the two
processes is much the same. Such similarities are there because the
differences in the PDFs for the valence quarks do not manifest
themselves so much for the chosen observables. The fact that the QCD
corrections are not flavour sensitive does not destroy the picture
when NLO corrections in QCD are added. Furthermore, it justifies the
use of correlated scales later on. For both processes, our findings
are not modified when the fixed scale choice $\mu_R=\mu_F=\mu_0
=m_t+m_W/2$ is used instead or when different PDF sets are
employed. We further note that, the ratio is built of integrated cross
section and the peak region is very similar in both distributions. One
could still point out that the normalisation matters. On the other
hand, the normalization is driven by the same power of $\alpha_s$ in
both processes, which necessarily cancels. Hence, it matters very much
that the NLO corrections may be calculated with correlated scales.

Finally, in the appendix \label{appendix} we review the
Kolmogorov-Smirnov (KS) test, that may be used to provide a
quantitative measure of similarity between  two given
processes. Not only do we argue there for the usefulness of the KS
test at great length but we also discuss its shortcomings and
advantages. This novel approach is not used as a theoretical
argument for the correlation of the tested processes.  It is rather an
interesting way of showing that two processes are similar in various
differential distributions. Thus, used alone the KS test is not a
sufficient argument to prove the correlation between $t\bar{t}W^+$ and
$t\bar{t}W^-$. However, used together with all other arguments
mentioned in this section it increases the confidence that
$t\bar{t}W^+$ and $t\bar{t}W^-$ are indeed very similar.

%
\section{Cross Section Ratios}
\label{sec:ratio}
%

Once we established that $pp\to t\bar{t}W^+$ and $t\bar{t}W^-$ are
correlated we can look at their ratio with the goal of increasing the
precision of NLO QCD predictions for both processes. We take the ratio
of $t\bar{t}W^+$ and $t\bar{t}W^-$ at the next-to-leading order in
QCD.  We do not concentrate on  ${\cal R} =\sigma^{\rm
LO}_{t\bar{t}W^+}/\sigma^{\rm LO}_{t\bar{t}W^-}$, as in the lowest
order in perturbation expansion the dependence on the strong coupling
constant cancels out completely resulting in highly underestimated
theoretical uncertainties for the cross section ratio. Consequently,
the NLO in QCD is the first order where the theoretical uncertainties
are meaningful for the ${\cal R}$ observable.
Indeed, for the fixed scale choice, $\mu_0=m_t+m_W/2$, we have
\begin{equation}
  \begin{split}
    \sigma^{\rm LO}_{t\bar{t}W^+} & =
106.88^{\, +27.75\, (26\%)}_{\,-20.53\, (19\%)}
{}^{\,+4.45\,(4\%)}_{\,-4.45 \, (4\%)} \, {\rm [ab]}, \\
\sigma_{t\bar{t}W^-}^{\rm LO} &={57.24}^{\,+14.92\, (26\%)}_{\,-11.04\,
  (19\%)} {}^{\, +2.79\, (5\%)}_{\,-2.79\, (5\%)} \,  {\rm [ab]}.
\end{split}
\end{equation}
The first sub- and super-scripts indicate the scale variation while the
second ones the PDF uncertainties.  The LO cross section ratio reads
\begin{equation}
{\cal R}=\sigma^{\rm LO}_{t\bar{t}W^+} /\sigma_{t\bar{t}W^-}^{\rm LO} =
1.867^{\, +0.002\,(\,\,\, 0.1\%)}_{\,-0.001\, (0.05\%)} {}^{\,+0.057\,
  (3\%)}_{\,-0.057\,(3\%)}.
\end{equation}
Thus, the 
scale dependence is at the $0.1\%$ level. The NLO
QCD corrections to ${\cal R}$ are negative and of the order of
$3\%$. Consequently, the LO scale uncertainties underestimate the size
of the NLO corrections by a factor of $30$. Similar behaviour is
observed for the dynamical scale setting. Specifically, for
$\mu_0=H_T/3$ we obtain
\begin{equation}
  \begin{split}
    \sigma^{\rm LO}_{t\bar{t}W^+} & =
115.10^{\, +30.50\, (26\%)}_{\,-22.45\, (20\%)}
{}^{\,+4.80\,(4\%)}_{\,-4.80 \, (4\%)} \, {\rm [ab]}, \\
\sigma_{t\bar{t}W^-}^{\rm LO} &={62.40}^{\,+16.67\, (27\%)}_{\,-12.27\,
  (20\%)} {}^{\, +3.05\, (5\%)}_{\,-3.05\, (5\%)} \,  {\rm [ab]}.
\end{split}
\end{equation}
The LO cross section ratio for the dynamical scale setting yields 
\begin{equation}
{\cal R}=\sigma^{\rm LO}_{t\bar{t}W^+} /\sigma_{t\bar{t}W^-}^{\rm LO} =
1.844^{\, +0.004\,(0.2\%)}_{\,-0.003\, (0.2\%)} {}^{\,+0.056\,
  (3\%)}_{\,-0.056\,(3\%)}.
\end{equation}
In this case the NLO QCD corrections to ${\cal R}$ are also negative
but slightly smaller at only $2\%$.

We choose the
renormalisation and factorisation scales in the numerator and
denominator in a correlated way and we always choose the same values
for the scales in both the numerator and the denominator. This
approach is justified by the outcomes of the previous section, i.e. by
observing that various NLO distributions for $t\bar{t}W^+$ and
$t\bar{t}W^-$ have very similar shapes.  In other words, the ratio
${\cal R}$ is rather constant across the dominant parts of phase
space. This form of kinematic correlation indicates that NLO
predictions for the two processes are dominated by the same
topologies.
%
\begin{table}[t!]
  \renewcommand{\arraystretch}{1.6}
\begin{center}
  \begin{tabular}{c|c|c|c}
  $\mu_0=m_t+m_W/2$ &$\sigma^{\rm NLO}_{t\bar{t}W^+}\pm \delta_{\rm
scale}\pm\delta_{\rm PDF}$ &$\sigma^{\rm NLO}_{t\bar{t}W^-}\pm
\delta_{\rm scale}\pm\delta_{\rm PDF}$ & ${\cal R}\pm \delta_{\rm
                                         scale}\pm\delta_{\rm PDF}$
    \\ NNPDF3.0& [ab]&[ab]& ${\cal R}
=\sigma^{\rm
NLO}_{t\bar{t}W^+} / \sigma^{\rm NLO}_{t\bar{t}W^-}$\\ 
  \hline 
$p_{T,\,b}>25$ GeV & $123.2^{\,+6.3\, (5\%)}_{\,-8.7\,
(7\%)}{}^{\,+2.1\, (2\%)}_{\,-2.1\,(2\%)}$ & $68.0^{\,+4.8 \,
(7\%)}_{\,-5.5 \, (8\%)}{}^{\,+1.2\,(2\%)}_{\,-1.2\,(2\%)}$ &
$1.81^{\, +0.02\, (1\%)}_{\, -0.03\, (2\%)} 
{}^{\,+0.03 \, (2\%)}_{\, -0.03\, (2\%)}$
\\ 
$p_{T,\,b} >30$ GeV & $113.1^{\,+5.4\, (5\%)}_{\,-7.8
\, (7\%)}{}^{\,+1.9\,(2\%)}_{\,-1.9\,(2\%)}$& $62.3^{\,+4.2 \,
(7\%)}_{\,-4.9\, (8\%)}{}^{\,+1.1\,(2\%)}_{\,-1.1\,(2\%)}$ &
$ 1.81^{\,+0.02\,(1\%)}_{\,-0.04\,(2\%)} {}^{\, +0.03 \,
 (2\%)}_{\,-0.03\,(2\%)}$
\\ 
$p_{T,\,b}>35$ GeV & $102.6^{\,+4.7 \, (5\%)}_{\,-6.8
\, (7\%)}{}^{\,+1.7\,(2\%)}_{\,-1.7\,(2\%)}$ & $56.3^{\,+3.7
\,(7\%)}_{\,-4.4 \, (8\%)}{}^{\,+1.0\,(2\%)}_{\,-1.0\,(2\%)}$ &
 $1.82^{\,+0.02\, (1\%)}_{\,-0.04\,(2\%)} 
{}^{\,+0.03\, (2\%)}_{\, -0.03\,(2\%)}$
\\ 
$p_{T,\,b}>40$ GeV & $92.0^{\,+4.0 \,
(4\%)}_{\,-6.1 \, (7\%)}{}^{\,+1.6\,(2\%)}_{\,-1.6\,(2\%)}$ &
$50.3^{\,+3.3 \, (6\%)}_{\,-3.9\,
(8\%)}{}^{\,+0.9\,(2\%)}_{\,-0.9\,(2\%)}$ & 
 $1.83^{\,+0.02\,(1\%)}_{\, -0.04\,(2\%)} 
{}^{\,+0.03\,(2\%)}_{\, -0.03\, (2\%)}$ \\
 \end{tabular}
\end{center}
\vspace{0.4cm}
\caption{\label{tab:ratio-fixed} \it
  NLO QCD integrated fiducial cross sections for $pp \to
t\bar{t}W^\pm$ in the multi-lepton final state at the LHC with
$\sqrt{s}=13$ TeV. Also shown are results for ${\cal R}=\sigma^{\rm
NLO}_{t\bar{t}W^+}/\sigma^{\rm NLO}_{t\bar{t}W^-}$. Theoretical
uncertainties as estimated from the scale variation and from the PDFs
are listed as well.  Four different values of the $p_{T,\,b}$ cut are
used. The NNPDF3.0 PDF set is employed and $\mu_R=\mu_F=\mu_0$ where
$\mu_0=m_t+m_W/2$.}
\end{table}
\begin{table}[t!]
  \renewcommand{\arraystretch}{1.6}
\begin{center}
\begin{tabular}{c|c|c|c}
  $\mu_0=H_T/3$ &$\sigma^{\rm NLO}_{t\bar{t}W^+}\pm \delta_{\rm
scale}\pm\delta_{\rm PDF}$ &$\sigma^{\rm NLO}_{t\bar{t}W^-}\pm
\delta_{\rm scale}\pm\delta_{\rm PDF}$ & ${\cal R} \pm \delta_{\rm
scale}\pm\delta_{\rm PDF}$
\\ NNPDF3.0
&[ab]&[ab]& ${\cal R}=\sigma^{\rm
            NLO}_{t\bar{t}W^+}/\sigma^{\rm NLO}_{t\bar{t}W^-}$
\\ 
\hline $p_{T,\,b}>25$ GeV & $124.4^{\, +4.3
\, (3\%)}_{\, -7.7 \, (6\%)} {}^{\,+2.1\,(2\%)}_{\,-2.1\,(2\%)}$
&$68.6^{\, +3.5\, (5\%)}_{\,-4.8\,
(7\%)}{}^{\,+1.2\,(2\%)}_{\,-1.2\,(2\%)}$ & 
 $1.81^{\, +0.02\, (1\%)}_{\, -0.03\,(2\%)} 
{}^{\,+0.03\, (2\%)}_{\, -0.03\, (2\%)}$
\\
$p_{T,\,b} >30$ GeV & $113.9^{\,+3.5 \, (3\%)}_{\,-6.8\,
(6\%)}{}^{\,+1.9\,(2\%)}_{\,-1.9\,(2\%)}$ & $62.7^{\,+3.0 \,
(5\%)}_{\,-4.3 \, (7\%)}{}^{\,+1.1\,(2\%)}_{\,-1.1\,(2\%)}$ & 
 $1.82^{\,+0.02\, (1\%)}_{\, -0.03\, (2\%)} 
{}^{\,+0.03\, (2\%)}_{\, -0.03\, (2\%)}$
\\ 
$p_{T,\,b}>35$ GeV & $103.1^{\,+3.1 \,
(3\%)}_{\,-6.0\, (6\%)} {}^{\,+1.7\,(2\%)}_{\,-1.7\,(2\%)}$ &
$56.5^{\,+2.6 \, (5\%)}_{\,-3.8\,
(7\%)}{}^{\,+1.0\,(2\%)}_{\,-1.0\,(2\%)}$ & 
$1.82^{\, +0.02 \, (1\%)}_{\,-0.03\, (2\%)} 
{}^{\, +0.03\, (2\%)}_{\, -0.03\, (2\%)}$
\\
$p_{T,\,b}>40$ GeV & $92.3^{\,+2.8 \, (3\%)}_{\,-5.3 \, (6\%)}
{}^{\,+1.5\,(2\%)}_{\,-1.5\,(2\%)}$ & $50.4^{\,+2.3 \, (5\%)}_{\,-3.4
\, (7\%)}{}^{\,+0.9\,(2\%)}_{\,-0.9\,(2\%)}$ & 
 $1.83^{\,+0.02\, (1\%)}_{\, -0.03\, (2\%)}
{}^{\, +0.03\, (2\%)}_{\,-0.03\,(2\%)}$ 
\\
 \end{tabular}
\end{center}
\vspace{0.4cm}
\caption{\label{tab:ratio-dynamic} \it As in Table
  \ref{tab:ratio-fixed} but for $\mu_0=H_T/3$. }
\end{table}
%

In Tables \ref{tab:ratio-fixed} and \ref{tab:ratio-dynamic} we present
integrated fiducial cross sections at NLO in QCD for $pp\to
t\bar{t}W^+$ and $t\bar{t}W^-$ in the multi-lepton decay channel
together with the theoretical uncertainties due to scale dependence
for $\mu_F=\mu_R=\mu_0$ where $\mu_0=m_t+m_W/2$ or $\mu_0=H_T/3$. The
uncertainty on higher orders for the integrated fiducial cross
sections is estimated by varying $\mu_R$ and $\mu_F$ independently
around a central scale $\mu_0$ in the range $1/2 \le \mu_R/\mu_0 ,
\mu_F/\mu_0 \le 2$ with the additional condition $1/2 \le \mu_R/\mu_F
\le 2$. As it is always done we search for the minimum and maximum of
the resulting cross sections. For the PDF error we use
the corresponding prescription from the NNPDF3.0 group to provide the
$68\%$ confidence level (C.L.) PDF uncertainties. Specifically, the
NNPDF3.0 group uses the Monte Carlo sampling method in conjunction
with neural networks and the PDF uncertainties are obtained with the
help of the replicas method, see e.g. \cite{Buckley:2014ana}. Also
given in Tables \ref{tab:ratio-fixed} and \ref{tab:ratio-dynamic} is
the ${\cal R}$ cross section ratio and its systematic uncertainties
$\delta_{\rm scale}$ and $\delta_{\rm PDF}$. To properly account for
the cross-correlations between the two processes the latter
is evaluated in the similar fashion as $\delta_{\rm PDF}$ for
$\sigma^{\rm NLO}_{t\bar{t}W^+}$ and $\sigma^{\rm
NLO}_{t\bar{t}W^-}$.  First we examine the stability of ${\cal R}$
with respect to the $p_{T,\,b}$ cut. To this end, we show results for
four different values of the $p_{T,\,b}$ cut.  We observe very stable
cross section ratio results both in terms of the central value and
theoretical uncertainties.  Furthermore, we notice that the scale
choice does not play any role for such an inclusive
observable. The PDF
uncertainties, which for $pp \to t\bar{t}W^+$ and $pp \to t\bar{t}W^-$
are consistently at the $2\%$ level, do not cancel out substantially
in the cross section ratio. The final theoretical error receives similar
contributions from $\delta_{\rm scale}$ and $\delta_{\rm PDF}$. These
uncertainties are both at the $2\%$ level.  Such precise theoretical
predictions have normally been obtained only once the NNLO QCD
corrections are incorporated. Thus, ${\cal R}$ at NLO in QCD
represents a very precise observable to be measured at the LHC. A few
comments are in order here.  NLO correlations for $t\bar{t}W^+$ and
$t\bar{t}W^-$ are completely insensitive to potentially large
$gg$-induced NNLO corrections to the ${\cal R}=\sigma^{\rm
NLO}_{t\bar{t}W^+}/\sigma^{\rm NLO}_{t\bar{t}W^-}$ ratio. For example,
due to the fact that the $gg$ initial state is the same for the two
processes we can assume that the $t\bar{t}W^+$ and $t\bar{t}W^-$ cross
sections receive the same $gg$-channel correction, which we denote as
$\delta \sigma_{gg}^{\rm NNLO}$. As a result, the ratio would be
shifted by a relative factor $\delta {\cal R}/{\cal R}\approx ({\cal
R} -1) \delta \sigma_{gg}^{\rm NNLO} /\sigma^{\rm NLO}_{t\bar{t}W^+}
\approx \delta \sigma_{gg}^{\rm NNLO} /\sigma^{\rm NLO}_{t\bar{t}W^+}
$, which could in principle amount to several percent. Therefore, it
could be well above the reported uncertainty estimate of
 $1\%-2\%$. The judgment of the scale variation
prescription for ${\cal R}$ would of course be much easier in the
presence of NNLO QCD
calculations. Unfortunately such calculations are out of reach even
for the simplest case of $pp \to t\bar{t}W^\pm$ production with stable
top quarks and $W^\pm$ gauge bosons. Adding decays of the unstable
particles and incorporating the complete off-shell effects is simply
difficult to imagine at the current stage of NNLO QCD
calculations. Nevertheless, in the absence of the NNLO calculations
for the process at hand we can assess the $gg$-channel correction
$\delta \sigma_{gg}^{\rm NNLO}$ by performing a LO study for $gg \to
e^+ \nu_e \,\mu^- \bar{\nu}_\mu \, e^+ \nu_e \, b \bar{b} \, \bar{u}
d$ and $gg \to e^+ \nu_e \, \mu^- \bar{\nu}_\mu \, e^+ \nu_e\,
b\bar{b} \, \bar{c} s$ with the same input parameters, cuts and for
example by employing the fixed scale choice. The size of $\delta
\sigma_{gg}^{\rm NNLO}/\sigma^{\rm NLO}_{t\bar{t}W^+}$ correction
estimated in this way amounts to $0.3\%$. Similar studies can be
performed for $gg \to e^- \bar{\nu}_e \, \mu^+ \nu_\mu \, e^-
\bar{\nu}_{e} \, b \bar{b} \, u \bar{d}$ and $gg \to e^- \bar{\nu}_e
\, \mu^+ \nu_\mu \, e^- \bar{\nu}_{e} \, b \bar{b} \, c \bar{s}$. To
evaluate LO cross sections, however, we have used cuts on light jets
that are not there when the true NNLO QCD corrections are
calculated. To remedy this in the next step we have used the SecToR
Improved Phase sPacE for real Radiation (\textsc{Stripper}) library
\cite{Czakon:2010td,Czakon:2011ve,Czakon:2014oma,Czakon:2019tmo}
that implements a general subtraction scheme for the evaluation of
NNLO QCD contributions from double-real radiation to processes with at
least two particles in the final state at LO. By employing
\textsc{Stripper}\footnote{Courtesy of M. Czakon.} we were able to
calculate the actual contribution $\delta \sigma_{gg}^{\rm
NNLO}/\sigma^{\rm NLO}_{t\bar{t}W^+}$ for the $pp \to t\bar{t}W^+$
process with stable top quarks and $W^+$ gauge boson.  This
contribution is of the order of $0.2\%$, thus, it is similar in size
to $\delta \sigma_{gg}^{\rm NNLO}/\sigma^{\rm NLO}_{t\bar{t}W^+}$
estimated by the LO studies with the complete top quark and $W^\pm$
off-shell effects included. Notably, it is also well below the
reported uncertainty estimate for the ${\cal R}=\sigma^{\rm
NLO}_{t\bar{t}W^+}/\sigma^{\rm NLO}_{t\bar{t}W^-}$ ratio $(1\% -
2\%)$. Because of that, the impact of the $gg$-channel on the ratio at
hand is quite small.

 We note here, that  there are different approaches in the
  literature for  handling of uncertainties in ratios.  For
  example one can take the relative size of the last considered order
  compared to the previous one as an estimate of the perturbative
  uncertainty, see e.g. Ref. \cite{Duhr:2020sdp}. Specifically,
  following \cite{Duhr:2020sdp} one can define 
  \begin{equation}
    \label{delta_new}
  \delta_{pert.} = \pm \left| 1
    -\frac{{\cal R}^{\rm NLO} (\mu_0)}{{\cal R}^{\rm LO}(\mu_0)}
  \right|\times 100\%\,,
\end{equation}
where the values of $\mu_0$ are chosen in a correlated way in the
numerator and the denominator of ${\cal R}$. This error estimator
assumes that the sub-leading terms should be smaller than the last
known correction. The obvious downside to this approach is that it
gives a vanishing result whenever two consecutive perturbative orders
provide identical numerical predictions. Furthermore, this
prescription leads to the rather suspicious shapes of the uncertainty
bands for the differential cross section ratios.  Thus, as clearly
stated in Ref. \cite{Duhr:2020sdp} this approach on its own can not serve as a
good estimator of perturbative uncertainties. Nevertheless, it can be
used in conjunction with traditional methods to further ensure that
the correlated scale dependence is indeed a reasonable approach. Had
we used the prescription from Eq.~\eqref{delta_new} we would obtain
$\delta_{pert.} =3\%$ for $\mu_0=m_t+m_W/2$ and $\delta_{pert.} =2\%$
for the dynamical scale setting. Thus, the uncertainty estimate for
the ${\cal R}=\sigma^{\rm NLO}_{t\bar{t}W^+}/\sigma^{\rm
NLO}_{t\bar{t}W^-}$ ratio would rather be $\delta_{pert.}= 2\%-3\%$, which is only
slightly larger than $\delta_{\rm scale}= 1\%-2\%$. 

In the next step we examine the impact of the top quark production and
decay modelling on the cross section ratio. To this end we present
results for the full NWA and for the NWA${}_{\rm LOdecay}$ case. The
former comprises NLO QCD corrections to the production and to the
subsequent top quark decays, the latter NLO QCD corrections to the
production of $t\bar{t}W^\pm$ and LO top quark decays. Should we use
the NLO QCD results in the full NWA for the $pp\to t\bar{t}W^\pm$
process our findings for $\mu_0=m_t+m_W/2$ would be as follows
\begin{equation}
{\cal R} = \frac{\sigma^{\rm NLO, \,NWA}_{t\bar{t}W^+}}{\sigma^{\rm NLO,\,
  NWA}_{t\bar{t}W^-}}  = 1.81 \pm 0.04 \, (2\%)\,,
\end{equation}  
where the quoted theoretical error results only from the scale
dependence as the PDF uncertainties will not be affected by
changes in the modeling of the top quark decays.  On the
other hand for the dynamical scale choice $\mu_0=H_T/3$ we would
obtain
\begin{equation}
{\cal R} = \frac{\sigma^{\rm NLO, \,NWA}_{t\bar{t}W^+}}{\sigma^{\rm NLO,\,
  NWA}_{t\bar{t}W^-}}  = 1.81 \pm 0.03 \, (2\%)\,.
\end{equation}  
We can observe that the full NWA approach does not modify either the
value or the size of the theoretical error for the integrated cross
section ratio. The latter result is not surprising taking into account
that the impact of the top quark off-shell effects on the integrated
fiducial $t\bar{t}W^\pm$ cross section is negligible.  Furthermore,
theoretical uncertainties for the full NWA and full off-shell case are
similar independently of the scale choice \cite{Bevilacqua:2020pzy}.

Finally, we have employed the NWA$_{\rm LOdecay}$ case. For
$\mu_0=m_t+m_W/2$ we obtained  
\begin{equation}
{\cal R} = \frac{\sigma^{\rm NLO,
    \,NWA_{LOdecay}}_{t\bar{t}W^+}}{\sigma^{\rm NLO,\,
  NWA_{LOdecay}}_{t\bar{t}W^-}}  = 1.82 \pm 0.02 \, (1\%)\,,
\end{equation}  
whereas for $\mu_0=H_T/3$ we can report
\begin{equation}
{\cal R} = \frac{\sigma^{\rm NLO,
    \,NWA_{LOdecay}}_{t\bar{t}W^+}}{\sigma^{\rm NLO,
    \,NWA_{LOdecay}}_{t\bar{t}W^-}}  = 1.81 \pm 0.02 \, (1\%)\,.
\end{equation}
Even for this case the cross section ratios are very stable and rather
insensitive to the details of the modelling of the top quark
production and decays. Let us note here, that for the absolute $pp\to
t\bar{t}W^\pm$ integrated cross sections the difference between the
NWA$_{\rm LOdecay}$ approach and the full off-shell one is at the
level of $5\%$. In addition, theoretical uncertainties due to the
scale dependence are higher in the former case, up to $11\%-13\%$
\cite{Bevilacqua:2020pzy}. Yet in the cross section ratio these
differences cancel out making ${\cal R}=\sigma^{\rm
NLO}_{t\bar{t}W^+}/\sigma^{\rm NLO}_{t\bar{t}W^-}$ very precise and
an extremely interesting theoretical observable to be measured at the
LHC.

To conclude this part, we note that for the cross
section ratio at NLO in QCD the residual perturbative uncertainties
are reduced to $1\%-2\%$ and are similar in size to the PDF
uncertainties. The theoretical uncertainties associated with the top
quark decay modelling are negligible. This suggests that the ${\cal
R}=\sigma^{\rm NLO}_{t\bar{t}W^+}/\sigma^{\rm NLO}_{t\bar{t}W^-}$
observable can be employed either for the precision SM measurements or
to shed some light on possible new physics scenarios that might reveal
themselves only once sufficiently precise theoretical predictions are
available. For example in the case of BSM physics the presence of two
same-sign leptons in the final state, a relatively rare phenomenon at
the LHC, constitutes an optimal signature for many new physics models
from supersymmetry, supergravity and Majorana neutrinos to models with
the modified Higgs boson sector.  Given the final accuracy of ${\cal
R}$, it should be used to put more stringent constrains on the
parameter space of these models.

%
\section{Charge Asymmetries in
  $\boldsymbol{t\bar{t}W^\pm}$
  Production}
\label{sec:assym}
%

The $pp$ initial state at the LHC is expected to produce top quark and
antiquark rapidity distributions that are symmetric about $y = 0$ in
$t\bar{t}$ production. However, since the quarks in the initial state
can be from valence quarks, while the antiquarks are from the sea, the
larger average momentum-fraction of quarks leads to an excess of top
quarks produced in the forward directions. The rapidity distribution
of top quarks in the SM is therefore broader than that of the more
centrally produced top antiquarks. This suggests that $\Delta|y|
=|y_t| -|y_{\bar{t}}|$, which is the difference between the absolute
value of the top quark rapidity $|y_t|$ and the absolute value of the
anti-top quark rapidity $|y_{\bar{t}}|$, is a suitable observable to
measure the top quark charge asymmetry at the LHC. This asymmetry is
nevertheless very small, see
e.g. \cite{Bernreuther:2012sx,Czakon:2017lgo}. For the $pp\to
t\bar{t}W^\pm$ process the presence of the $W^\pm$ gauge boson results
in the top quark charge asymmetry that is significantly larger than
in $t\bar{t}$ production \cite{Maltoni:2014zpa}.  The main reason
behind this is the absence of the symmetric $gg$ channel that is not
accessible until NNLO. Furthermore, the emission of the $W^\pm$ gauge
boson from the initial state polarises the initial quark line and in
turn the $t\bar{t}$ pair. As a result, the charge asymmetries for the
top quark decay products are large and already present at the leading
order. In the following we calculate afresh the top quark charge
asymmetry in the $t\bar{t}W^\pm$ process in the multi-lepton final
state using the state-of-the-art NLO QCD calculations with the full
top quark off-shell effects included. Additionally, the asymmetries
for the top quark decay products, $A^b_c$ and $A_c^\ell$, will be
examined. In this part of the paper, one of our main goals is to
carefully assess the impact of the top quark modelling on $A_c^t,
A^b_c$ and $A_c^\ell$.  We start with asymmetries at the integrated
level albeit in the fiducial regions of the phase space as defined in
Section \ref{sec:setup}. For $A_c^\ell$ we will additionally calculate
the differential and cumulative asymmetry with respect to the
following observables: $p_T(\ell_t\ell_{\bar{t}})$,
$|y(\ell_t\ell_{\bar{t}})|$ and $M({\ell_t\ell_{\bar{t}}})$, where
$\ell_{t}, \ell_{\bar{t}}$ stands for the charged leptons stemming
from the top and anti-top quark decay respectively. For the two
processes under consideration $pp \to e^+ \nu_e\, \mu^- \bar{\nu}_\mu
\, e^+ \nu_e \, b\bar{b}$ and $pp \to e^-\bar{\nu}_e\, \mu^+ \nu_\mu
\, e^- \bar{\nu}_e \, b\bar{b}$ the reconstruction of the (anti-)top
quark momentum through its decay products is required.  As we are
dealing with identical leptons in the final state, however, we must
employ an additional mechanism to properly assign the positron
(electron) and the corresponding neutrino $\nu_e$ (anti-neutrino
$\bar{\nu}_e$) to the correct top (anti-top) quark.  In the case of
$t\bar{t}W^+$ production we use the following four different resonance
histories (a similar procedure is applied in the $t\bar{t}W^-$ case)
%
%
\begin{equation}
\begin{split}
t&\to e^+_1 \, \nu_{e,1} \,b \quad \quad \quad \quad 
{\rm and} \quad \quad \quad\quad \bar{t}
\to \mu^- \, \bar{\nu}_\mu \, \bar{b}\,,\\[0.2cm]
t&\to e^+_1 \, \nu_{e,2} \,b \quad \quad \quad \quad 
{\rm and} \quad \quad \quad\quad \bar{t}
\to \mu^- \, \bar{\nu}_\mu \, \bar{b}\,,\\[0.2cm]
t&\to e^+_2 \, \nu_{e,1} \,b \quad \quad \quad \quad 
{\rm and} \quad \quad \quad\quad \bar{t}
\to \mu^- \, \bar{\nu}_\mu \, \bar{b}\,,\\[0.2cm]
t&\to e^+_2 \, \nu_{e,2} \,b \quad \quad \quad \quad 
{\rm and} \quad \quad \quad\quad \bar{t}
\to \mu^- \, \bar{\nu}_\mu \, \bar{b}\,. 
\end{split}
\end{equation}
%
%
These four resonant histories are not sufficient if NLO QCD
calculations are considered. In the case of the subtracted real
emission part we additionally take into account the  extra light jet if
resolved. Specifically, to closely mimic what is done on the
experimental side only the light jet that passes all the cuts, that
are  also required  for the two $b$-jets, is added to the resonance
history. Thus, in such a case a total of twelve different resonant
histories have to be considered. We compute for each history the
following quantity, see Ref.~\cite{Bevilacqua:2019quz}
%
\begin{figure}[t!]
\begin{center}
  \includegraphics[width=0.48\textwidth]{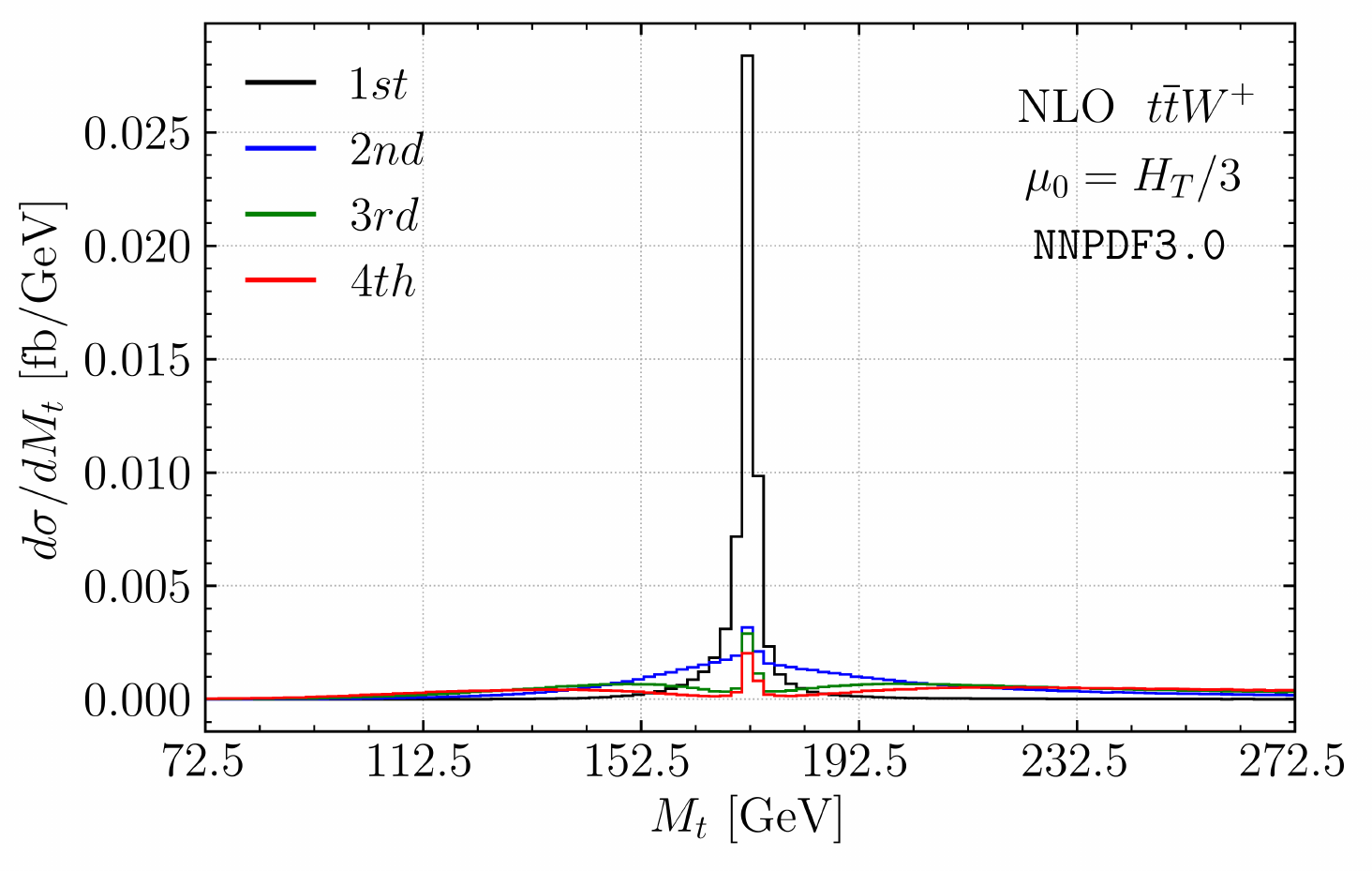}
    \includegraphics[width=0.49\textwidth]{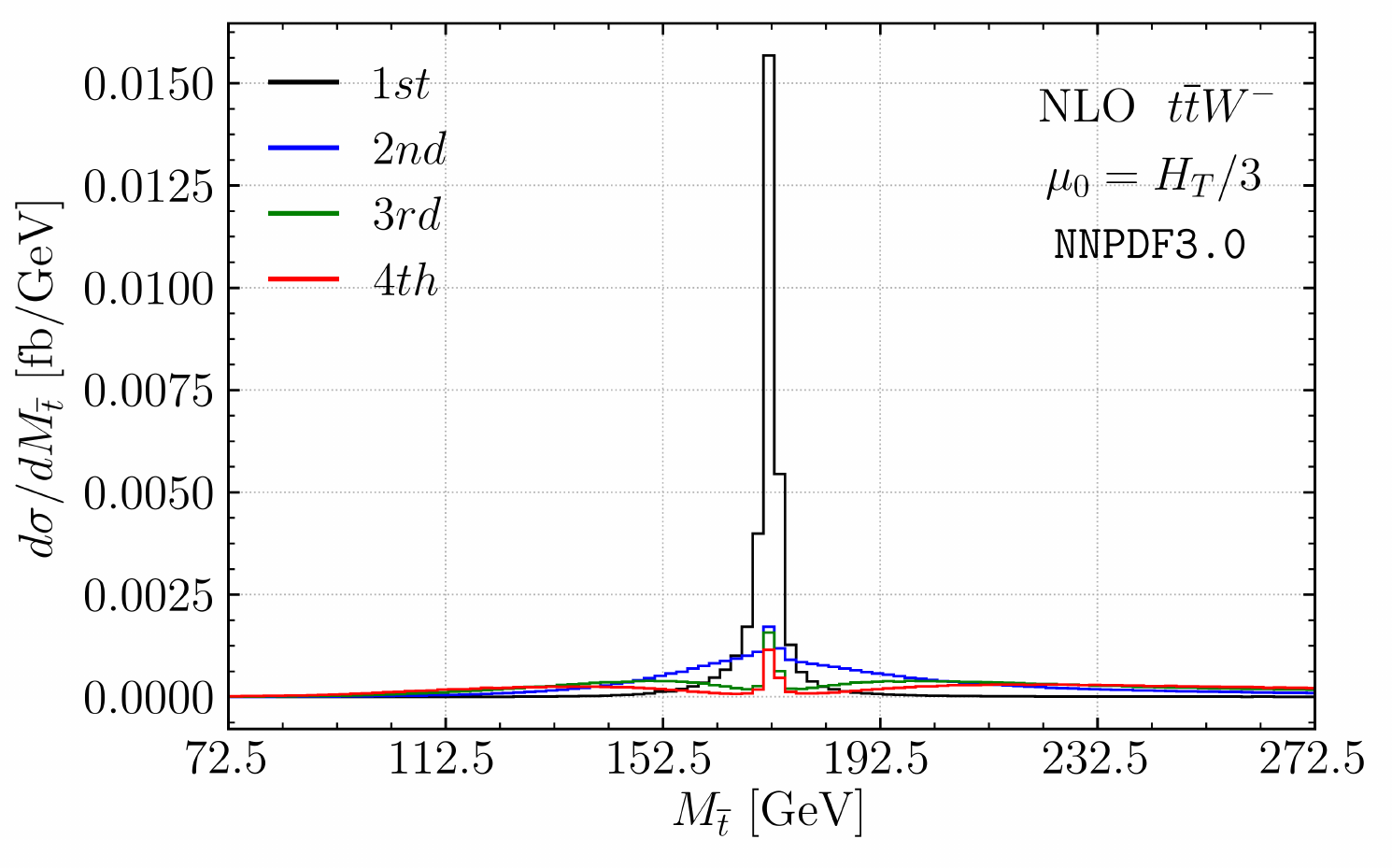}
  \end{center}
\caption{\label{fig-reconstruction} \it Reconstructed invariant mass of
  the top quark and anti-top quark at NLO in QCD  for $pp \to
  t\bar{t}W^+$ and $pp\to t\bar{t}W^-$ in the multi-lepton final
  state. Results are given for  the LHC with $\sqrt{s}=13$
TeV. The NLO NNPDF3.0 PDF set is employed and
$\mu_R=\mu_F=\mu_0$ where $\mu_0=H_T/3$.}
\end{figure}
\begin{figure}[t!]
  \begin{center}
    \includegraphics[width=0.49\textwidth]{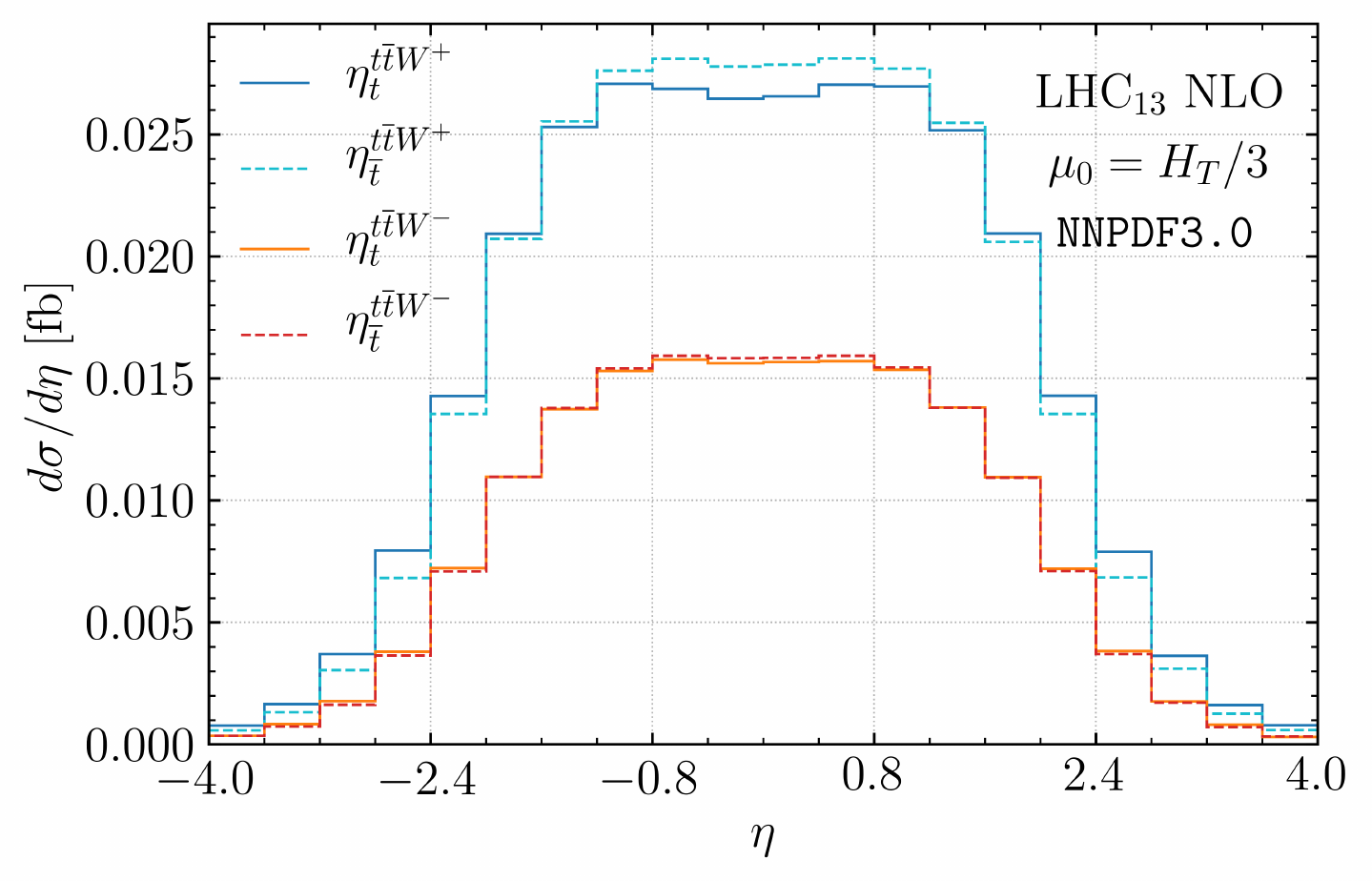}
    \includegraphics[width=0.48\textwidth]{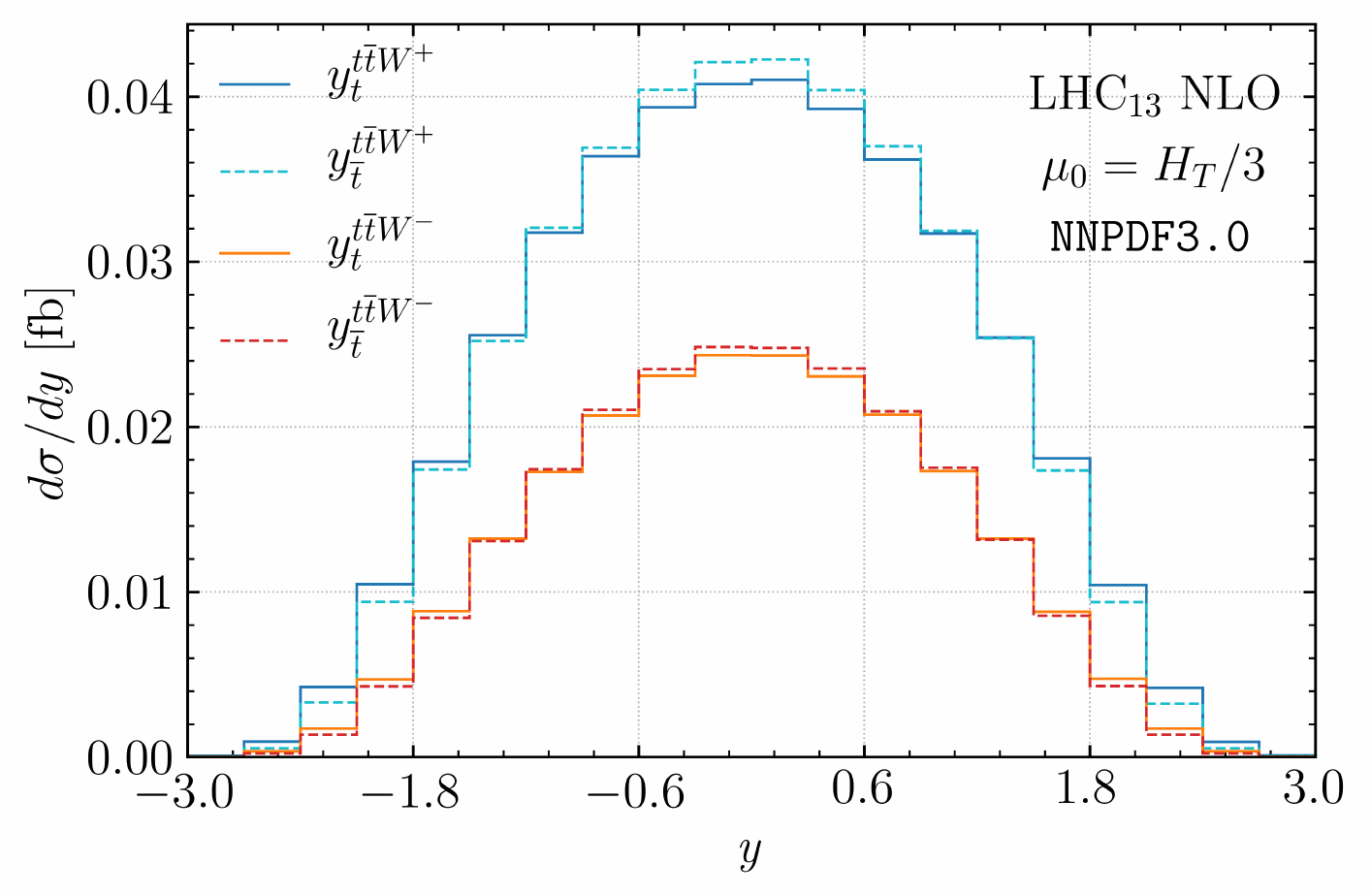}
    \includegraphics[width=0.49\textwidth]{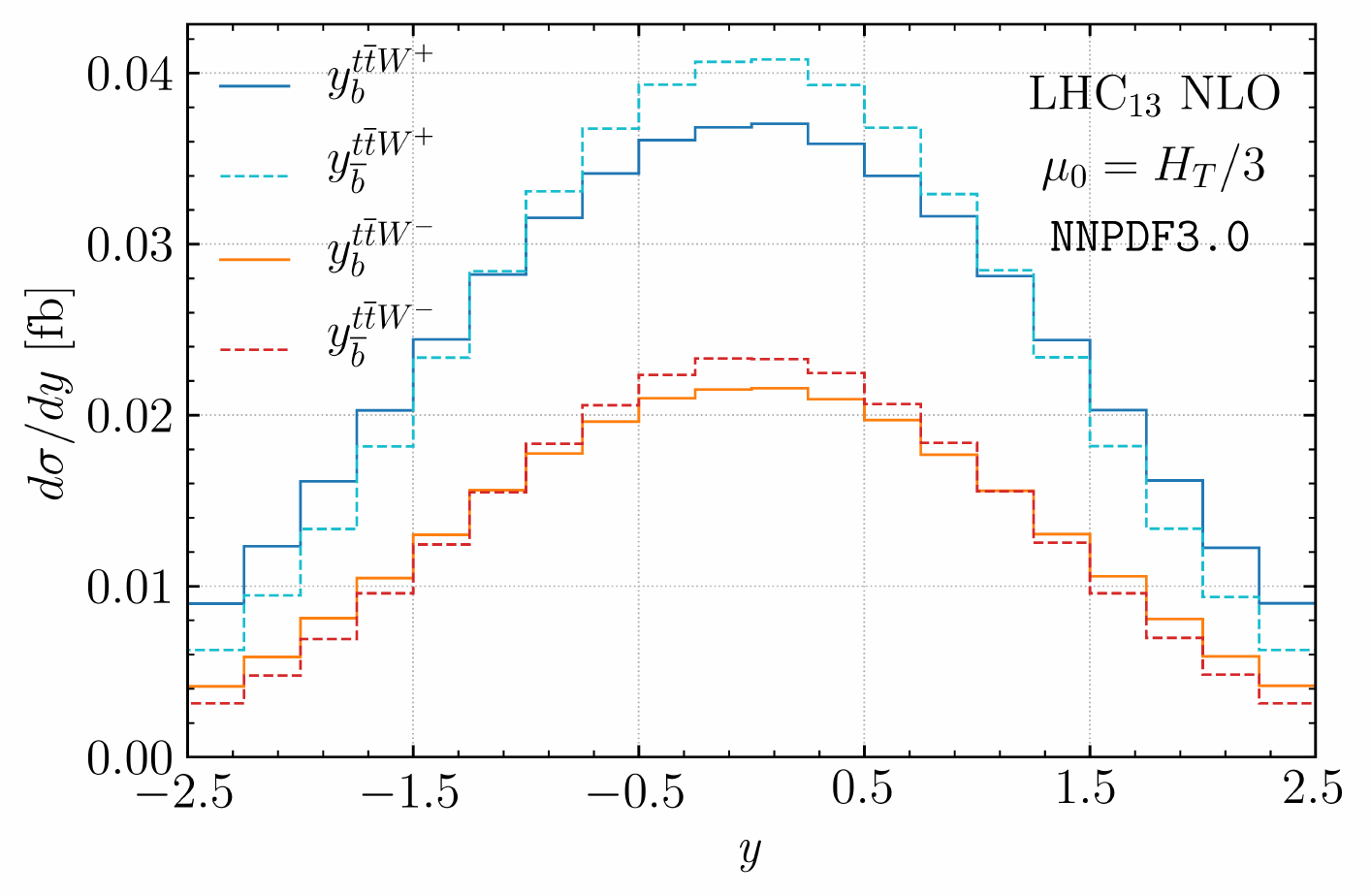}
    \includegraphics[width=0.49\textwidth]{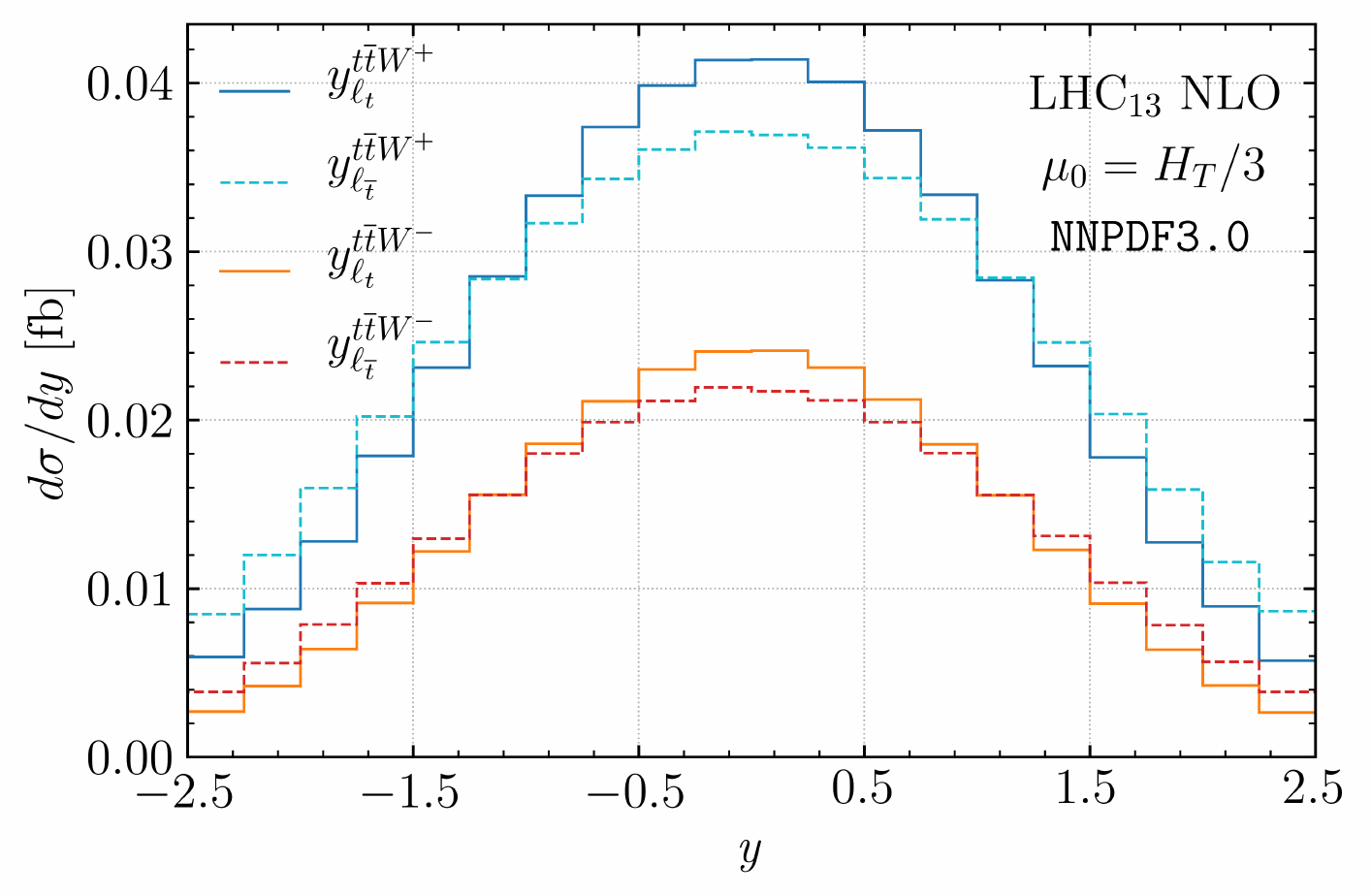}
  \end{center}
\caption{\label{fig:etay} \it Comparison of the rapidity and
pseudo-rapidity distributions of the $t$ and $\bar{t}$ quarks at the
NLO QCD level for $pp \to t\bar{t}W^+$ and $pp \to t\bar{t}W^-$ in the
multi-lepton final state. Also shown are rapidity distributions of the
charged leptons and $b$-jets from $t$ and $\bar{t}$ decays.  Results
are given for the LHC with $\sqrt{s}=13$ TeV. The NLO NNPDF3.0 PDF set
is employed and $\mu_R=\mu_F=\mu_0$ where $\mu_0=H_T/3$ is used.}
\end{figure}
%
\begin{equation}
  {\cal Q}=|M_t-m_t|+|M_{\bar{t}} -m_t|\,,
\end{equation}
where $M_t$ and $M_{\bar{t}}$ are the (reconstructed) invariant masses
of the top and anti-top quark respectively and $m_t=172.5$ GeV.  For
each phase space point we pick the history that minimises the ${\cal
Q}$ value. In this way all the (anti-)top quark decay
products are identified. They are employed in the definition of $A_c^t$,
$A_c^\ell$ and $A_c^b$.  To show how well such a reconstruction
works in Figure~\ref{fig-reconstruction} we display the reconstructed
invariant mass of the top (anti-top) quark at NLO in QCD for the $pp
\to t\bar{t}W^+$ ($pp \to t\bar{t}W^-$) process in the multi-lepton
channel. Out of all twelve histories the four histories with the
smallest ${\cal Q}$ value are shown. Clearly one can see that the
reconstruction works very well.

Using the notation of
Ref.~\cite{Czakon:2014xsa,Czakon:2016ckf,Czakon:2017lgo} we define the
top quark charge asymmetry as follows
\begin{equation}
  A_c^t= \frac{\sigma_{\rm bin}^+-\sigma_{\rm bin}^-}{
    \sigma_{\rm bin}^++\sigma_{\rm bin}^-}\,,
\quad \quad \quad \quad 
\sigma^\pm_{\rm bin} = \int \theta(\pm \Delta|y|)\, \theta_{\rm bin}
\, d\sigma\,,
\label{def:asymmetry}
\end{equation}  
where $\Delta|y| =|y_t| -|y_{\bar{t}}|$ and $d\sigma$ is the
differential fiducial $t\bar{t}W^\pm$ cross section calculated at NLO
in QCD. The binning function $\theta_{\rm bin}$ can take the values zero
or one. Its purpose is to restrict to a given bin the kinematics of
the $t\bar{t}W^\pm$ process in one of the kinematic variables that is
considered. The integrated asymmetry is obtained by setting
$\theta_{\rm bin}=1$. We note here that the charge-symmetric $gg$
initial state, that is the dominant mechanism for $t\bar{t}$
production at the LHC, is not present for $t\bar{t}W^\pm$
production. Therefore, unlike for $pp\to t\bar{t}$, it will not
contribute to the denominator of Eq.~\eqref{def:asymmetry} to dilute
the asymmetry. The LHC measurements for the top quark charge asymmetry
in $pp\to t\bar{t}$ production have been carried out in terms of
rapidity as well as pseudorapidity differences, see e.g.
\cite{Chatrchyan:2011hk,ATLAS:2012an,Khachatryan:2016ysn,
Aad:2016ove,Sirunyan:2017lvd,ATLAS-CONF-2019-026}. Even-though, the
top quark charge asymmetry based on rapidity and pseudorapidity has
the same features its value can differ quite
substantially. Consequently, we shall provide results for $A^t_c$ for
both cases. In the case of the top quark decay products $A_c^\ell$ and
$A_c^b$ are based on $\Delta|y| =|y_{\ell_t}|-|y_{\ell_{\bar{t}}}|$
and $\Delta|y| =|y_b| -|y_{\bar{b}}|$ respectively.

The top quark charge asymmetry can be visualised by superimposing the
rapidity (or the pseudo-rapidity) of $t$ and $\bar{t}$ for the
$t\bar{t}W^+$ process. The same can be done separately for
$t\bar{t}W^-$. Similarly, we can plot together the top and
anti-top quark decay products, $b$ and $\bar{b}$ as well as $\ell_t$
and $\ell_{\bar{t}}$.  In Figure \ref{fig:etay} we present
such a comparison at the NLO QCD level for the $t\bar{t}W^\pm$
process.  We can observe that all spectra are symmetric about $y=0$
$(\eta=0)$, as it should be, and that the anti-top quark is more
central with respect to the top quark. The same is visible for the
$b$-jet. This can be directly translated into the positive value of
$A_c^t$ and $A_c^b$. The situation is reversed for the charged
leptons. In the later case the charged lepton from the top quark decay
is more central, which will manifest itself in the negative value of
$A_c^\ell$.

 In Tables \ref{table:ttW+_offshell_vs_NWA_fixedscale_1}
and \ref{table:ttW+_offshell_vs_NWA_fixedscale_2}  we present our
findings for $A_c^t$, $A_c^\ell$ and $A_c^b$ at NLO QCD for
$t\bar{t}W^+$ production in the multi-lepton channel at the LHC with
$\sqrt{s}=13$ TeV. Results are given for the fixed scale choice
$\mu_R=\mu_F=m_t+m_W/2$. The top quark charge asymmetry calculated in
terms of rapidities (pseudo-rapidities) is denoted as $A^t_{c,y}$
$(A^t_{c,\eta})$. We present the results with the full off-shell
effects included as well as for the full NWA and for the NWA${}_{\rm
LOdecay}$ case. For all three approaches theoretical uncertainties due
to the scale dependence are also given. They are estimated by varying
the renormalisation and factorisation scales in $\alpha_s$ and PDFs up
and down by a factor of $2$ around the central scale of the process
$\mu_0$. We show theoretical predictions for the unexpanded and
expanded version of the asymmetry.  As the ratio in
Eq.~\eqref{def:asymmetry} generates contributions of ${\cal
O}(\alpha_s^2)$ and higher, which in principle can be affected by the
unknown NNLO contributions, we expand $A^i_{c}$ to first order in
$\alpha_s$.  The expanded version of $A^i_c$ at NLO in QCD, where $i$
stands for $i=t,\ell,b$, is defined as
\begin{equation}
  A^i_{c,exp} = \frac{\sigma_{\rm LO}^{-}}{\sigma_{\rm LO}^+} \left(
1+ \frac{\delta \sigma^-_{\rm NLO}}{\sigma^-_{\rm LO}} - \frac{\delta
  \sigma_{\rm NLO}^+}{\sigma_{\rm LO}^+}
    \right)\,.
\end{equation}  
where $\sigma^\pm$ stands for $\sigma^\pm=\sigma^+_{\rm bin} \pm
\sigma^-_{\rm bin}$ and $\delta \sigma^\pm_{\rm NLO}$ are the NLO
contributions to the fiducial cross section. Furthermore,
$\sigma^\pm_{\rm LO}$ are evaluated with NLO input parameters. In
Tables \ref{table:ttW+_offshell_vs_NWA_fixedscale_1} and
\ref{table:ttW+_offshell_vs_NWA_fixedscale_2} we include in
parenthesis the Monte Carlo (MC) integration errors to show that the
latter are smaller than or at least similar in size to the theoretical
errors from the scale dependence. Since the PDF dependence of the
asymmetry is very small (at the per-mill level) we do not quote the
PDF errors in our predictions.

 Before we analyse our results we remind the reader that the results
presented in Ref.~\cite{Maltoni:2014zpa} were generated for NLO+PS and
for a different setup and input parameters.  Thus, a direct comparison
for the absolute values for $A^t_{c,\eta}$, $A^b_{c,y}$,
$A^{\ell}_{c,y}$ is not possible, but we are rather interested in the
relative size of their theoretical uncertainties. In
Ref.~\cite{Maltoni:2014zpa} LO spin correlations are incorporated with
the help of \textsc{MadSpin}. The latter is employed before events can
be passed to \textsc{Herwig6} \cite{Corcella:2000bw}. The two charged
leptons coming from top and anti-top quark decays are chosen to be
respectively positrons and electrons, while the extra $W^\pm$ bosons
decay into muons. Consequently, leptons and $b$-jets coming from the
top and anti-top quark decays can be uniquely identified.  Issues
related to the top quark reconstruction are, therefore, not
considered. The results given in Ref.~\cite{Maltoni:2014zpa} will
serve us as a guideline in the study of the impact of the top quark
reconstruction. In particular for $A^t_{c,\eta}$ theoretical uncertainties
quoted in Ref.~\cite{Maltoni:2014zpa} are of the order of $\delta_{\rm
scale} ={}^{+19\%}_{- 14\%}$.  For $A^b_{c,y}$ and $A^\ell_{c,y}$, on
the other hand, they are $\delta_{\rm scale} ={}^{+2.5\%}_{-2.2\%}$
and $\delta_{\rm scale} ={}^{+8.5\%}_{- 6.0\%}$ respectively. The
results for $A^t_{c,\eta}$, $A^b_{c,y}$ and $A^\ell_{c,y}$ have been
given for the combined $pp\to t\bar{t}W^\pm$ case and for
$\mu_R=\mu_F=2m_t$. Furthermore, for the computation of $A^b_{c,y}$,
events that do not feature two $b$-jets coming from the top quark
decays were discarded.
%
%
\begin{table}[t!]
\renewcommand{\arraystretch}{1.6}
\begin{center}
  \begin{tabular}{ r | c | c | c }
 $t\bar{t}W^+$ & \textsc{Off-shell}   &  \textsc{Full NWA}
    &  $\textsc{NWA}_{\scriptsize \mbox{LOdecay}}$   \\
    $\mu_0=m_t+m_W/2$& & & \\
\hline
$A_{c,y}^{t}$  $[\%]$ & 
 {$2.09 (8)^{+1.06 \, (51\%)}_{-0.70\,(33\%)}$ } &
 {$1.68(4)^{+1.00(60\%)}_{-0.67(40\%)}{} $} &
 {$0.86(3)^{+0.66\, (77\%)}_{-0.43\, (50\%)}$ } \\
$A_{c,exp,y}^{t}$ $[\%]$& 
 {$2.62(10)^{+0.39\, (15\%)}_{-0.34\,(13\%)}$}  &  
 {$2.19(4)^{+0.38(17\%)}_{-0.34(16\%)}{} $} &
 {$1.94(5)^{+0.46\,(24\%)}_{-0.32\, (16\%)}$ } \\
\hline
  $A_{c,\eta}^{t}$  $[\%]$&  {$3.10(8)^{+1.21\,(39\%)}_{-0.80\,(26\%)}$} &  
 {$2.58(4)^{+1.31(51\%)}_{-0.75(29\%)}$} &
 {$1.16(4)^{+0.71\,(61\%)}_{-0.44\,(38\%)}$} \\
$A_{c,exp,\eta}^{t}$ $[\%]$& 
 {$3.70(10)^{+0.46\,(12\%)}_{-0.40\,(11\%)}$}  &  
 {$3.18(5)^{+0.56(18\%)}_{-0.34(11\%)}$} &
  {$2.25(5)^{+0.51\,(23\%)}_{-0.32\,(14\%)}$ } \\
  \end{tabular}
  \vspace{0.4cm}
\caption{\it  Unexpanded and expanded $A_c^t$ 
asymmetry at NLO in QCD for $pp \to t\bar{t}W^+$ in multi-lepton
channel at the LHC with $\sqrt{s}=13$ TeV.  Various approaches for the
modelling of the top quark production and decays are considered: the
full off-shell case, the full NWA and the NWA${}_{\rm LOdecay}$
case. Also given are Monte Carlo (in parenthesis) integration and
theoretical errors.  The NNPDF3.0 PDF set is employed and
$\mu_R=\mu_F=\mu_0$ where $\mu_0 = m_t+m_W/2$.}
\label{table:ttW+_offshell_vs_NWA_fixedscale_1}
\end{center}
\end{table}
\begin{figure}[t!]
  \begin{center}
    \includegraphics[width=0.49\textwidth]{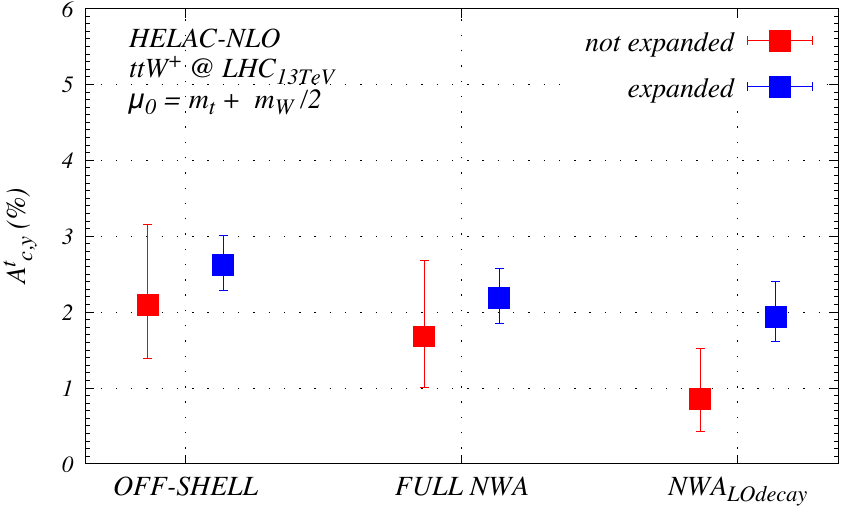}
    \includegraphics[width=0.49\textwidth]{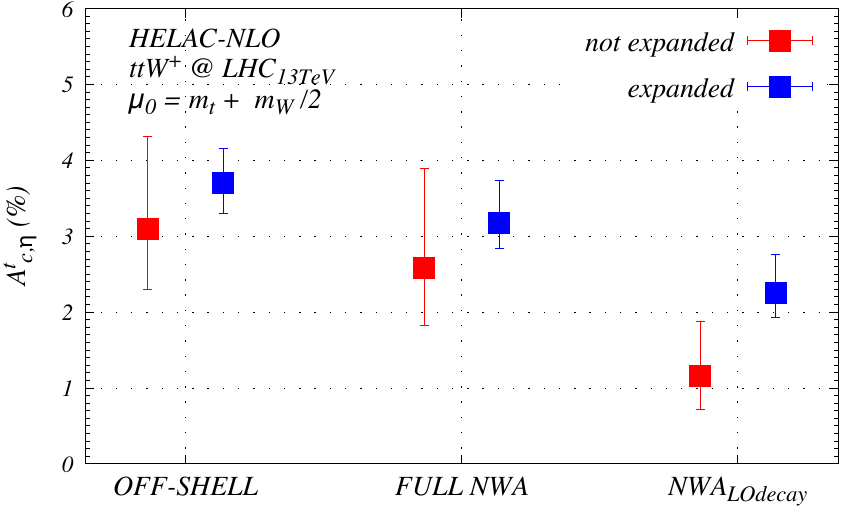}
  \end{center}
\caption{\label{fig:top-asymm} \it Unexpanded and expanded $A^t_{c}$
asymmetry at NLO in QCD for $pp \to t\bar{t}W^+$ in
the multi-lepton channel at the LHC with $\sqrt{s}=13$ TeV. Various
approaches for the modelling of the top quark production and decays
are considered: the full off-shell case, the full NWA and the
NWA${}_{\rm LOdecay}$ case. The NNPDF3.0 PDF set is employed and
$\mu_R=\mu_F=\mu_0$ where $\mu_0 = m_t+ m_W/2$.}
\end{figure}
%
%

In the following we analyse our findings for $A^t_{c}$ as
reconstructed from the top quark decay products for
$\mu_R=\mu_F=m_t+m_W/2$. They are presented in Table
\ref{table:ttW+_offshell_vs_NWA_fixedscale_1} and graphically depicted
in Figure \ref{fig:top-asymm}. First we notice that the difference
between the unexpanded NLO asymmetry $(A^t_{c})$ and the one with a
consistent expansion in $\alpha_s$ $(A^t_{c,exp})$ expressed either in
terms of rapidity $(A^t_{c,y})$ or pseudo-rapidity differences
$(A^t_{c,\eta})$, is rather moderate for the full off-shell and the
full NWA case. Specifically, the absolute value of $A^t_{c}$ increases
by $0.5\%-0.6\%$  (by about $20\%-30\%$ in the relative
terms) when the expansion is introduced. For the NWA${}_{\rm
LOdecay}$ approach, on the other hand, we observe a larger increase by
$1.1\%$. In addition, theoretical uncertainties due to the scale
dependence are substantially reduced for the expanded version of
$A^t_{c}$. In that case we have estimated that the uncertainties are
of the order of $15\%$, $18\%$ and $24\%$ depending on the approach
used. These uncertainties are similar in size to uncertainties given
in Ref. \cite{Maltoni:2014zpa} for $A^t_{c,\eta}$. Therefore, the two
definitions, $A^t_{c}$ and $A^t_{c,exp}$, give reasonably consistent
results especially in the full off-shell and full NWA case as it can
be nicely visualised in Figure \ref{fig:top-asymm}. We can also note
that the results for $A^t_{c,y}$ differ by almost $1\%$ (by about
$40\%-50\%$ in the relative terms) from those for $A^t_{c,\eta}$. Only
in the case of NWA${}_{\rm LOdecay}$ is the difference smaller,
i.e. $0.3\%$ ($15\%-35\%$ in the relative terms).  Finally, similar
remarks apply to the charge asymmetry in $t\bar{t}W^-$ production.

A few comments are in order here. The $A^t_{c}$ asymmetry in
$t\bar{t}W^\pm$ production is rather small so any effect can have
substantial influence on its absolute value. From
Ref. \cite{Broggio:2019ewu} we already know, that the inclusion of the
electroweak corrections increases the asymmetry only by a small
amount, i.e. by about $0.16\%$. On the other hand, NNLO QCD
corrections, which will stabilise the size of the theoretical
uncertainties, can play a crucial role for $A^t_{c}$. Although NNLO
QCD corrections to $t\bar{t}W^\pm$ production with $gg \to
t\bar{t}W^\pm q\bar{q}^\prime$ processes are completely symmetric and
can contribute only to the denominator of $A^t_{c}$, they might still
alter the value of $A^t_{c}$. First studies presented in
Ref. \cite{Broggio:2019ewu}, with the approximate NNLO\footnote{The
approximate NNLO predictions from Ref. \cite{Broggio:2019ewu} are
evaluated by adding to the NLO results the ${\cal O} (\alpha_s^2)$
term of the expansion of the NNLL soft gluon resummation of the cross
section. The soft gluon resummation concerns, however, only the Born
process.}  to $t\bar{t}W^\pm$ where stable top quarks and $W^\pm$
gauge bosons are considered, have shown, however, that $A^t_{c}$
increases by less by $0.6\%$ and $0.8\%$ percent for the
$m(t\bar{t}W^\pm )$ and $H_T$ based scale choices. By including these
higher order effects the scale dependence is reduced to $\delta_{\rm
scale}={}^{+6.0\%}_{-3.0\%}$ \cite{Broggio:2019ewu}. Finally, similar
effects have been observed for the $pp\to t\bar{t}$ process at the LHC
for $\sqrt{s}=8$ TeV \cite{Czakon:2017lgo} where among others the
inclusive top quark charge asymmetry $A_{c,y}^t$ and $A_{c,exp,y}^t$
at NLO QCD and NNLO QCD has been studied together with NLO electroweak
corrections.  Even-though there is no close analogy between the higher
order corrections to $t\bar{t}$ and $t\bar{t}W^\pm$ production as
these processes are dominated by different partonic channels and
should thus be regarded as uncorrelated/dissimilar, it is interesting
to see the similar (relative) size of the various effects for
$A^t_{c,y}$ and $A^t_{c,exp,y}$. The absolute value of the top quark
charge asymmetry for the $pp\to t\bar{t}$ process at the LHC is of
course much smaller.

\begin{figure}[t!]
  \begin{center}
    \includegraphics[width=0.49\textwidth]{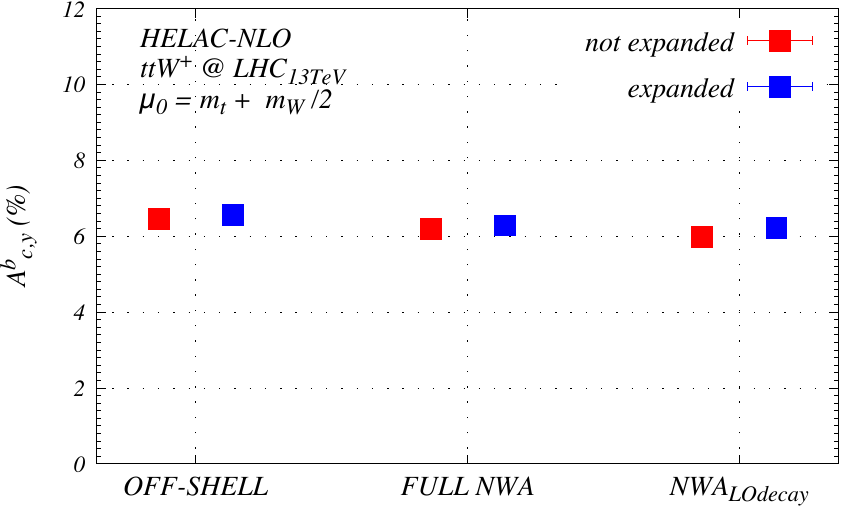}
    \includegraphics[width=0.49\textwidth]{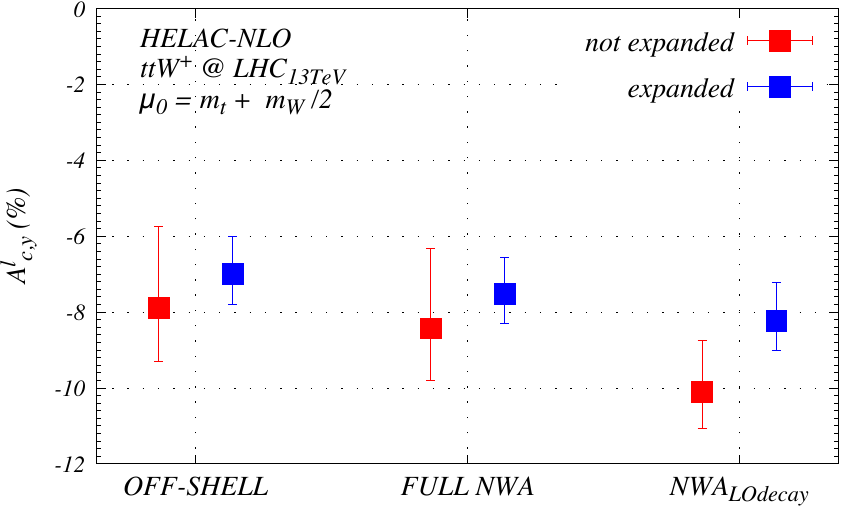}
  \end{center}
  \caption{\label{fig:lep_b-asymm} \it As in Figure
    \ref{fig:top-asymm} but for $A^b_{c}$ and $A^\ell_{c}$.}
    \end{figure}

In Table \ref{table:ttW+_offshell_vs_NWA_fixedscale_2}
and in Figure \ref{fig:lep_b-asymm} we present results for the charge
asymmetries of the top quark decay products $A^\ell_{c}$ and $A^b_{c}$
for $\mu_R=\mu_F=m_t+m_W/2$. Not only are these asymmetries much
larger, but the reconstruction of the top quarks is also not
required. Moreover, the advantage of the $A^\ell_{c}$ observable in
comparison to $A^b_{c}$ lies in the fact that measurements of leptons
are particularly precise at the LHC due to the excellent lepton energy
resolution of the ATLAS and CMS detectors. For $A^b_{c}$, on the other
hand, good $b$-jet tagging efficiency and low light jet misstag rate
would be mandatory. For the full off-shell and full NWA case the
difference between $A^{\ell}_{c}$ and  $A^{\ell}_{c,exp}$ is $0.9\%$
in absolute terms, thus within theoretical uncertainties of
$A^{\ell}_{c,exp}$ . Only in the NWA$_{\rm LOdecay}$ case does it increase
to $1.9\%$, which is above the theoretical scale uncertainty
$\delta_{scale}$ even for $A^{\ell}_{c}$. Overall, theoretical
uncertainties for $A^\ell_{c,exp}$ are below $15\%$ independently of
the approach  employed. Thus, they are slightly higher than in
Ref. \cite{Maltoni:2014zpa}. For $A^b_c$ the situation is very
stable. We observe small $0.1\%-0.2\%$ changes between $A^b_c$ and
$A^b_{c,exp}$. Theoretical uncertainties of the order of $1\%$ are
estimated, which is similar in size to theoretical errors quoted in
\cite{Maltoni:2014zpa}. We can conclude this part by saying that the
full NWA description is sufficient to describe $A^t_{c}$, $A^b_{c}$
and $A^\ell_{c}$. The inclusion of the complete off-shell
effects for the $pp \to t\bar{t}W^+$ process increases the central
values of the asymmetries while at the same time the theoretical
errors are kept almost unchanged. We have also shown that
in the case of the top quark charge asymmetry and the asymmetries of
the top quark decay products the NLO QCD corrections to the top quark
decays play a crucial role.
%
%
\begin{table}[t!]
\renewcommand{\arraystretch}{1.6}
\begin{center}
  \begin{tabular}{ r | c | c | c }
 $t\bar{t}W^+$ & \textsc{Off-shell}   &  \textsc{Full NWA}
    &  $\textsc{NWA}_{\scriptsize \mbox{LOdecay}}$   \\
    $\mu_0=m_t+m_W/2$& & & \\
\hline
$A_{c,y}^{b}$  $[\%]$& 
 {$6.46(8)^{+0.05\,(0.8\%)}_{-0.05\,(0.8\%)}$}  &  
 {$6.18(4)^{+0.13(2.1\%)}_{-0.05(0.8\%)}{} $}  &
 {$5.99(3)^{+0.10\,(1.7\%)}_{-0.01\,(0.2\%)}$ } \\
$A_{c,exp,y}^{b}$ $[\%]$& 
 {$6.56(10)^{+0.02\,(0.3\%)}_{-0.07\,(1.1\%)}$}  &  
 {$6.28(4)^{+0.03(0.5\%)}_{-0.01\,(0.1\%)}$} &
 {$6.21(5)^{+0.06\,(1.0\%)}_{-0.01\,(0.2\%)}$ }\\
\hline
$A_{c,y}^{\ell}$ $[\%]$& 
 {$-7.90(10)^{+2.15\,(27\%)}_{-1.39\,(17\%)}$}  &  
 {$-8.43(4)^{+2.10(25\%)}_{-1.37(16\%)}{} $} &
 {$-10.11(3)^{+1.36\, (13\%)}_{-0.95\,(9.4\%)}$ } \\
$A_{c,exp,y}^{\ell}$ $[\%]$& 
 {$-7.00(12)^{+1.00\,(14\%)}_{-0.80\,(11\%)}$}  &  
 {$-7.52(4)^{+0.95(13\%)}_{-0.78(10\%)}$} &
 {$-8.23(5)^{+1.01\,(12\%)}_{-0.79\,(9.6\%)}$ }\\
  \end{tabular}
  \vspace{0.4cm}
\caption{\it  Unexpanded and expanded $A_c^\ell$ and $A_c^b$
asymmetries at NLO in QCD for $pp \to t\bar{t}W^+$ in multi-lepton
channel at the LHC with $\sqrt{s}=13$ TeV.  Various approaches for the
modelling of the top quark production and decays are considered: the
full off-shell case, the full NWA and the NWA${}_{\rm LOdecay}$
case. Also given are Monte Carlo (in parenthesis) integration and
theoretical errors.  The NNPDF3.0 PDF set is employed and
$\mu_R=\mu_F=\mu_0$ where $\mu_0 = m_t+m_W/2$.}
\label{table:ttW+_offshell_vs_NWA_fixedscale_2}
\end{center}
\end{table}
%
%

Our conclusions are not changed when the dynamical scale choice,
$\mu_0=H_T/3$, is employed instead. Results for $A^t_{c,}$,
$A^\ell_{c}$ and $A^b_{c}$ at NLO in QCD with $\mu_0=H_T/3$ are shown
in Table \ref{table:ttW+_offshell_vs_NWA_dynamicscale_1} and Table
\ref{table:ttW+_offshell_vs_NWA_dynamicscale_2}. When comparing to the
theoretical predictions for $\mu_0=m_t+m_W/2$ we can notice an overall
agreement, within $0.1\sigma-0.7\sigma$, between all central values of
the asymmetries. In addition, similar theoretical uncertainties due to
the scale dependence are estimated for both scale choices.
%
\begin{table}[t!]
\renewcommand{\arraystretch}{1.6}
\begin{center}
\begin{tabular}{ r | c | c | c }
  $t\bar{t}W^+$& \textsc{Off-shell}   &  \textsc{Full NWA}  &
                                            $\textsc{NWA}_{\scriptsize
                                                              \mbox{LOdecay}}$      \\
   $\mu_0=H_T/3$& &&    \\
\hline
$A_{c,y}^{t}$ $[\%]$& 
 {$2.36(8)^{+1.19 \,(50\%)}_{-0.77\, (33\%)}$}  &  
 {$1.93(5)^{+1.23 \,(64\%)}_{-0.72 \, (37\%)}$} &
 {$1.11(3)^{+0.55\, (49\%)}_{-0.53 \, (48\%)}$ } \\ 
$A_{c,exp,y}^{t}$ $[\%]$& 
 {$2.66(10)^{+0.38\,(14\%)}_{-0.34\, (13\%)}$}  &  
 {$2.20(5)^{+0.45(20\%)}_{-0.31(14\%)}$} &
 {$2.08(5)^{+0.24\, (11\%)}_{-0.40\, (19\%)}$ }  \\
  \hline
$A_{c,\eta}^{t}$ $[\%]$ & 
 {$3.46(9)^{+1.41\,(41\%)}_{-0.90\,(26\%)}$} &  
 {$3.02(5)^{+1.44(48\%)}_{-0.93(31\%)}$} &
 {$1.42(4)^{+0.59\, (41\%)}_{-0.56\, (39\%)}$}  \\ 
$A_{c,exp,\eta}^{t}$ $[\%]$ & 
 {$3.81(10)^{+0.46\,(12\%)}_{-0.40\,(10\%)}$}  &  
 {$3.36(5)^{+0.48(14\%)}_{-0.43(13\%)}$} &
 {$2.42(5)^{+0.27\,(11\%)}_{-0.44\, (18\%)}$ }  \\ 
\end{tabular}
\vspace{0.4cm}
\caption{\it As in Table \ref{table:ttW+_offshell_vs_NWA_fixedscale_1}
  but for $\mu_0=H_T/3$.}
\label{table:ttW+_offshell_vs_NWA_dynamicscale_1}
\end{center}
\end{table}
\begin{table}[t!]
\renewcommand{\arraystretch}{1.6}
\begin{center}
\begin{tabular}{ r | c | c | c }
  $t\bar{t}W^+$& \textsc{Off-shell}   &  \textsc{Full NWA}  &
                                            $\textsc{NWA}_{\scriptsize
                                                              \mbox{LOdecay}}$      \\
   $\mu_0=H_T/3$& &&    \\
\hline
$A_{c,y}^{b}$ $[\%]$ & 
 {$6.48(9)^{+0.04\,(0.6\%)}_{-0.05\, (0.8\%)}$} &
 {$6.16(4)^{+0.07(1.1\%)}_{-0.01 \, (0.2\%)} $}  &
 {$6.05(3)^{+0.02\,(0.3\%)}_{-0.01\,(0.2\%)}$ } \\ 
$A_{c,exp,y}^{b}$ $[\%]$ & 
 {$6.53(10)^{+0.03\,(0.4\%)}_{-0.08\,(1.2\%)}$}  &  
 {$6.21(5)^{+0.09(1.4\%)}_{-0.05(0.8\%)}$} &
 {$6.23(5)^{+0.02\,(0.3\%)}_{-0.04\,(0.6\%)}$ } \\ 
\hline
$A_{c,y}^{\ell}$ $[\%]$ & 
 {$-7.46(11)^{+2.46\, (33\%)}_{-1.55\, (21\%)}$}  &  
 {$-7.94(4)^{+2.45(31\%)}_{-1.54(19\%)}$} &
 {$-9.81(4)^{+1.46\, (15\%)}_{-1.03\, (10\%)}$ } \\ 
$A_{c,exp,y}^{\ell}$ $[\%]$ & 
 {$-6.93(13)^{+1.01\, (14\%)}_{-0.81\,(12\%)}$}  &  
 {$-7.43(5)^{+0.99(13\%)}_{-0.79(11\%)}{} $} &
 {$-8.14(5)^{+1.00\, (12\%)}_{-0.81\, (10\%)}$ }  \\ 
\end{tabular}
\vspace{0.4cm}
\caption{\it As in Table \ref{table:ttW+_offshell_vs_NWA_fixedscale_2}
  but for $\mu_0=H_T/3$.}
\label{table:ttW+_offshell_vs_NWA_dynamicscale_2}
\end{center}
\end{table}

Our state-of-the art results for the top quark charge asymmetry and
for the charge asymmetries of the top quark decay products are
summarised in Table \ref{table:ttWpm_offshell}. We
provide the NLO QCD results for $A_{c,exp}^t$, $A_{c,exp}^\ell$ and
$A^b_{c,exp}$.  They are calculated from the theoretical predictions,
which include the full top quark off-shell effects. We present results
for $pp \to t\bar{t} W^+$ and $pp \to t\bar{t}W^-$ in the multi-lepton
channel at the LHC with $\sqrt{s}=13$ TeV.  We additionally present
the combined results for the $pp \to t\bar{t} W^\pm$ process.  Also in
this case the results for the top quark charge asymmetry are given in
terms of rapidities, $\Delta|y|=|y_t|-|y_{\bar{t}}|$, and
pseudo-rapidities, $\Delta |\eta|=|\eta_t|-|\eta_{\bar{t}}|$. A
comment on the difference in size of asymmetries for $pp \to
t\bar{t}W^+$ and $pp \to t\bar{t}W^-$ is in order.  The asymmetries
are larger for $pp \to t\bar{t}W^+$ than for $pp \to
t\bar{t}W^-$. Otherwise, however, they behave similarly. As pointed
out in Ref.~\cite{Maltoni:2014zpa} this can be understood by applying
an argument based on parton luminosities.  At the LO the $t\bar{t}W^+$
process is produced predominantly via $u\bar{d}$ whereas for
$t\bar{t}W^-$ the $\bar{u}d$ subprocess is the most relevant one. The
longitudinal momenta of the initial partons are on average $p_u > p_d>
p_{\bar{u}}\approx p_{\bar{d}}$.  In both cases the momentum of the
top and anti-top quarks is connected to the momentum of the $q$ and
$\bar{q}$ respectively. The large longitudinal momentum transferred to
the top quark from the initial $u$ quark in the $t\bar{t}W^+$ case
increases the corresponding $|y_t|$ value. Consequently, the charge
asymmetry of the top quark is enhanced compared to the one calculated
for $t\bar{t}W^-$.  When analysing the combined results for $pp \to
t\bar{t}W^\pm$ we can observe that the theoretical uncertainties due
to the scale dependence reach up to $15\%$. The scale choice plays no
role here as for $\mu_0=m_t+m_W/2$ and $\mu_0=H_T/3$ similar results
are obtained.
%
\begin{table}[h!]
\renewcommand{\arraystretch}{1.6}
\begin{center}
\begin{tabular}{ c | c | c | c }
  $\mu_0=m_t+m_W/2$& $t\bar{t}W^+$  &  $t\bar{t}W^-$  & $t\bar{t}W^\pm$     \\
\hline
$A_{c,exp,y}^{t}$ $[\%]$ & {$~~~2.62^{+0.39\, (15\%)}_{-0.34\,(13\%)}$} 
 & $~~~1.97^{+0.31\,(16\%)}_{-0.25\,(13\%)}$
& $~~~2.40^{+0.37\,(15\%)}_ {-0.31\,(13\%)}$\\
$A_{c,exp,\eta}^{t}$ $[\%]$ &  {$~~~3.70^{+0.46\,(12\%)}_{-0.40\,(11\%)}$} 
 &   $~~~1.31^{+0.32\,(24\%)}_{-0.25\,(19\%)}$
& $~~~2.87^{+0.41\,(14\%)}_{-0.35\,(12\%)}$ \\ 
$A_{c,exp,y}^{b}$ $[\%]$ &  {$~~~6.56^{+0.02\,(0.3\%)}_{-0.07\,(1.1\%)}$}
 & $~~~4.80^{+0.05\,(1.0\%)}_{-0.05\,(1.0\%)}$&
 $~~5.93^{+0.03\,(0.5\%)}_{-0.08\,(1.3\%)}$\\ 
$A_{c,exp,y}^{\ell}$ $[\%]$ &  {$-7.00^{+1.00\,(14\%)}_{-0.80\,(11\%)}$} 
  & $-5.68^{+0.78\,(14\%)}_{-0.61\,(11\%)}$ &
  $-6.51^{+0.93\,(14\%)}_{-0.74\,(11\%)}$\\
  &&&\\
$\mu_0=H_T/3$& $t\bar{t}W^+$  &  $t\bar{t}W^-$  & $t\bar{t}W^\pm$   \\
\hline
$A_{c,exp,y}^{t}$ $[\%]$ & 
 {$~~~2.66^{+0.38\,(14\%)}_{-0.34\, (13\%)}$}  &
$~~~2.05^{+0.33\,(16\%)}_{-0.27\,(13\%)}$
& $~~~2.45^{+0.37\,(15\%)}_{-0.31\,(13\%)}$ \\
$A_{c,exp,\eta}^{t}$ $[\%]$ & 
{$~~~3.81^{+0.46\,(12\%)}_{-0.40\,(10\%)}$}
 &   $~~~1.31^{+0.33\,(25\%)}_{-0.26\,(20\%)}$
& $~~~2.94^{+0.42\,(14\%)}_{-0.35\,(12\%)}$ \\ 
$A_{c,exp,y}^{b}$ $[\%]$ & 
 {$~~~6.53^{+0.03\,(0.4\%)}_{-0.08\,(1.2\%)}$}  &
 $~~~4.80^{+0.06\,(1.2\%)}_{-0.11\,(2.3\%)}$
& $~~~5.91^{+0.04\,(0.7\%)}_{-0.09\,(1.5\%)}$\\ 
$A_{c,exp,y}^{\ell}$ $[\%]$ & 
{$-6.93^{+1.01\, (14\%)}_{-0.81\,(12\%)}$}
 & $-5.67^{+0.81\,(14\%)}_{-0.63\,(11\%)}$
 & $-6.46^{+0.95\,(15\%)}_{-0.75\,(12\%)}$ \\ 
\end{tabular}
\vspace{0.4cm}
\caption{\it Expanded $A_c^t$, $A_c^\ell$ and $A_c^b$ at NLO in QCD
for $pp \to t\bar{t}W^+$ and $pp \to t\bar{t}W^-$ in the multi-lepton
channel at the LHC with $\sqrt{s}=13$ TeV.  Results are obtained with
the full off-shell effects included. Also given are combined results
for $pp \to t\bar{t} W^\pm$ and theoretical uncertainties. The
NNPDF3.0 PDF set is employed and $\mu_R=\mu_F=\mu_0$ where
$\mu_0=m_t+m_W/2$ and $\mu_0=H_T/3$.}
\label{table:ttWpm_offshell}
\end{center}
\end{table}

%
\section{Differential  and
  cumulative asymmetry} 
\label{sec:diff}
%

\begin{figure}[t!]
  \begin{center}
    \includegraphics[width=0.49\textwidth]{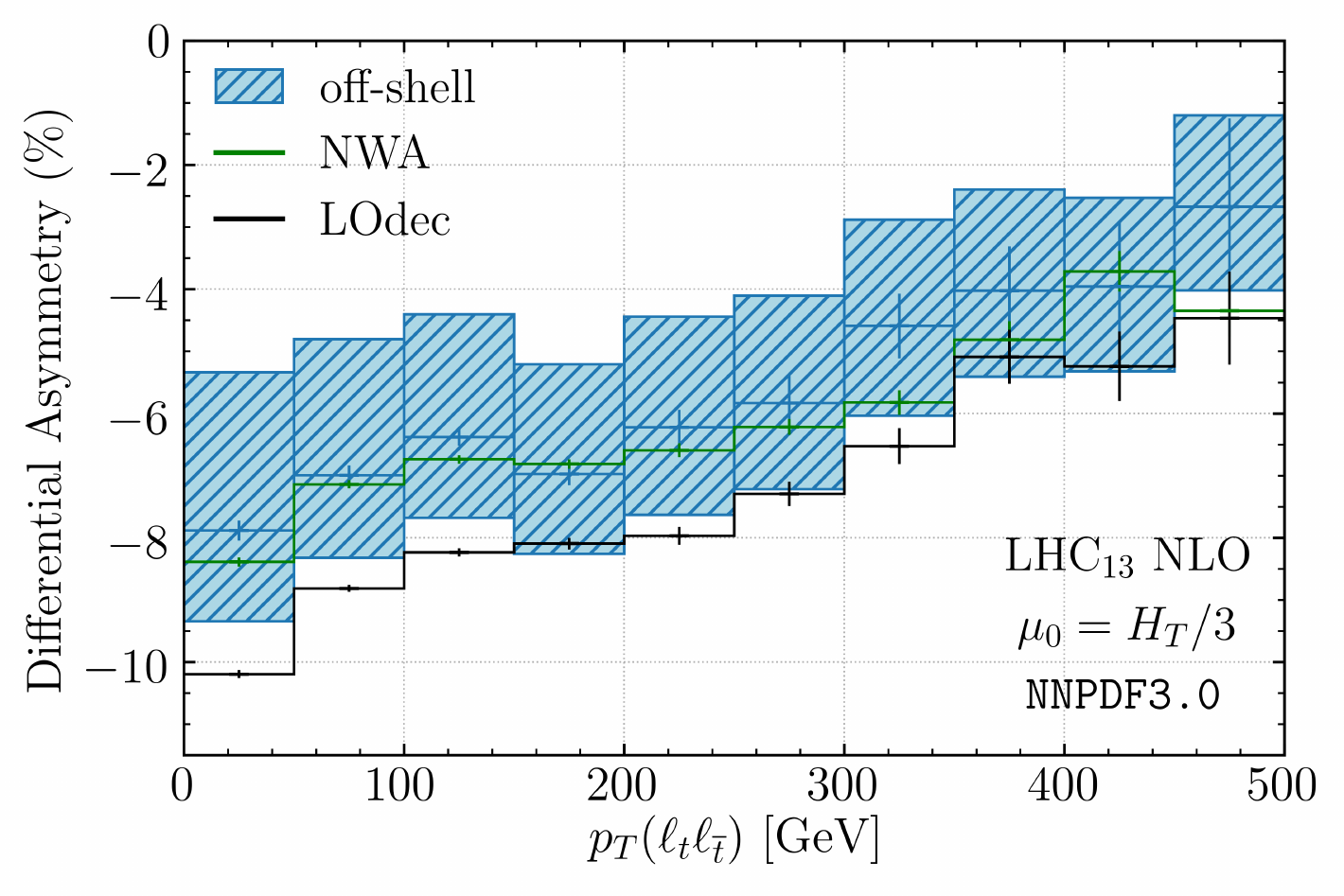}
    \includegraphics[width=0.49\textwidth]{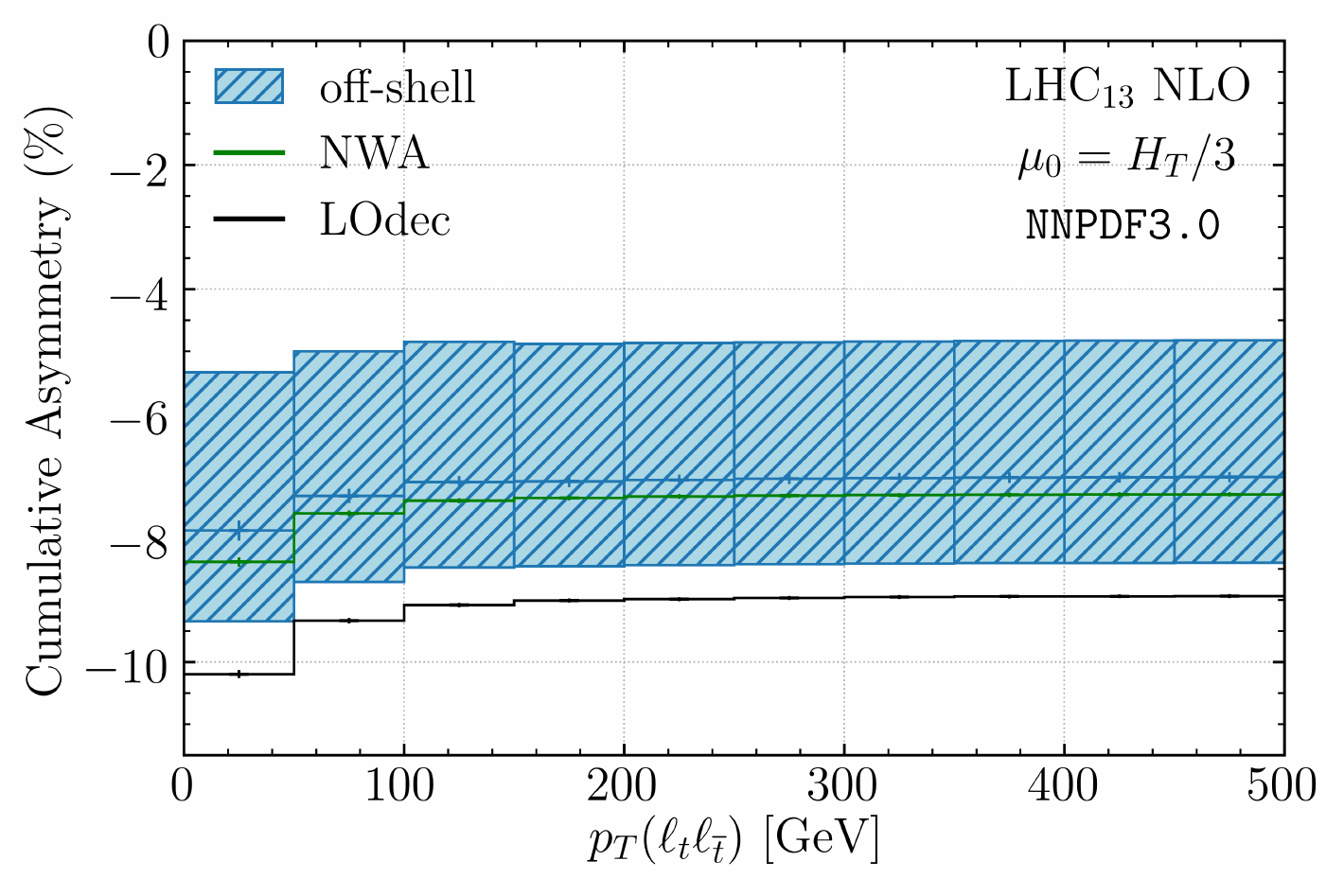}
  \end{center}
\caption{\label{fig:diff_assymetries_ttwpm_ptll} \it The
$p_{T}(\ell_t\ell_{\bar{t}})$-dependent differential (left panel) and
cumulative (right panel) $A_c^\ell$ asymmetry at NLO QCD for $pp \to
t\bar{t}W^\pm$ in the multi-lepton channel at the LHC with $\sqrt{s}=13$
TeV. Various approaches for the modelling of the top quark production
and decays are considered. Also given are theoretical uncertainties
for the full off-shell case. For all approaches Monte Carlo errors are
provided for both differential and cumulative asymmetries.  The
NNPDF3.0 PDF set is employed and $\mu_R=\mu_F=\mu_0$ where
$\mu_0=H_T/3$.}
\end{figure}
\begin{figure}[t!]
  \begin{center}
    \includegraphics[width=0.49\textwidth]{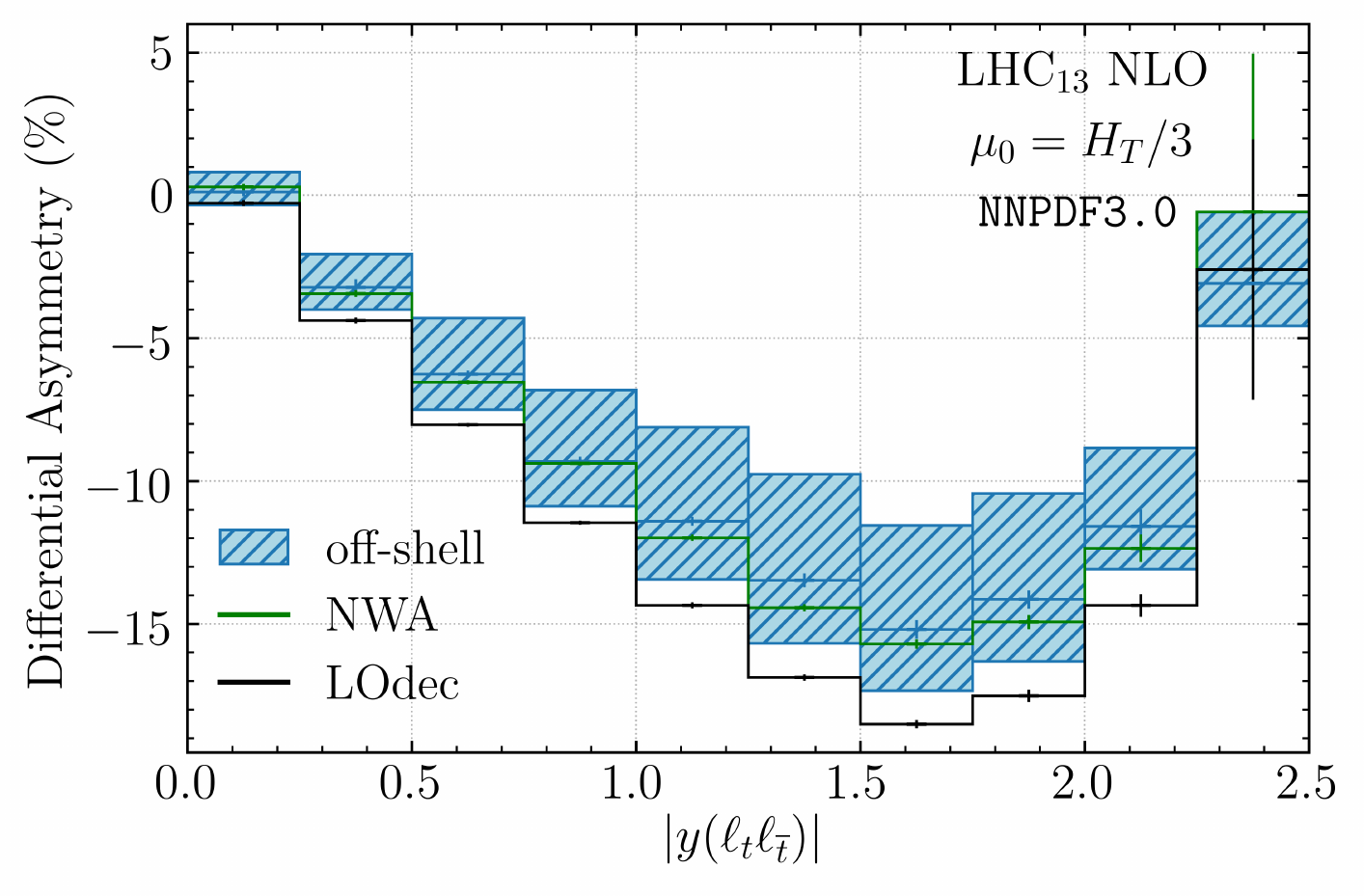}
    \includegraphics[width=0.49\textwidth]{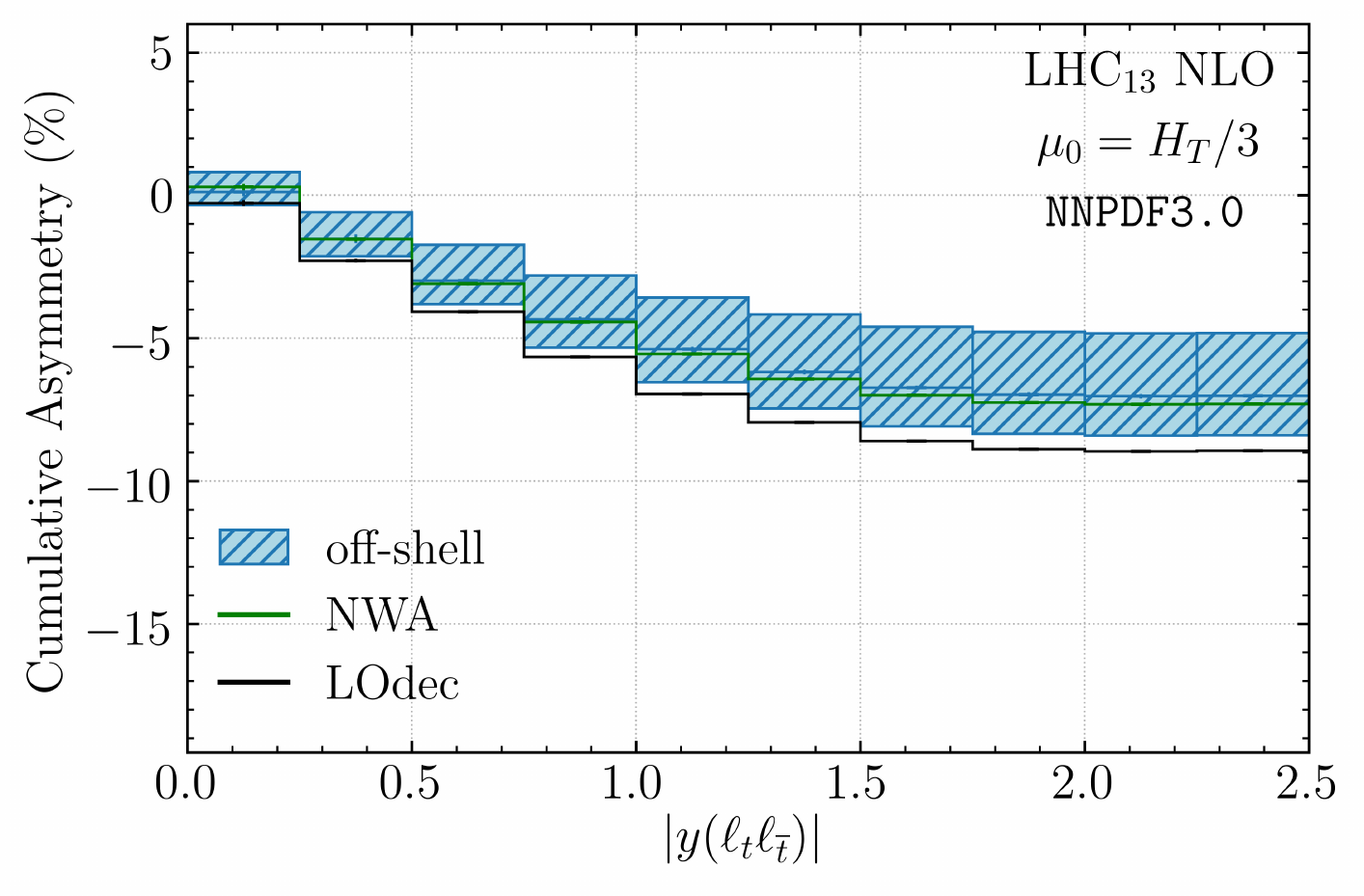}
  \end{center}
\caption{\label{fig:diff_assymetries_ttwpm_yll} \it  As in Figure
  \ref{fig:diff_assymetries_ttwpm_ptll} but for $|y(\ell_t\ell_{\bar{t}})|$.}
\end{figure}
\begin{figure}[t!]
  \begin{center}
    \includegraphics[width=0.49\textwidth]{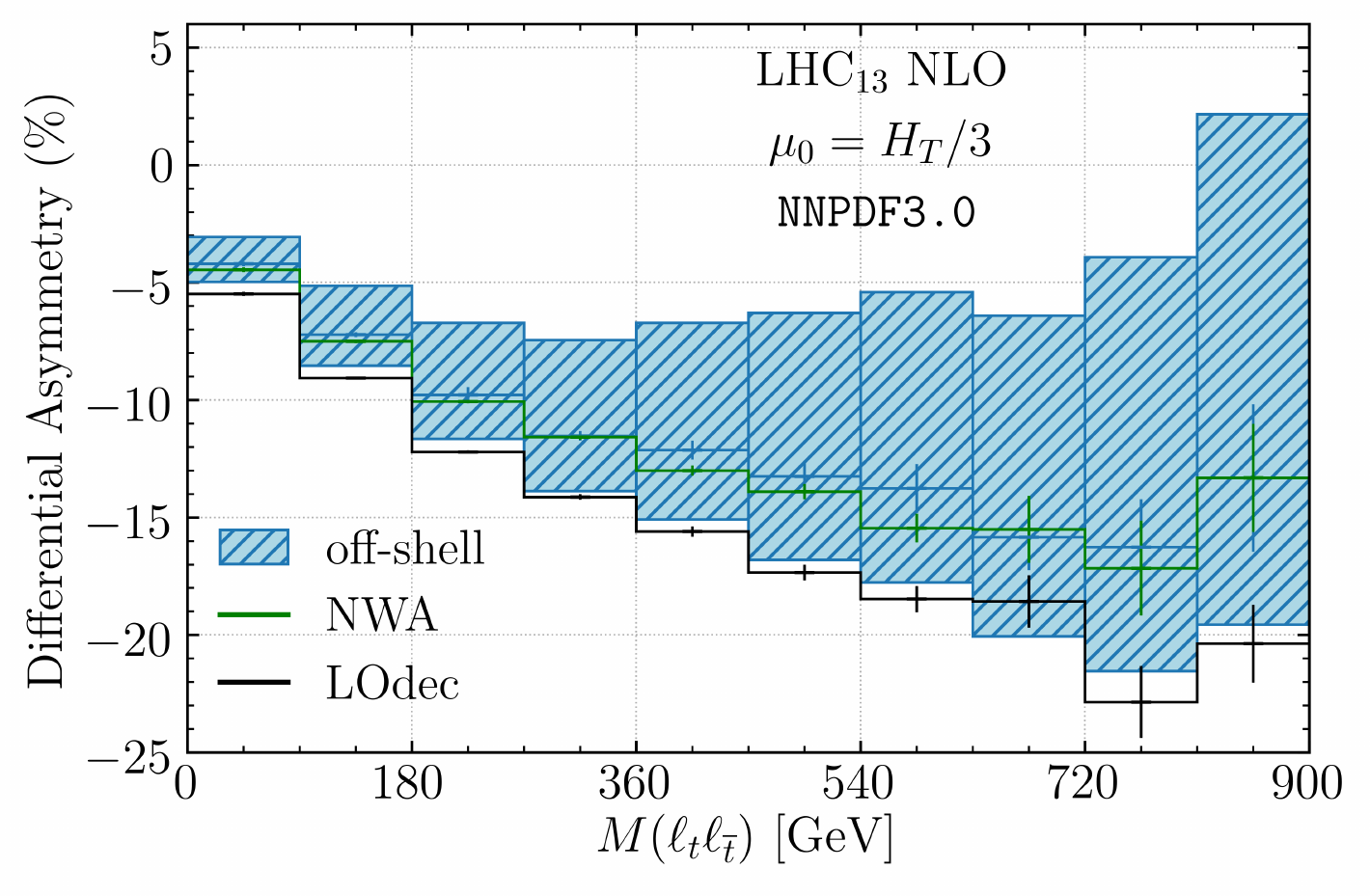}
    \includegraphics[width=0.49\textwidth]{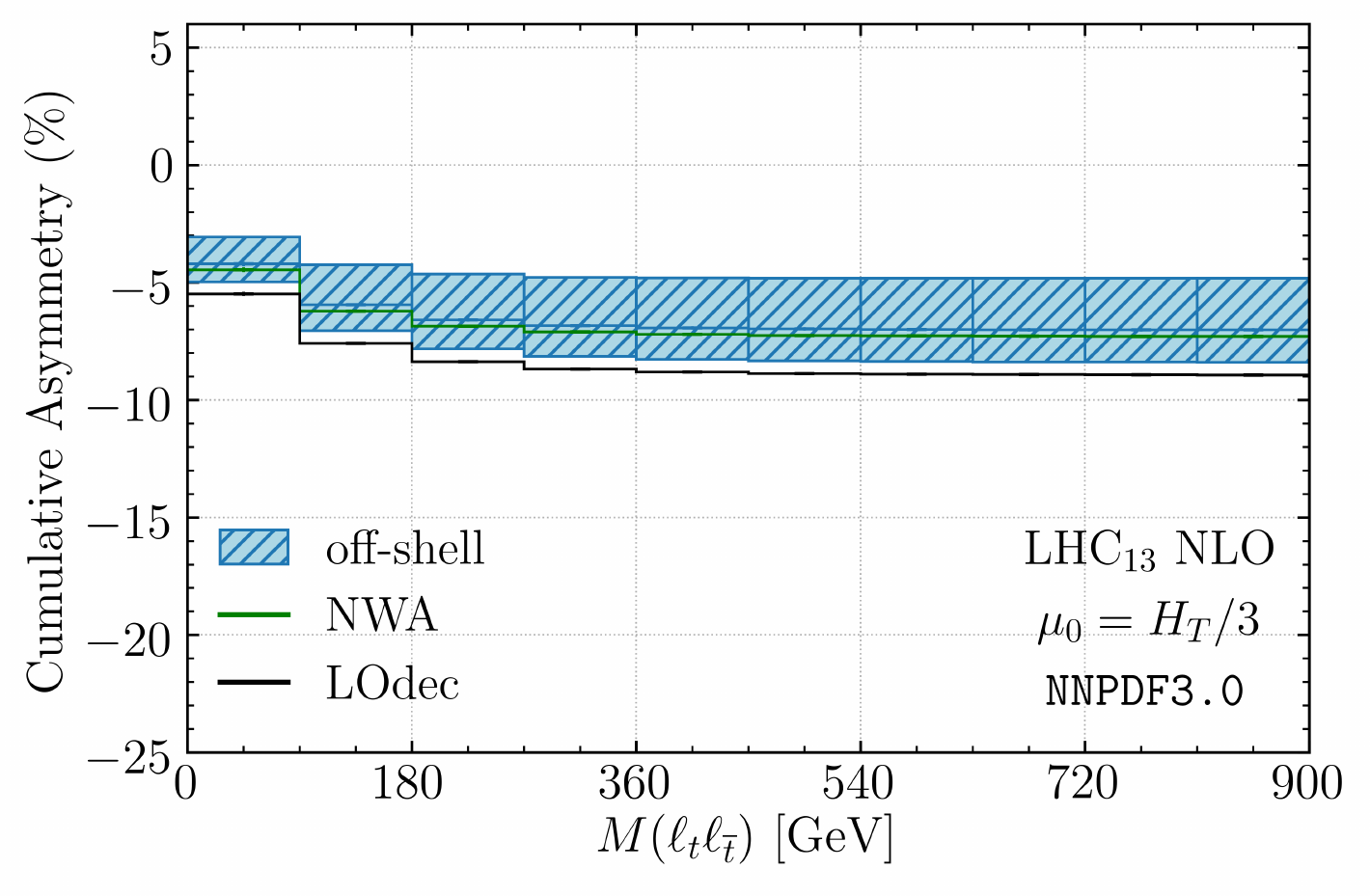}
  \end{center}
\caption{\label{fig:diff_assymetries_ttwpm_mll} \it  As in Figure
  \ref{fig:diff_assymetries_ttwpm_ptll} but for $M(\ell_t\ell_{\bar{t}})$.}
\end{figure}

In this part of the paper we present predictions for differential
$A_c^\ell$ asymmetry with respect to the following observables:
transverse momentum of the two charged leptons,
$p_T(\ell_t\ell_{\bar{t}})$, rapidity of the two charged leptons,
$|y(\ell_t\ell_{\bar{t}})|$, and invariant mass of the two charged
leptons, $M({\ell_t\ell_{\bar{t}}})$, where $\ell_t\ell_{\bar{t}}$
originate from the $t\bar{t}$ pair. The differential results are given
using the unexpanded definition from Eq.~\eqref{def:asymmetry}. We
also present predictions for cumulative asymmetries, that are closely
related to the corresponding differential asymmetries. One can employ
the same definition as in Eq.~\eqref{def:asymmetry}, however, this
time for a given value of the kinematic variable for which we compute
the asymmetry the bin ranges from zero to that value. Even though
differential and cumulative asymmetries contain the same information,
the latter one behaves better simply because it is more inclusive,
i.e.  the higher order corrections are distributed more uniformly over
the whole kinematic range. In addition, the cumulative asymmetry
should give the integrated one in the last bin assuming that
the plotted range of the corresponding differential distribution
covers the whole available phase space. In practice, we shall see that
if the left-over phase space region is negligible the integrated
asymmetry can be recovered very accurately. Let us note that
differential asymmetries have been studied at the LHC for the $pp\to
t\bar{t}$ production process by both experimental collaborations ALTAS
and CMS, see e.g. \cite{Khachatryan:2015oga,Sirunyan:2017lvd}.

In Figure \ref{fig:diff_assymetries_ttwpm_ptll} the
$p_{T}(\ell_t\ell_{\bar{t}})$-dependent differential and cumulative
$A_c^\ell$ asymmetry at NLO QCD for $pp \to t\bar{t}W^\pm$ in
multi-lepton channel at the LHC with $\sqrt{s}=13$ TeV is displayed.
Various approaches for the modelling of the top quark production and
decays are considered. Also given are theoretical uncertainties for
the full off-shell case. For all approaches Monte Carlo integration
errors are provided for both differential and cumulative asymmetries.
For the $p_T(\ell_t\ell_{\overline{t}})$-dependent differential
asymmetry the MC error is smaller than the theoretical one in all bins
but the last.  In the last bin both uncertainties are comparable in
size.  The NNPDF3.0 PDF set is employed and $\mu_R$ as well as $\mu_F$
are set to the common value $\mu_R=\mu_F=\mu_0=H_T/3$. For the
differential $A_c^\ell$ asymmetry the difference between the full
off-shell result and the full NWA case is in the $5\%-30\%$ range
depending on the bin, yet within theoretical uncertainties, that are
of the order of $30\%$. We notice that this is not the case for the
last bin where the top quark off-shell effects affect $A_c^\ell$
substantially. Specifically, they are above $60\%$. Also theoretical
uncertainties increase in that bin and are of the order of
$50\%$.  Both values, however,  are  harder to specify more
precisely due to the large statistical errors.  The NWA${}_{\rm LOdecay}$
case, on the other hand, is outside the
scale dependence bands almost in the whole plotted range. The
difference to the full off-shell approach is larger, even up to
$70\%$. A similar effect is also visible for the cumulative asymmetry
where a rather constant $30\%$ difference is noted for the NWA${}_{\rm
LOdecay}$ case. Finally, we note that the last bin of the cumulative
$A_c^\ell$ asymmetry gives $A_c^\ell = -7.02(8)$ for the complete
off-shell case, $A_c^\ell = -7.30(4)$ for the full NWA and $A_c^\ell =
-8.94(3)$ for the NWA${}_{\rm LOdecay}$ case, where in parentheses the
MC error is displayed.  All three results are indeed in perfect
agreement within the MC errors with the corresponding results for the
unexpanded leptonic charge asymmetry for the combined $pp\to
t\bar{t}W^\pm$ process.

Similar observations can be made for the other two differential and
cumulative $A_c^\ell$ asymmetries.  The $|y(\ell_t\ell_{\bar{t}})|$-
and $M(\ell_t\ell_{\bar{t}})$-dependent versions of $A_c^\ell$ are
exhibited in Figure~\ref{fig:diff_assymetries_ttwpm_yll} and
Figure~\ref{fig:diff_assymetries_ttwpm_mll} respectively. Since in
each case the $y$ axis is chosen to be the same
for the differential and cumulative version of $A_c^\ell$ we can
distinctly observe that the cumulative asymmetry has smaller
fluctuations of the theoretical errors and it is smoother due to
better statistical errors.  Taking the differential
$M(\ell_t\ell_{\bar{t}})$-dependent version of $A_c^\ell$ as an
example we observe large, of the order of $100\%$, theoretical
uncertainties at the tails. On the other hand, the cumulative
$M(\ell_t\ell_{\bar{t}})$-dependent $A_c^\ell$ asymmetry has stable
theoretical uncertainties of the order of $30\%$ in the whole plotted
range.

We summarise by noting, that several processes beyond the SM can alter
$A_c$, see e.g.
\cite{Jung:2011zv,Ferrario:2008wm,AguilarSaavedra:2011hz,
Maltoni:2014zpa,Alvarez:2020ffi}, either with anomalous vector or
axial-vector couplings or via interference with SM
processes. Different models also predict different asymmetries as a
function of the invariant mass and the transverse momentum, see
e.g. \cite{AguilarSaavedra:2011ci}.  Of course due to the much smaller
cross section the $pp\to t\bar{t}W^\pm$ process will not replace the
use of the asymmetries in $t\bar{t}$ production, however, it can
provide a complementary tool as it is uniquely sensitive to the chiral
nature of possible new physics that might manifest itself in this
channel.  This motivates our interest in the top quark charge
asymmetry as well as the asymmetries of its decay products and their
sensitivity to the top quark production and decay
modelling. Furthermore, using our NLO QCD results with the full 
 off-shell effects included, we are able to provide more precise
theoretical predictions for $A_c^t$, $A_c^\ell$ and $A_c^b$ in the
$t\bar{t}W^\pm$ production process at the LHC with $\sqrt{s}=13$
TeV. Finally, having at hand the full theory with no approximations
included we are able to study the real size of theoretical
uncertainties due to the scale dependence. In other words to verify
whether they are under- or overestimated in the presence of
various approximations.

%
\section{Summary}
\label{sec:summary}
%

In this paper we provided the state-of-the-art theoretical predictions
for observables, which might be used to constrain numerous new physics
scenarios in the $t\bar{t}W^\pm$ channel.  We considered the
$t\bar{t}W^\pm$ production process in the multi-lepton decay channel
for the LHC Run II energy of $\sqrt{s}=13$ TeV for which discrepancies
in the overall normalisation and in the modelling of the top quark
decays have been recently reported by the ATLAS collaboration. Without
the need of including terms beyond NLO in the perturbative expansion
in $\alpha_s$ we obtained $1\%-2\%$ theoretical uncertainties due to
the scale dependence for this process by calculating the following
cross section ratio ${\cal
R}=\sigma_{t\bar{t}W^+}/\sigma_{t\bar{t}W^-}$.  The PDF
uncertainties for ${\cal R}$ are similar in size.  Fully realistic NLO
QCD calculations have been employed in our studies for both
$t\bar{t}W^+$ and $t\bar{t}W^-$. Specifically, we use $e^+\nu_e \,
\mu^-\bar{\nu}_\mu \, e^+\nu_e\, b\bar{b}$ and $e^-\bar{\nu}_e \,
\mu^+{\nu}_\mu \, e^-\bar{\nu}_e\, b\bar{b}$ matrix elements in our
NLO QCD calculations. They include all resonant and non-resonant top
quark and $W$ gauge boson Feynman diagrams, their interference effects
as well as off-shell effects of $t$ and $W$. We examined the fixed and
dynamical scale choice for $\mu_R$ and $\mu_F$ to assess their impact
on the cross section ratio. We noticed that the scale choice does not
play any role for such an inclusive observable. Indeed, although the
scale variation is taken correlated in both cases, the fact that the
errors come out the same in both cases means that tails of
distributions, where the processes show more differences do not matter
for the analysis. Otherwise, the error estimate for the fixed scale
should be much larger, or the value of the ratio and asymmetries
should be shifted outside the error bands. In the next step we
examined the impact of the top quark production and decay modelling on
the cross section ratio.  We observed that the full NWA approach does
not modify either the value or the size of the theoretical error for
the integrated cross section ratio. Even for the simplified version of
the NWA, i.e. for the NWA${}_{\rm LOdecays}$ case, no changes have
been observed. Thus, the ${\cal
R}=\sigma_{t\bar{t}W^+}/\sigma_{t\bar{t}W^-}$ observable is very
stable and insensitive to the details of the modelling of the top
quark decay chain. As such, it can be safely exploited at the LHC
either for the precision SM measurements or in searches for BSM
physics. The ${\cal R}$ observable can be used, for example, to
provide valuable input for the up and down quark parton distribution
functions.  In the case of new physics searches the presence of two
same-sign leptons in the final state offers a very interesting
signature, that has been highly scrutinised in many new physics
models. The latter range from supersymmetry and supergravity to the
more specific scenarios with the Majorana neutrinos and the modified
Higgs boson sector.  Given the final accuracy of ${\cal R}$ and its
insensitivity to the top quark modelling, the ${\cal R}$ observable
might be used at the LHC, for example to achieve more stringent limits
on the parameter space of these models.

In the second part of the paper we reexamined the top
quark charge asymmetry and the charge asymmetries of the top quark
decay products for $t\bar{t}W^+$, $t\bar{t}W^-$ and
$t\bar{t}W^\pm$ production in the fiducial phase-space regions.  Also in
this case theoretical predictions with the full off-shell effects were
utilised.  We presented predictions for the expanded and unexpanded
asymmetries. Overall, good agreement has been found for $A^t_{c}$,
$A^\ell_{c}$ and $A^b_{c}$ when comparing the full off-shell case with
the full NWA approach. For the NWA${}_{\rm LOdecay}$ case, however,
discrepancies between central values of the asymmetries even up to
$2\sigma$ have been found. The later fact indicates that NLO QCD
corrections to the top quark decays play a crucial role here.
Generally, the inclusion of the complete description for
the $pp \to t\bar{t}W^\pm$ process in the multi-lepton final state has
increased the central values of the asymmetries keeping at the same
time the theoretical errors unchanged and below $15\%$.   The scale
choice has played no role as for $\mu_0 = m_t + m_W /2$ and $\mu_0 =
H_T /3$ similar results have been obtained for $A_c^t$, $A_c^\ell$ and
$A_c^b$.

As a bonus of our study, we presented predictions for
the differential and cumulative $A_c^\ell$ asymmetry with respect to
$p_T(\ell_t \ell_{\bar{t}})$, $|y(\ell_t \ell_{\bar{t}})|$ and
$M(\ell_t \ell_{\bar{t}})$.  The advantage of choosing $A_c^\ell$ lies
in the fact that the measurements of the charged leptons are
particularly precise at the LHC due to the excellent lepton energy
resolution of the ATLAS and CMS detectors. We note here that for these
studies the unexpanded version of $A_c^\ell$ has been examined.
Depending on the bin the differences between the full off-shell
results and the full NWA ones have been in the $5\%- 30\%$ range.
However, this is well within theoretical uncertainties, that are of
the order of $30\%$. On the other hand, large differences have been
noticed for the NWA$_{\rm LOdecay}$ case even up to $70\%$. Similarly
for the cumulative asymmetry the NWA$_{\rm LOdecay}$ curves are lying
outside the uncertainty bands independently of the observable and the
considered bin. We would like to add here that even though
differential and cumulative asymmetries contain the same information,
the latter behaves better simply because it is more inclusive. In
other words, the higher order corrections are distributed more
uniformly over the whole kinematic range.

Last but not least,  we would like to mention at this
point that, several BSM physics scenarios can alter the top quark
charge asymmetry.  Thus, theoretical predictions for the $A_c^t$,
$A_c^\ell$ and $A_c^b$ observables  should be as
accurate as possible. Using our NLO QCD results with the full
off-shell effects included not only are we able to provide the
state-of-the-art theoretical predictions for $A_c^t$, $A_c^\ell$ and
$A_c^b$ in the $t\bar{t}W^\pm$ production process but 
also by the explicit comparison to various NWA approaches we could
carefully examine  the impact of different top-quark
  decay modelling accuracies on the scale uncertainties.

\acknowledgments

This research of H.Y.B., J.N. and M.W was supported by the Deutsche
Forschungsgemeinschaft (DFG) under the following grants: 400140256 -
GRK 2497: {\it The physics of the heaviest particles at the Large
Hardon Collider} and 396021762 - TRR 257: {\it P3H - Particle Physics
Phenomenology after the Higgs Discovery}. Support by a grant of the
Bundesministerium f\"ur Bildung und Forschung (BMBF) is additionally
acknowledged.

The work of G.B. was supported by grant K 125105 of the National
Research, Development and Innovation Office in Hungary.

H.B.H. has received funding from the European Research Council (ERC)
under the European Union's Horizon 2020 Research and Innovation
Programme (grant agreement no. 683211 and 772099). Furthermore, the
work of H.B.H has been partially supported by STFC consolidated HEP
theory grant ST/T000694/1.

Simulations were performed with computing resources granted by RWTH
Aachen University under project {\tt rwth0414}.

\appendix
\section{Kolmogorov-Smirnov test}
\label{appendix}

In section \ref{sec:correl} we have only used visual inspection to see
whether two given one-dimensional normalised cross section
distributions are similar or not. We stress that this is independent
of other arguments that we provided to argue for the similarity of the
processes.  Even though this is an excellent place to start with, we
would like to find a more quantitative approach to analyse the
issue. In statistics literature several standard procedures exist for
this task. Typically, the similarity of histograms is measured by a
test statistic. The latter provides the quantitative expression of the
distance between the two histograms that are compared. The smaller the
distance the more similar are the compared histograms. There are
several definitions of the test statistics in specialist literature on
statistical methods. In the following we shall concentrate on the
Kolmogorov-Smirnov test (KS test) statistics. The purpose of the
(two-sample) KS test is to look for differences in the shape of two
one-dimensional probability distributions. It is based on comparing
two cumulative distribution functions (CDFs). The KS test reports on
the maximum difference between the two CDFs and calculates a $p$-value
from that and the sample sizes.  If the two tested histograms are
indeed identical then they would have the same CDF. However, in
reality two samples that are compared are randomly taken from their
corresponding probability distributions.  Therefore, even for the two
truly identical histograms the corresponding CDFs will be slightly
different. We can use this fact to test the two distribution equality
by comparing the KS test statistic to $0$. If the latter is
significantly larger than $0$ and close to $1$, then we might conclude
that the distributions are not equal and the two processes considered
are not similar. We begin with the differential cross section
distribution for $pp \to t\bar{t} W^+$ and $pp \to t\bar{t} W^-$ as a
function of the variable $x$, where $x$ for example is $x= p_{T,
\,e_1}, M_{e_1e_2}$. When comparing both histograms, we use the same
number of bins.  We would like to verify the hypothesis that the two
histograms are similar. To this end we calculate the KS test
statistics according to
\begin{equation}
  {\rm KS}_{\rm statistic} = \sup_{x}|F^1_{n_1}(x)-F^2_{n_2}(x)|\,,
  \end{equation}
where $F^1_{n_1}$ and $F^2_{n_2}$ are the CDFs, $n_1$ and $n_2$ are
the sizes of the first and second sample respectively and $\sup$ is
the supremum function.  We assume approximately $2000$ events for
$pp \to t\bar{t} W^+$ and about $1000$ for $pp \to t\bar{t}W^-$, which
correspond to the integrated LHC luminosity of ${\cal L}=500$
fb${}^{-1}$ including a lepton-flavour factor of $8$. After finding
the maximum distance, we use the following condition
%
\begin{figure}[t!]
  \begin{center}
    \includegraphics[width=0.49\textwidth]{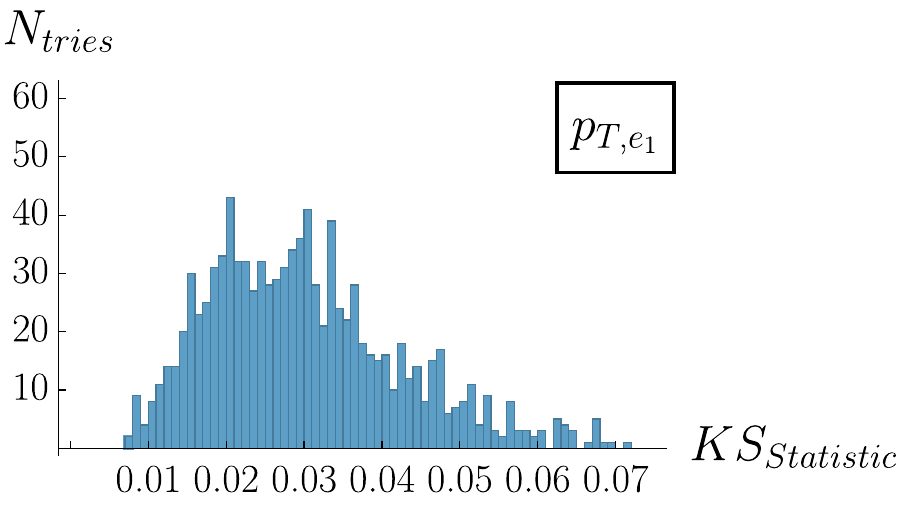}
      \includegraphics[width=0.49\textwidth]{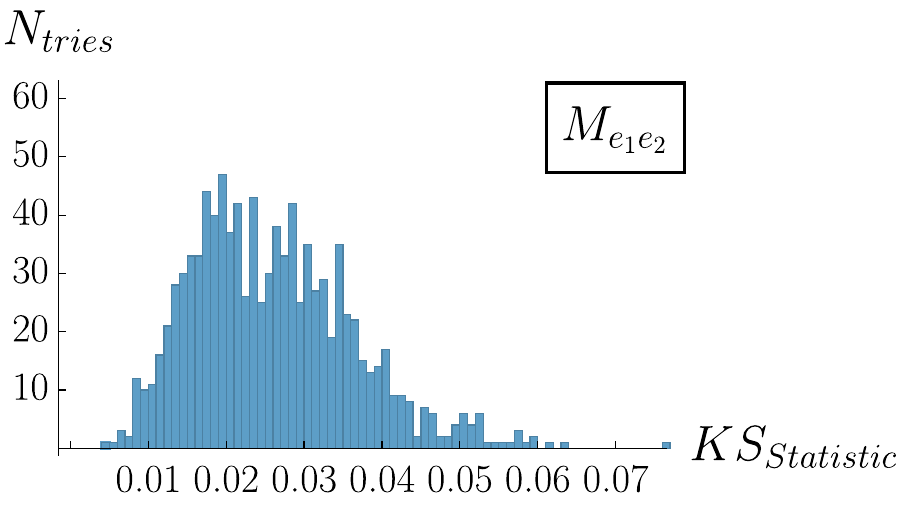}
           \includegraphics[width=0.49\textwidth]{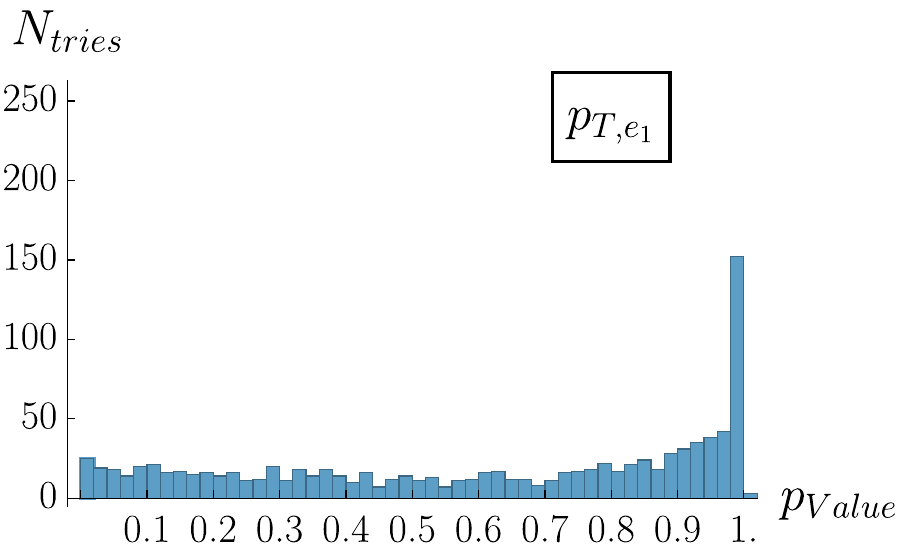}
      \includegraphics[width=0.49\textwidth]{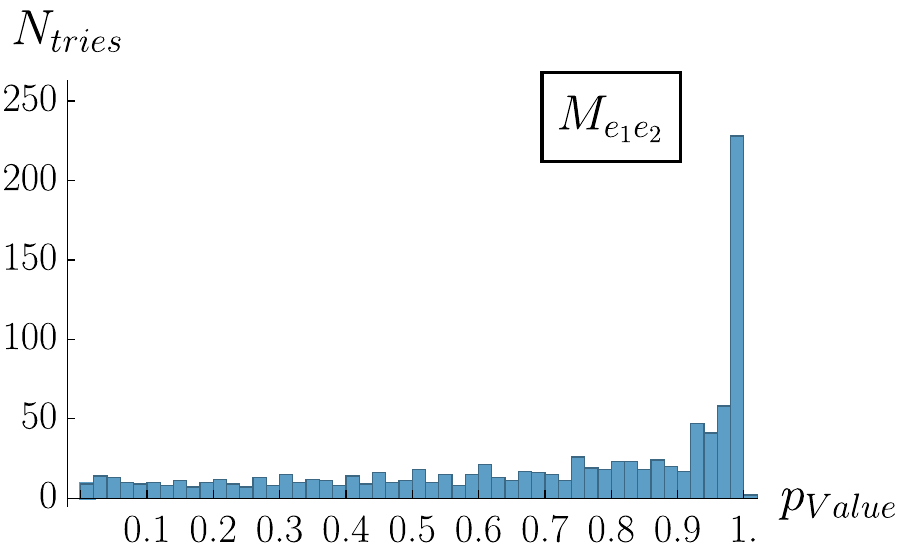}   
      \includegraphics[width=0.49\textwidth]{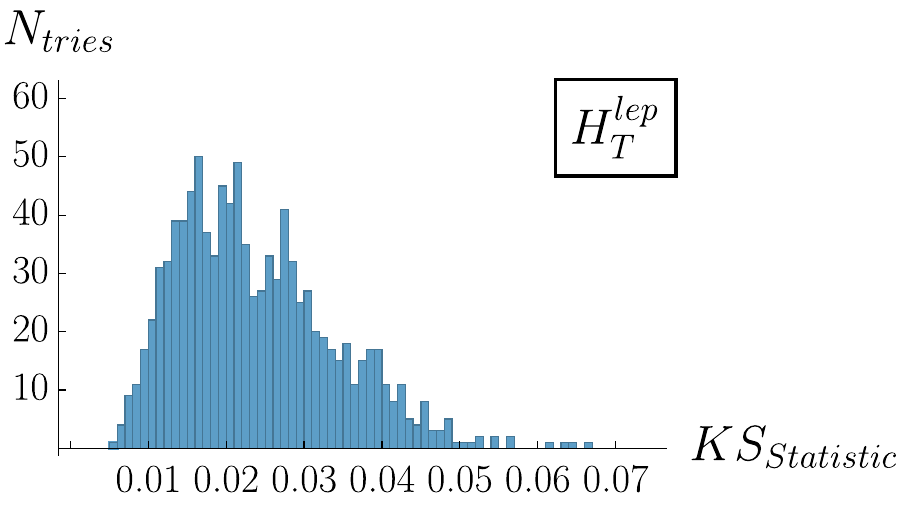}
      \includegraphics[width=0.49\textwidth]{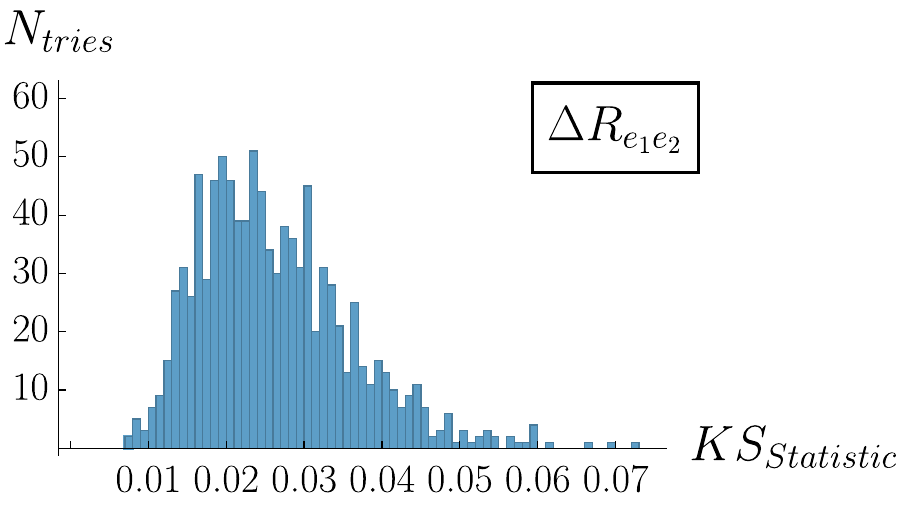}
       \includegraphics[width=0.49\textwidth]{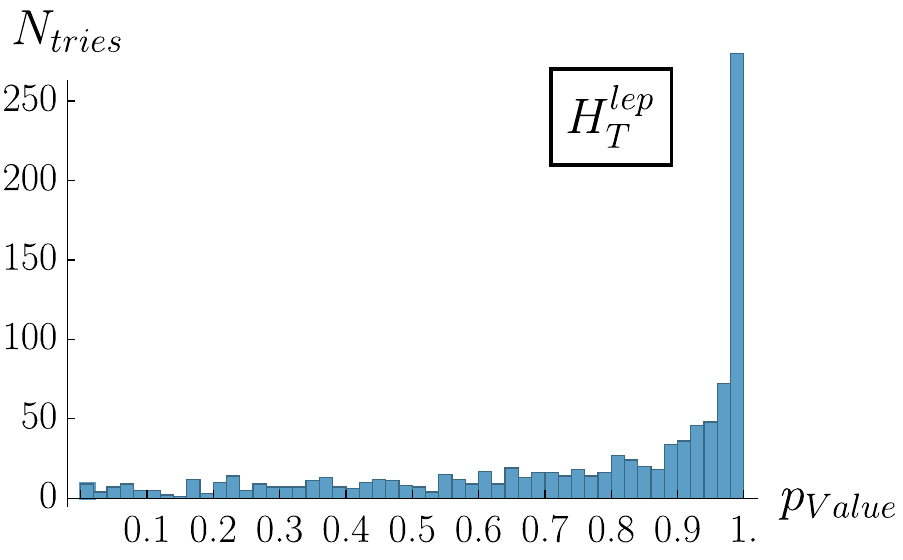}
\includegraphics[width=0.49\textwidth]{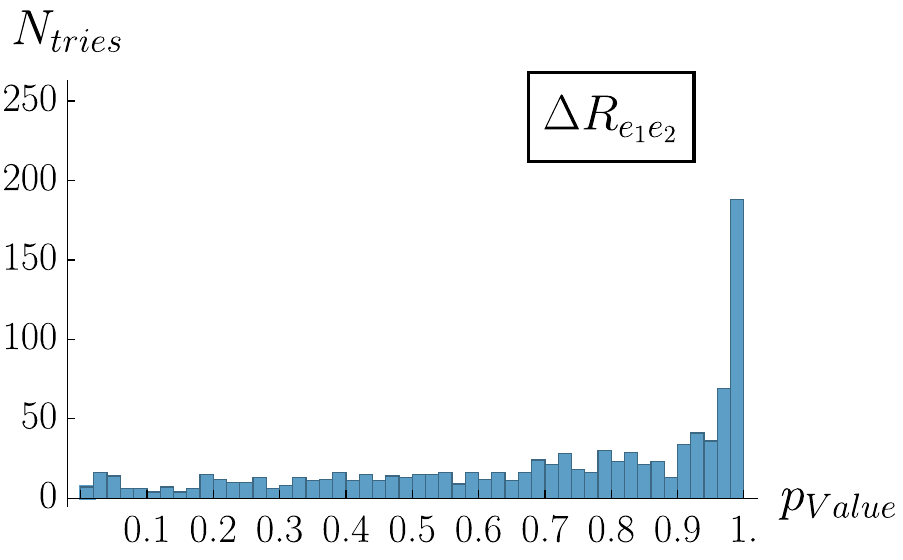}
\end{center}
\caption{\label{fig:statistic} \it  The distribution of the
Kolmogorov-Smirnov test statistic (distance) for the null hypothesis
of equality of the histogram shapes.  NLO QCD differential cross
section distributions for $pp \to t\bar{t}W^+$ and $pp \to
t\bar{t}W^-$ in the multi-lepton final state are employed as a function of
$p_{T,\,e_1}$, $M_{e_1e_2}$, $H_T^{lep}$ and $\Delta R_{e_1e_2}$ for
the LHC with $\sqrt{s}=$13 TeV. Also shown are the distributions of the
corresponding $p$-values.  The total number of $N_{tries}$ is set to
1000.}
\end{figure}
\begin{figure}[t!]
  \begin{center}
    \includegraphics[width=0.49\textwidth]{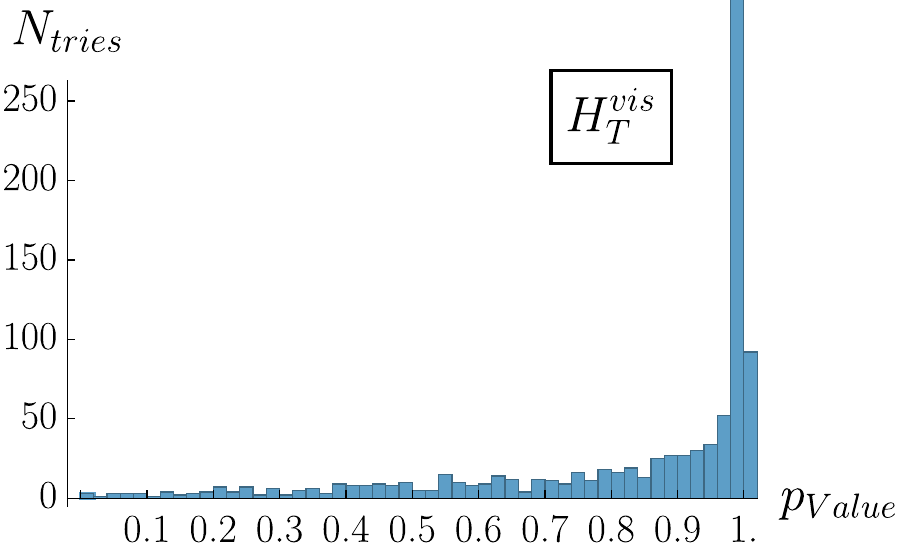}
        \includegraphics[width=0.49\textwidth]{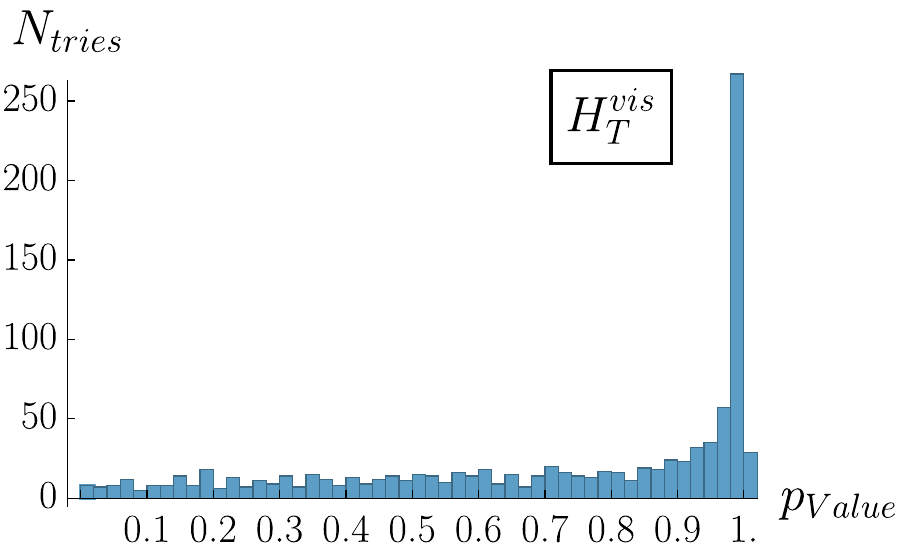}
        \includegraphics[width=0.49\textwidth]{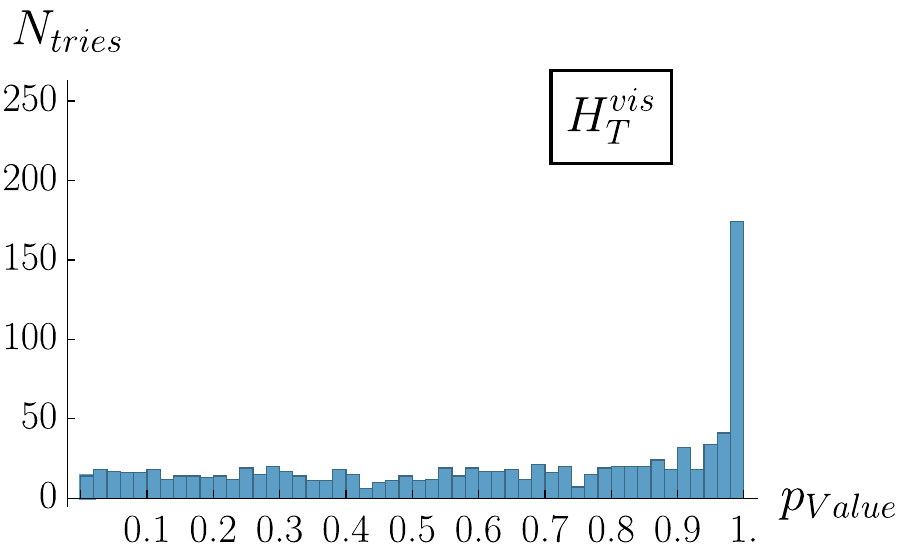}
                \includegraphics[width=0.49\textwidth]{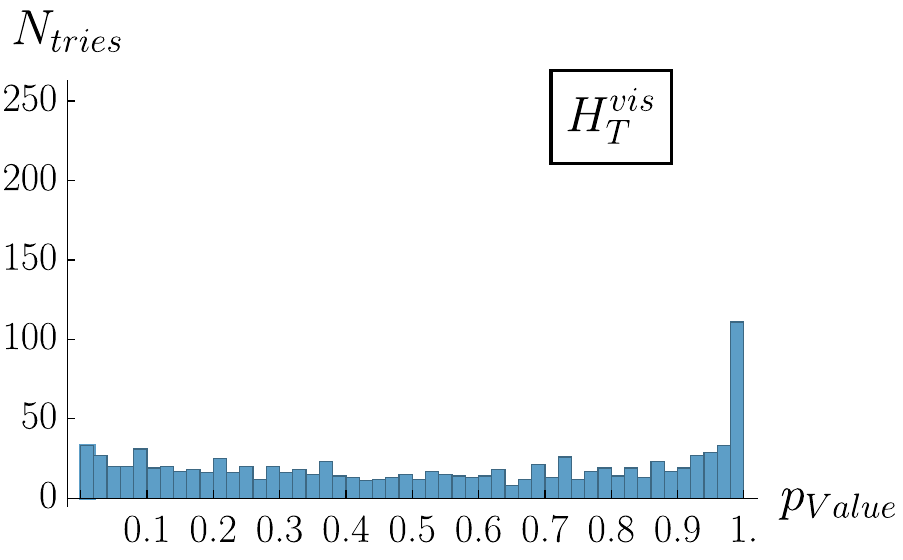}
              \end{center}
\caption{\label{fig:statistic-binning} \it  The distribution of
$p$-values for the Kolmogorov-Smirnov test statistic for the NLO QCD
differential cross section for $pp \to t\bar{t}W^+$ and $pp \to
t\bar{t}W^-$ in the  multi-lepton final state as a function of
$H_T^{vis}$ for the LHC with $\sqrt{s}=13$ TeV. A different number of
bins is assumed for each plot, however,  the
integrated luminosity is kept fixed. Specifically, we use the
following four cases: 5 bins (upper left), 10 bins (upper right), 20
bins (lower left) and 40 bins (lower right). The total number of
$N_{tries}$ is set to 1000.}
\end{figure}
%
%
\begin{equation}
  \sqrt{n}\, \, {\rm KS}_{\rm statistic} \,
 > \lambda(\alpha)\,,
\end{equation}  
where
\begin{equation}
  n= \frac{n_1\cdot n_2}{n_1+n_2}\,,
\end{equation}
with $n_1=2000$, $n_2=1000$ and  $\lambda(\alpha)$  is the threshold
value  that depends on the level of significance $\alpha$. It   can be
found from the following condition
\begin{equation}
  {\cal P} \left( \sqrt{n}\, \, {\rm KS}_{\rm statistic} > \lambda(\alpha)
  \right) = 1-{\cal Q}_{\rm KS}\left(\lambda(\alpha) \right) = \alpha\,,
\end{equation}  
where ${\cal
P}$ denotes probability and ${\cal Q}_{\rm KS}(x)$ stands for the
Kolmogorov-Smirnov distribution.  We reject the hypothesis that the
two distributions are similar if
\begin{equation}
 \sqrt{n}\, \, {\rm KS}_{\rm statistic} > \lambda(\alpha)\,,
\end{equation}  
and accept it when
\begin{equation}
 \sqrt{n}\, \, {\rm KS}_{\rm statistic} \le \lambda(\alpha)\,.
\end{equation}  
We would normally start to question the hypothesis of the similarity
of the histograms only if we find a difference larger than $2\sigma$
(the $p$-value smaller than $0.0455$). If the difference is smaller
than $2\sigma$ (the $p$-value larger than $0.0455$) then we assume
that the two tested distributions are indeed similar.  Results that
differ more  than $3\sigma$ (the $p$-value smaller than
$0.0027$) can be directly translated into having enough evidence to
reject the hypothesis, i.e. saying that there is a real
difference between the two samples that are being studied.  Note that
the KS test does not identify the source of the difference between
histograms. It is a robust way of saying that there is a difference,
however, the origin of such a difference must be identified by other
means.

As the example in Figure~\ref{fig:statistic} we present the
distribution of the KS test statistic for the following NLO QCD
differential cross sections: $p_{T,\,e_1}$, $M_{ee}$, $H_T^{lep}$ and
$\Delta R_{ee}$ for $pp\to t\bar{t}W^+$ and $pp \to t\bar{t}W^-$.  The
total number of tries is set to $N_{tries}=1000$. All KS test
statistic values are distributed within the $0.01-0.07$ range,
i.e. very close to zero, which suggests that $pp\to t\bar{t}W^+$ and
$pp \to t\bar{t}W^-$ are indeed correlated.  Also shown in
Figure~\ref{fig:statistic} are the distributions of the corresponding
$p$-values for the KS test statistic. We can observe that the
$p$-values are mostly distributed in the vicinity of $1$, again
supporting the hypothesis that $pp\to t\bar{t}W^+$ and $pp \to
t\bar{t}W^-$ are highly correlated.  We note here, that similar
results have been obtained for the kinematics of the $b$-jet and for
the $H_T^{vis}$ observable.

We would like to stress at this point, that for higher integrated
luminosity or for an increased number of bins, the sensitivity of the
KS test increases as well. As an example we present in
Figure~\ref{fig:statistic-binning} the distribution of
$p$-values for the KS test statistic for the $H_T^{vis}$ observable
for $pp\to t\bar{t}W^+$ and $pp \to t\bar{t}W^-$. We use four
different values for the number of histogram bins, keeping the number
of total events fixed for both processes. Specifically, we employ 5,
10, 20 and 40 bins respectively.  We can observe that the percentage
of $N_{tries}$  with the $p$-value close to $1$  is getting
lower as the number of bins increases.

We summarise this part by noting, that there are many test
statistics for the comparison of the shapes of two one-dimensional
histograms. The most popular are: the Pearson-$\chi^2$ test, the
Anderson-Darling test or the Cramer-von-Mises test, see e.g.
\cite{Porter:2008mc}. Each of these tests has its pros and cons and it
is not possible to choose the one test that is the best for all
applications. Overall, the more we know about what we really want to
compare and test, the more reliable the test we can choose for our
particular problem. We have examined all the above-mentioned tests and
have decided to use the Kolmogorov-Smirnov test of the equality.  The
two sample KS test assumes continuous distributions. It is one of the
most general nonparametric\footnote{The nonparametric test does not
assume that data points are sampled from the Gaussian distribution or
any other defined distribution for that matter.} tests for comparing
two samples, as it is sensitive to differences in shape of the
empirical cumulative distribution functions of the two samples.  It is
also the most robust test as it tests for any violation of the null
hypothesis.  However, it requires a relatively large number of data
points in each bin.  We further notice, that the KS test is more
sensitive to the regions near the peak of the tested distributions
rather than to their tails. For the latter the Anderson-Darling test
would do a better job.  This observation is very useful in our case as
for many dimensionful observables tails are usually plagued by larger
statistical fluctuations and are, therefore, not really reliable for
such comparisons.

Finally, we stress that the distributions of
observables observed in the two processes are not identical. Hence,
the outcome of the test statistic does depend on the number of
events. With large numbers of events, the test statistic would
obviously discover that the distributions are different. However, we
are here interested in the question of "how similar are the
distributions?" and not "are the distributions
identical?". Quantifying similarity can therefore be done by choosing
a number of events. Had we taken processes with very dissimilar
distributions the $p$-value for the same number of events as chosen
here, would be much smaller, and we would conclude that the
distributions are less similar than in this particular case.



\end{document}